\documentclass[12pt,a4paper,final]{revtex4-1}
\usepackage{amsthm}
\usepackage{amssymb}
\usepackage{multirow}
\usepackage{float}
\usepackage{graphicx}
\usepackage{placeins}
\usepackage{psfrag}
\usepackage{color}
\usepackage{caption2}
\usepackage{flafter}
\usepackage{newlfont}
\usepackage{subeqn}
\usepackage{subfigure}
\usepackage{bbding}
\usepackage{pifont}
\usepackage{wasysym}
\usepackage{dcolumn}
\usepackage{bm}
\usepackage{braket}
\usepackage{wrapfig}

\begin{document}

\title{Continuous-wave solutions and modulational instability in spinor
condensates of positronium }
\author{Ishfaq Ahmad Bhat$^1$, T. Mithun$^{2,3}$, B. A. Malomed$^{4,5}$, and
K. Porsezian$^1$}
\address{$^1$Department of Physics, Pondicherry University, Puducherry
605014, India\\
$^2$Department of Physics, SP Pune University, Pune 411007, India\\
$^3$Center for Theoretical Physics of Complex Systems, Institute for
Basic Science (IBS), Daejeon 34051, Republic of Korea\\
$^4$Department of Physical Electronics, School of Electrical Engineering,
Faculty of Engineering, Tel Aviv University, Tel Aviv 69978, Israel\\
$^5$ITMO University, St. Petersburg 197101, Russia.}
\email{ponzsol@yahoo.com}

\begin{abstract}
We obtain general continuous-wave (CW)\ solutions in the model of a spinor
positronium condensate in the absence of magnetic field. The CW solutions
with both in-phase ($n=0$) and out-of-phase ($n=1$) spin components exist,
with their ranges limited by the total particle density, $\rho $. In the
limit of negligible population exchange between the spin components, the CW
solutions are found to be stable or unstable, depending on the particle
density of the para positronium. Ortho positronium, in the $F=1$ spinor
state, forms a ferromagnetic condensate with stable in-phase CW solutions
only. Subsequent examination of the modulational instability (MI) is carried
out both in the limit case of identical wavenumbers in the spin components, $%
\Delta k\equiv k_{1}-k_{-1}=0$, and in the more general case of $\Delta
k\neq 0$ too. The CW solutions with $n=0$ and $1$ solutions, which are
stable in the case of $\Delta k=0$, are unstable for $\Delta k\neq 0$, for
the natural repulsive sign of the nonlinearities. The total particle
density, $\rho $, in the limit of $\Delta k=0$ is found to have a
significant role for the stability of the condensate, which is determined by
the sign of the self-interaction nonlinearity.
\end{abstract}

\maketitle

\section{Introduction}

Positronium (Ps) is a commonly known bound state of an electron and a
positron, with the total angular momentum $F=0$ or $F=1$. The spin
configurations with $F=0$ and $F=1$, \textit{viz}., the para- $(^{1}S_{0})$
and ortho-Ps $(^{3}S_{1})$, are separated by an energy gap, $\Delta
E=8.44\times 10^{-4}$ eV, and annihilate, respectively, by emitting two or
three gamma-quanta \cite{Rich,Castanares}. This makes positronium a useful
source for $511$ keV gamma-ray lasers \cite{Mills1,Avetissian}. The lifetime
of the para-Ps is $0.125$ ns, while the ortho-Ps lives much longer, $142$
ns. Referring to its longer lifetime, Platzmann and Mills \cite{Mills2} had
proposed, in 1994, a possibility of making ortho-Ps Bose Einstein Condensate
(BEC) in a cold silicon cavity. The condensation temperature being inversely
proportional to the mass of the species under the consideration, and
proportional to its density, for positronium the condensation must occur at
much higher temperature and/or density compared to those for usual bosonic
atoms \cite{Shu,Pethick}. Presently, spin-polarized ensembles of Ps atoms
are available, but with densities at least two orders of magnitude less than
required to form the condensate \cite{Cassidy}. Nevertheless, a specific
method has been recently proposed for the realization of the Ps BEC \cite%
{Shu}. The method relies on original cooling of Ps atoms through interaction
with a cold silica cavity and Ps-Ps two-body collisions, to be followed by
laser cooling.\newline

Atomic condensates of rubidium $(^{87}$Rb$)$ \cite{Anderson}, sodium $(^{23}$%
Na$)$ \cite{Ketterle1}, and lithium $(^{7}$Li$)$ \cite{Bradley} were first
created in magnetic traps. In such BECs, the spin degree of freedom is
frozen due to its coupling to the field, hence the system is defined by a
scalar order parameter. Unlike magnetic traps, optical ones hold all spin
components of a given hyperfine state, without forcing the atoms to align
their spins in a specific direction. Spinor BECs with $2F+1$ spin components
have been experimentally created in optical traps \cite{Ketterle2}-\cite%
{Chapman}, thus providing an opportunity to explore intrinsic spin dynamics
in the condensate. Spinor BECs display various phenomena, such as spin
domains \cite{Ohmi}, skyrmions \cite{Marzlin}, magnetism and its dependence
on scattering lengths \cite{Ho}, interaction-dependent ferromagnetic
phase transitions \cite{Klemm}, modulational instability (MI) in the case of
repulsive nonlinearity \cite{Ueda}-\cite{Robins}, existence of
multicomponent solitons \cite{Wadati}, oscillatory coherent spin mixing \cite%
{Chang,Petit}, etc. Apart from these facts, spinor condensates
including nonlinear spin-exchange interactions find use in magnetometry \cite%
{Higbie} and atom interferometry \cite{Oberthaler}. \newline

The objective of the present work is to examine continuous-wave (CW)
solutions and their MI in the spinor-BEC model
of positronium. This is a relevant aim, as flat CW backgrounds support
various dynamical phenomena, including the modulational instability, dark
and anti-dark solitons, vortices, etc. \cite{new book}. MI is the
exponential growth of the Bogoliubov modes of the miscible condensates \cite%
{Aggarwal}. In scalar BECs, MI solely depends on the sign of the nonlinear
interactions, taking place for the attractive sign. However, in the case of $%
F=1$ \cite{Robins, Pu} and $F=2$ \cite{Mithun} spinor condensates, it was
reported that the MI depends not only on the interaction, but is also
sensitive to phase shifts between components and population ratios. Mixing
the $F=0$ and $F=1$ spin components components in positronium may lead to
new interesting properties. Detailed experimental studies of the MI in self-attractive
BEC were recently reported in Refs. \cite{Randy} and \cite{experiment-2}.

The presentation in the paper is organized as follows. Section II introduces
the theoretical model, based on the Gross-Pitaevskii (GP) equations
governing the mean-field dynamics of the spinor condensate, ignoring effects
produced by the finite lifetime (spontaneous annihilation) of the ortho-Ps.
Section III addresses CW solutions, focusing on conditions for their
existence and stability. Section IV discusses the MI and dispersion
relations in different cases. Section V summarizes the results of the
analysis of the MI in the spinor positronium, and Section VI concludes the
work. Section VI also includes a brief discussion of possible manifestations
of the obtained results in experimental studies of the MI.

\section{The model}

We consider a uniform spinor BEC in an optical trap, without external
magnetic fields. Accordingly, positronium is free to realize any of its four
spin states, labeled as $\ket{p}$ and $\ket{1}$,$\ket{0}$,$\ket{-1}$, with $%
\ket{p}$ representing the para state and $\ket{-1,0,1}$ standing for three
values of the magnetic quantum number $M$ in the ortho-state. Thus,
different states $\ket{F, M}$ of the ortho positronium are designated by
$\ket{1, M}$, while the para state corresponds to $\ket{0, 0}$. The
Hamiltonian of this system is an combination of the non-interacting
single-particle Hamiltonians and the interaction energy ($\hat{H}_{\mathrm{%
int}}$) \cite{Wang, Kawaguchi}:
\begin{equation}
\hat{H}=\int d^{3}\mathbf{r}\sum_{j=0,\pm 1,p}\psi _{j}^{\dagger }\left[
\frac{\mathbf{p}^{2}}{2m}+V_{\mathrm{ext}}(\mathbf{r})+\epsilon _{j}\right]
\psi _{j}+\hat{H}_{\mathrm{int}}~,  \label{eq:1}
\end{equation}%
where $\mathbf{p}$ is the momentum operator, $V_{\mathrm{ext}}(\mathbf{r})$
the trapping potential, and $\epsilon _{j}$ the internal energy of the spin
state \emph{j}. In the mean-field approximation, the interaction Hamiltonian
is (see its detailed derivation in Ref. \cite{Wang}):
\begin{equation}
\hat{H}_{\mathrm{int}}=\frac{1}{2}\int d^{3}\mathbf{r}(\tilde{g}_{0}\rho
^{2}+\tilde{g}_{1}|2\psi _{1}\psi _{-1}-\psi _{0}^{2}+\psi _{p}^{2}|^{2}),
\label{eq:2}
\end{equation}%
where $\rho =\sum_{j=0,\pm 1,p}|\psi _{j}|^{2}$ is the total atomic density,
$\tilde{g}_{0}$ and $\tilde{g}_{1}$ are nonlinearity coefficients, defined
by $\tilde{g}_{0}=4\pi \hbar ^{2}a_{1}/m$ and $\tilde{g}_{1}=\pi \hbar
^{2}(a_{1}-a_{0})/m$, while $a_{0}$ and $a_{1}$ are the scattering lengths
of the para ($S=0$) and ortho ($S=1$) positronium respectively (these
parameters were theoretically calculated in several theoretical works
\cite{Adhikari1}-\cite{Morandi}. An
independent para-ortho scattering length does not appear in Eq.(\ref{eq:2}),
as the interaction between the ortho and para positronium 
vanishes for an odd spin channel. This fact follows
from the presentation of the atomic spin in the form of $%
F_{\mathrm {colliding~pair}}=F_{\mathrm {ortho}}+F_{\mathrm {para}} \equiv 1$
\cite{Ho,Kawaguchi}. \newline

For the BEC confined in a cigar-shaped optical trap, the corresponding
Hamiltonian $\hat{H}$ with $\epsilon_{j} = \epsilon_{o}$ for $%
j=0,\pm 1$ produces a set of coupled one-dimensional GP equations, following
the usual procedure of the dimensional reduction \cite%
{Wang,Devassy,Kasamatsu}:
\begin{subequations}
\label{eq:subeqns3}
\begin{equation}
\emph{i$\hbar$}\dot{\psi_1}=(H_0+\epsilon_o + 2 g_{1}|\psi_{-1}|^2)\psi_1 +
g_{1}\psi^*_{-1}(\psi^2_p-\psi^2_0),  \label{eq:subeq3a}
\end{equation}%
\begin{equation}
\emph{i$\hbar$}\dot{\psi_0}=(H_0+\epsilon_o + g_{1}|\psi_{0}|^2)\psi_0 -
g_{1}\psi^*_{0}(2\psi_1\psi_{-1}+\psi^2_p),  \label{eq:subeq3b}
\end{equation}%
\begin{equation}
\emph{i$\hbar$}\dot{\psi_{-1}}=(H_0+\epsilon_o+ 2
g_{1}|\psi_{1}|^2)\psi_{-1} + g_{1}\psi^*_1(\psi^2_p-\psi^2_0),
\label{eq:subeq3c}
\end{equation}%
\begin{equation}
\emph{i$\hbar$}\dot{\psi_p}=(H_0+\epsilon_p+ g_{1}|\psi_{p}|^2)\psi_p +
g_{1}\psi^*_{p}(2\psi_1\psi_{-1}-\psi^2_0)  \label{eq:subeq3d}
\end{equation}
where, the overdot stands for $\partial /\partial t$, and
\end{subequations}
\begin{equation}
H_{0}=-\frac{\hbar ^{2}}{2m}\frac{\partial ^{2}}{\partial z^{2}}+V_{\mathrm{%
ext}}(z)+g_{0}\rho .  \label{eq:4}
\end{equation}%
The annihilation of the positronium is disregarded here, hence the GP
equations are valid on a limited time scale. The nonlinear coefficients are
now defined by $g_{0}=\frac{2\hbar ^{2}}{ma_{\perp }^{2}}a_{1}$ and $g_{1}=%
\frac{\hbar ^{2}}{2ma_{\perp }^{2}}(a_{0}-a_{1})$ with $a_{\perp }=\sqrt{%
\frac{\hbar }{m\omega _{\perp }}}$ and $\omega _{\perp }$ is the transverse
trap frequency. The terms $\sim g_{1}$ in Eqs.(\ref{eq:subeqns3}) govern the
population exchange between different spin states, while the total density
remains constant. Equations (\ref{eq:subeqns3}) may be normalized to a
dimensionless form by means of rescaling:
\begin{subequations}
\label{eq:subeqns5}
\begin{equation}
t^{\prime }=t\omega _{\perp }  \label{eq:subeq5a}
\end{equation}%
\begin{equation}
z^{\prime }=\frac{z}{a_{\perp }}  \label{eq:subeq5b}
\end{equation}%
\begin{equation}
V_{\mathrm{ext}}^{\prime }(z)=V_{\mathrm{ext}}(z)\hbar \omega _{\perp }
\label{eq:subeq5c}
\end{equation}%
\begin{equation}
\psi _{j}^{\prime }(z)=\frac{\psi _{j}(z)}{\sqrt{2|a_{\perp }|}}
\label{eq:subeq5d}
\end{equation}
The resulting dimensionless GP equations have the same form as above, but
with $\hbar =1$ and $m=1$ and new dimensionless nonlinear coefficients $%
g_{0}^{\prime }=a_{1}/a_{\perp }$ and $g_{1}^{\prime
}=(a_{0}-a_{1})/a_{\perp }$, while the internal energy of the \emph{j}$-$th
state is measured in units of $\hbar \omega_{\perp} $. In the rest of the
paper, the energies are shifted to $\epsilon _{p}=0$ and $\epsilon
_{o}\equiv \epsilon $. These scaled variables are used in figures displayed
below, while equations are written in dimensional units.

\section{Continuous-wave (CW)\ solutions}

We begin by considering the following general CW solutions \cite{Tasgal}:
\end{subequations}
\begin{equation}
\psi _{j}=A_{j}\exp [i(k_{j}z+\theta _{j}-\omega _{j}t)],  \label{eq:6}
\end{equation}%
where $j=0,\pm 1,p$ represent, as above, the respective spin components.
Wave numbers, phase shifts, and frequencies (chemical potentials) $%
k_{j},\theta _{j}$ and $\omega _{j}$ are real, while amplitudes $A_{j}$ are
positive.\newline

The substitution of the CW ansatz (\ref{eq:6}) for $j=0,\pm 1,p$ into
governing equations (\ref{eq:subeqns3}) yields
\begin{subequations}
\label{eq:subeqns7}
\begin{equation}
\hbar \omega _{1}=\frac{\hbar ^{2}k_{1}^{2}}{2m}+g_{0}\rho +\epsilon
+2g_{1}A_{-1}^{2}+(-1)^{n}g_{1}\frac{A_{-1}}{A-1}(A_{p}^{2}-A_{0}^{2}),
\label{eq:subeq7a}
\end{equation}%
\begin{equation}
\hbar \omega _{0}=\frac{\hbar ^{2}k_{0}^{2}}{2m}+g_{0}\rho +\epsilon
+g_{1}A_{0}^{2}-g_{1}\bigg(2(-1)^{n}A_{-1}A_{1}+A_{p}^{2}\bigg),
\label{eq:subeq7b}
\end{equation}%
\begin{equation}
\hbar \omega _{-1}=\frac{\hbar ^{2}k_{-1}^{2}}{2m}+g_{0}\rho +\epsilon
+2g_{1}A_{1}^{2}+(-1)^{n}g_{1}\frac{A_{1}}{A_{-1}}(A_{p}^{2}-A_{0}^{2}),
\label{eq:subeq7c}
\end{equation}%
\begin{equation}
\hbar \omega _{p}=\frac{\hbar ^{2}k_{0}^{2}}{2m}+g_{0}\rho
+g_{1}A_{p}^{2}+g_{1}\bigg(2(-1)^{n}A_{-1}A_{1}-A_{0}^{2}\bigg),
\label{eq:subeq7d}
\end{equation}%
where $n$ and $s$ are integers, which are defined below in Eq.(\ref%
{eq:subeq8c}), and the following relations between the wave numbers,
frequencies, and phase shifts of the different components must hold:
\end{subequations}
\begin{subequations}
\label{eq:subeqns8}
\begin{equation}
k_{p}=k_{0}=\frac{1}{2}(k_{1}+k_{-1}),  \label{eq:subeq8a}
\end{equation}%
\begin{equation}
\omega _{p}=\omega _{0}=\frac{1}{2}(\omega _{1}+\omega _{-1}),
\label{eq:subeq8b}
\end{equation}%
\begin{equation}
\theta _{p}+s\pi =\theta _{0}=\frac{1}{2}(\theta _{1}+\theta _{-1}+n\pi ),
\label{eq:subeq8c}
\end{equation}

For the compatibility of Eqs.(\ref{eq:subeqns8}) with the other equations,
amplitude $A_{0}$ of the $\psi _{0}$ component must satisfy the condition
\end{subequations}
\begin{equation}
A_{0}^{2}=\frac{1}{2}\Bigg[\rho -\big(A_{1}-(-1)^{n}A_{-1}\big)^{2}+\frac{%
2\gamma (-1)^{n}A_{1}A_{-1}}{(A_{1}^{2}+A_{-1}^{2})}\Bigg],  \label{eq:9}
\end{equation}%
where
\begin{equation}
\gamma \equiv \frac{\hbar ^{2}(k_{1}-k_{-1})^{2}}{8mg_{1}}+\frac{\epsilon }{%
2g_{1}}.  \label{eq:10}
\end{equation}

The second term on the right-hand side of Eq. (\ref{eq:10}) displays the
competition between the internal energy difference $\epsilon $ and the
spin-mixing interaction with the effective 1D strength $g_{1}$. The ortho to
para interconversion is substantial for $g_{1}\gg \epsilon $. \newline

\begin{figure}[tbp]
\begin{center}
\subfigure[]{\includegraphics[width=0.33\textwidth]{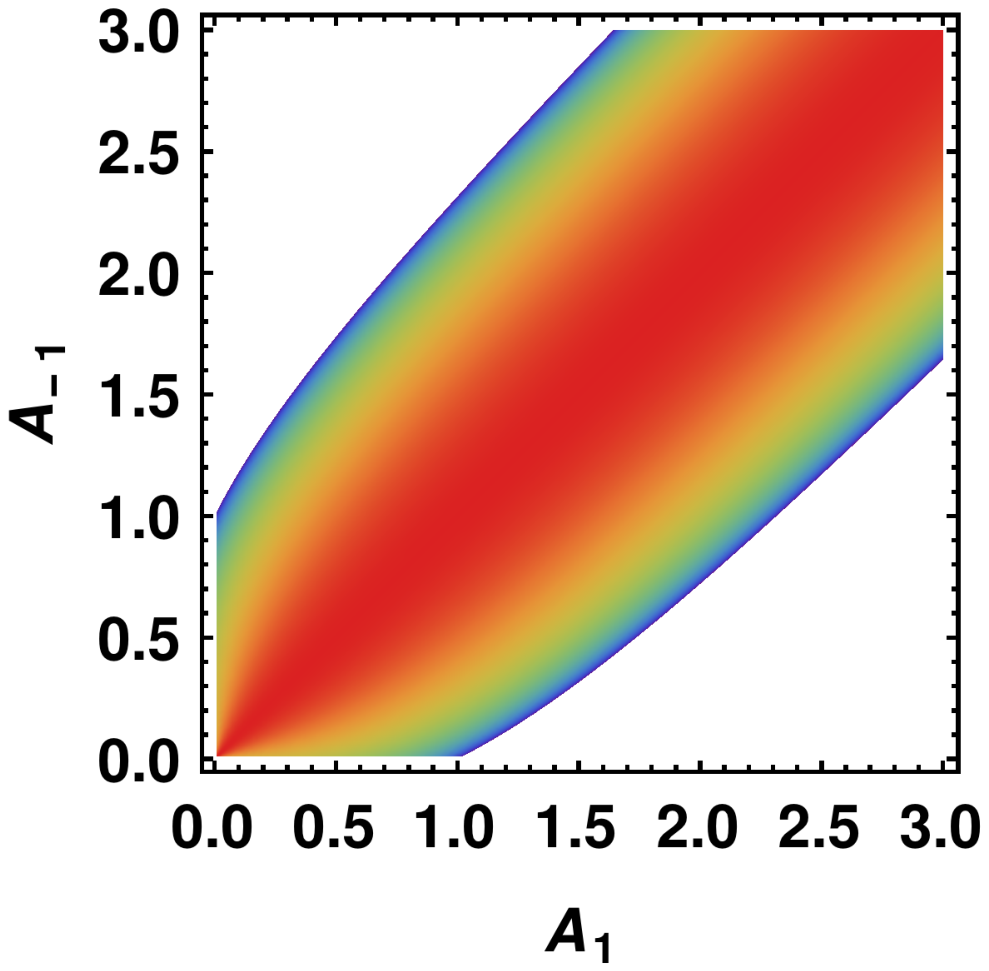}} %
\includegraphics[width=0.7 cm, height=5.1 cm]{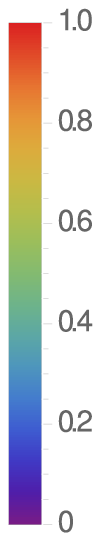} %
\subfigure[]{\includegraphics[width=0.33\textwidth]{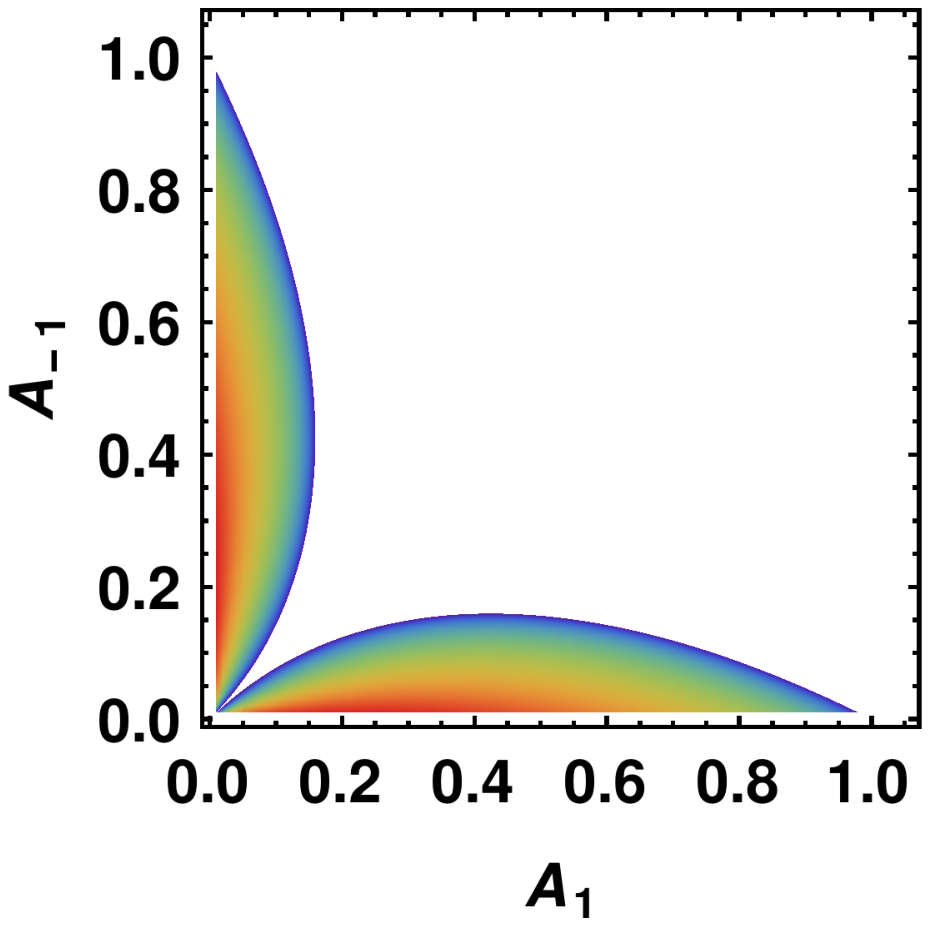}} %
\includegraphics[width=0.7 cm, height=5.2 cm]{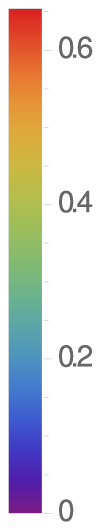}\\[0pt]
\end{center}
\caption{(Color online ) Existence ranges of CW solutions in positronium for
$\protect\rho =1$ and $\protect\gamma =1$ $(g_{1}^{\prime }>0)$. Different
colors represent the corresponding amplitude $A_{0}$ in the CW solution.
\newline
\textbf{(a)}: The existence ranges for even $n$, represented by $n=0$.
\newline
\textbf{(b)}: The existence ranges for odd $n$, represented by $n=1.$}
\label{Fig.1}
\end{figure}

From the CW ansatz it follows that the condition of $A_{0}$ being positive
makes the left-hand side of Eq. (\ref{eq:9}) real and non-negative, thus
giving a criterion for the existence of the CW solutions. Thus, for the
ground state of positronium with $g_{1}>0$, CW solutions for both even $n$
and odd $n$ exist, but with different existence ranges, as shown in Figure %
\ref{Fig.1}. These ranges for different CW solutions are found to depend on
the total number density, $\rho $, in addition to the magnitude and sign of $%
g_{1}$, which, in turn, can be tuned by means of the Feshbach-resonance
techniques \cite{Blatt}. In particular, the CW solutions with even $n$
(represented by $n=0$) exist if
\begin{equation}
\gamma \geq \frac{(A_{1}^{2}+A_{-1}^{2})}{2A_{1}A_{-1}}\big[%
(A_{1}-A_{-1})^{2}-\rho \big].  \label{eq:11}
\end{equation}%
There are CW solutions with odd $n$ (represented by $n=1$) if
\begin{equation}
\gamma \leq \frac{(A_{1}^{2}+A_{-1}^{2})}{2A_{1}A_{-1}}\big[\rho
-(A_{1}+A_{-1})^{2}\big].  \label{eq:12}
\end{equation}%
Clearly, values which $A_{1}$ and $A_{-1}$ must assume for the fulfilment of
the conditions for the existence of the respective CW solutions are limited
by the total density, $\rho $.\newline

\begin{figure}[tbp]
\begin{center}
\subfigure[]{\includegraphics[width=0.33\textwidth]{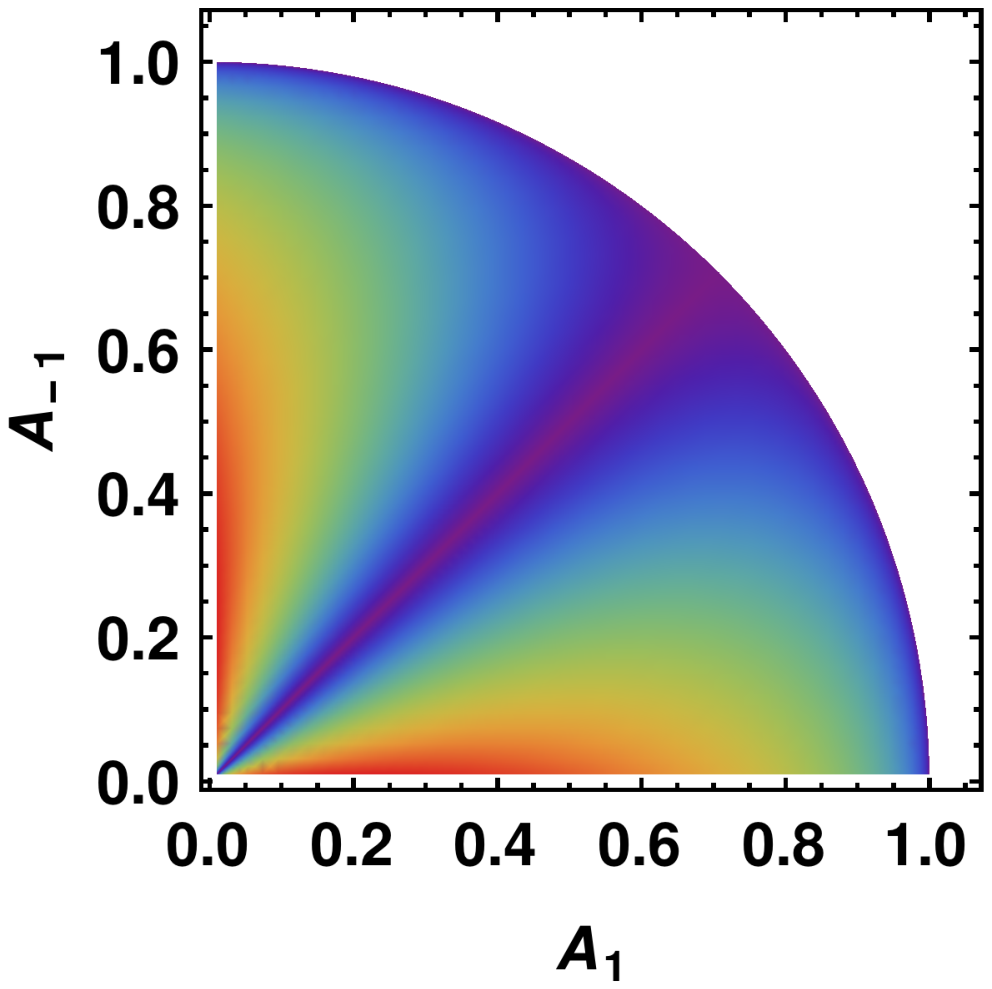}} %
\includegraphics[width=0.7 cm, height=5.2 cm]{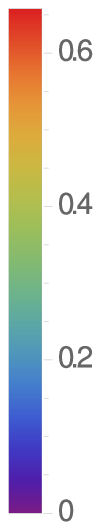} %
\subfigure[]{\includegraphics[width=0.33\textwidth]{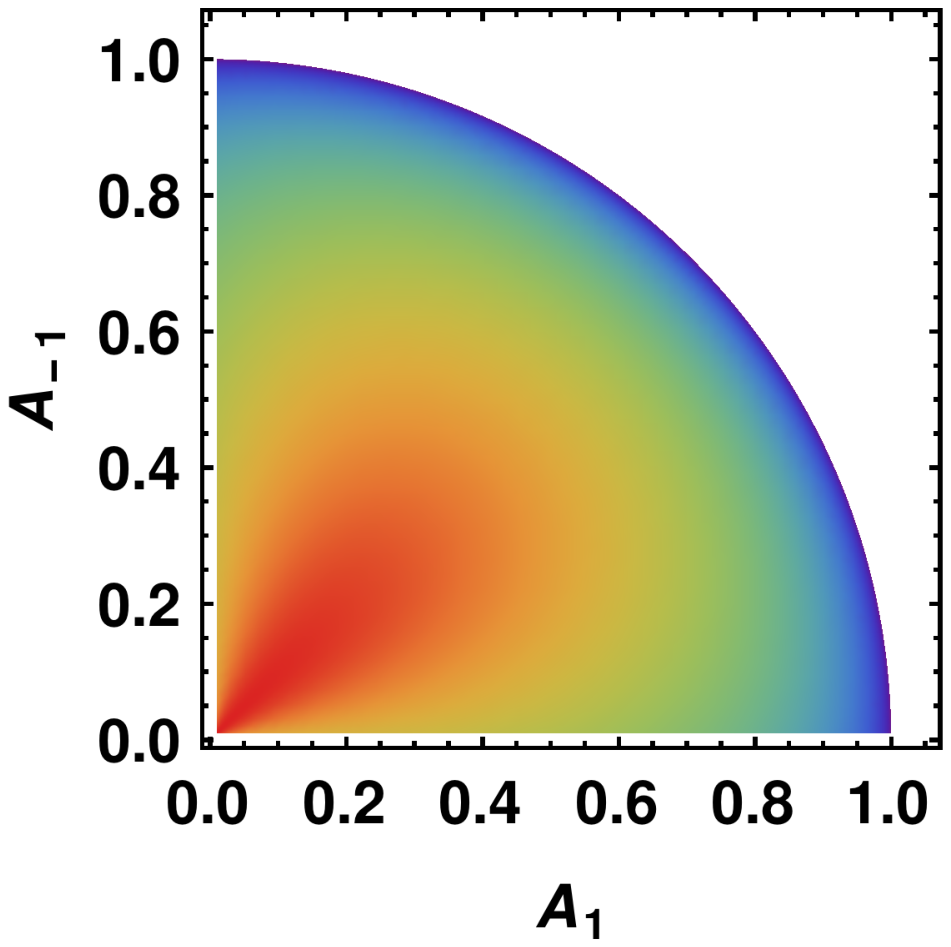}} %
\includegraphics[width=0.7 cm, height=5.25 cm]{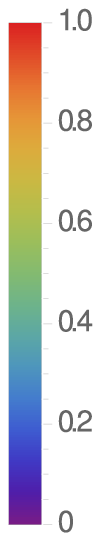}\\[0pt]
\end{center}
\caption{(Color online) Existence ranges of CW solutions in positronium for $%
\protect\rho =1$ and $\protect\gamma =-1$ $(g_{1}^{\prime }<0)$. Different
colors represent the corresponding amplitude $A_{0}$ in the CW solution.
\newline
\textbf{(a)}: The existence ranges for even $n$, represented by $n=0$.
\newline
\textbf{(b)}: The existence ranges for odd $n$, represented by $n=1.$}
\label{Fig.2}
\end{figure}

The case of $g_{1}<0$, which may be realized, as mentioned above, with the
help of the Feshbach resonance, modifies the existence ranges of the CW
solutions, as shown in Fig.\ref{Fig.2}. There are CW solutions for even $n$
when
\begin{equation}
\gamma \leq \frac{(A_{1}^{2}+A_{-1}^{2})}{2A_{1}A_{-1}}\big[\rho
-(A_{1}-A_{-1})^{2}\big].  \label{eq:13}
\end{equation}%
However, for odd $n$, CW solutions exist at
\begin{equation}
\gamma \leq \frac{(A_{1}^{2}+A_{-1}^{2})}{2A_{1}A_{-1}}\big[%
(A_{1}+A_{-1})^{2}-\rho \big].  \label{eq:14}
\end{equation}%
%
%
%
%

In accordance with Eq. (\ref{eq:subeq8b}), and combining Eqs. (\ref%
{eq:subeq7b}) and (\ref{eq:subeq7d}) in the limit of $g_{1}\gg \epsilon $,
we arrive at an equation similar to one obtained in Ref. \cite{Wang}, which
relates the ortho and para populations of the condensate:
\begin{equation}
A_{p}^{2}=A_{0}^{2}-2(-1)^{n} A_{1}^{2}A_{-1}^{2}.  \label{eq:15}
\end{equation}%
The ground state is found to have the maximum para population for odd $(n=1)$
CW solutions, which, for equal amplitudes of the $M=\pm 1$ components,
becomes equal to the density of the ortho component:
\begin{equation}
\rho _{p}=\rho _{o}-2\bigg(A_{1}+(-1)^{n}A_{-1}\bigg)^{2}  \label{eq:16}
\end{equation}%
where $\rho _{p}=A_{p}^{2}$ and $\rho _{o}=A_{1}^{2}+A_{0}^{2}+A_{-1}^{2}$
are densities of the para and ortho components, respectively. The ortho
sector of the condensate for equal densities of ortho and para components is
equivalent to a polar state of the $F=1$ condensate \cite{Ho,Wang}. \newline

For the stability analysis of the CW solutions, we neglect the ortho-to-para
interconversion, assuming $g_{1}\ll \epsilon $. Except for the wavenumbers $%
k_{j}$ with $j=0,\pm 1,p$, CW parameters, such as the amplitudes $A_{j}$ and
the phases $\theta _{j}$, are made time-dependent. The linearization of Eqs.
(\ref{eq:subeqns3}), taking into account the conservation of total density, $%
\rho =\sum_{j=0,\pm 1,p}A_{j}^{2}$, and magnetization, $\left(
A_{-1}^{2}-A_{1}^{2}\right) $, and subsequent linearization with respect to
amplitude and phase perturbations (the latter reduce to $\delta \theta (t)=%
\big(\theta _{1}(t)+\theta _{-1}(t)-2\theta _{0}(t)\big)$) shows that the CW
solutions are either stable or unstable, depending on the density of the
para-component relative to that of the one with $M=0$ in the
ortho-component. Oscillation eigenfrequencies of infinitesimal perturbations
are then given by
\begin{equation}
\omega _{n=0,1}^{2} =2\bigg(\frac{g_{1}A_{0}}{\hbar }\bigg)^{2}\bigg[%
(A_{1}+(-1)^{n}A_{-1})^{2}+ \\
\frac{(A_{-1}^{2}-A_{1}^{2})^{2}(A_{0}^{2}-A_{p}^{2})}{4A_{1}^{2}A_{-1}^{2}}%
\bigg].   \label{eq:17}
\end{equation}%
These frequencies are real, and hence the CW solutions are stable, if $%
A_{0}^{2}>A_{p}^{2}$. If $A_{0}^{2}<A_{p}^{2}$, then, for $n=0$, the CW
solutions are stable if
\begin{equation}
A_{p}^{2}<A_{0}^{2}+\bigg(\frac{2A_{1}A_{-1}}{A_{-1}-A_{1}}\bigg)^{2}.
\label{eq:18}
\end{equation}%
Likewise, for $n=1$ and $A_{1}\neq A_{-1}$, the CW solutions are stable if
\begin{equation}
A_{p}^{2}<A_{0}^{2}+\bigg(\frac{2A_{1}A_{-1}}{A_{-1}+A_{1}}\bigg)^{2}.
\label{eq:19}
\end{equation}%
The respective spin oscillations with real frequency $\omega $ represent the
exchange of populations between the different spin states \cite%
{Zhang,Bigelow}. \newline

In the limit of $\Psi _{p}=0$, we are left with the ortho-Ps condensate
only, the system being equivalent to the $F=1$ spinor condensate. The spinor
ortho-Ps condensate with $g_{1}>0$ supports solely even $(n=0)$ CW solutions
for all values of $A_{1}$ and $A_{-1}$ with
\begin{equation}
A_{0}^{2}=2(-1)^{n}A_{1}A_{-1}\bigg(1+\frac{\frac{\hbar ^{2}}{2mg_{1}}(\frac{%
\Delta k}{2})^{2}}{\big(A_{1}+(-1)^{n}A_{-1}\big)^{2}}\bigg),  \label{eq:20}
\end{equation}%
where $\Delta k\equiv k_{1}-k_{-1}$. This is a peculiarity of the
ferromagnetic states of the $F=1$ spinor condensate \cite{Tasgal}. Thus, the
CW states of the ortho-Pt only represent a ferromagnetic condensate. The
conditions for the existence of the possible CW solutions in spinor Ps are
summarized in Table I. \newline

The stability analysis of the CW solutions by means of the linearization
with respect to the amplitude and the phase perturbations show that the CW
solutions are stable against the infinitesimal perturbations, with real
perturbation eigenfrequencies
\begin{equation}
\omega _{n=0}^{2}=\bigg(\frac{g_{1}A_{0}^{2}}{\hbar }\bigg)^{2}{\bigg[\frac{%
(A_{-1}^{2}-A_{1}^{2})^{2}}{A_{1}^{2}A_{-1}^{2}}+\frac{8(A_{1}+A_{-1})^{2}}{%
A_{0}^{2}}\bigg].}  \label{eq:21}
\end{equation}%
These are frequencies of the coherent spin mixing, which account for the
exchange of populations between the different spin states of the ortho-Ps.

\begin{center}
\begin{table}[tph]
\caption{Possible CW solutions with respective conditions for their
existence in the spinor condensate of positronium for different values of $%
\protect\gamma=\frac{1}{8mg_1}[\hbar^2(k_{1}-k_{-1})^2 + 4 m\protect\epsilon%
] $. $A_{1}$ and $A_{-1}$ are allowed amplitudes of the respective spin
components, and $\protect\rho $ is the total density.}
\label{tab1}
{\footnotesize \centering
\begin{tabular}{p{2cm}p{1.5cm}p{5cm}p{5cm}c}
\hline\hline
$\gamma $ & $\psi_p$ & CW $(n=0)$ & CW $(n=1)$ &  \\[0.5ex] \hline
\multirow{2}{*}{$ \ge 0$} & $0$ & $A_1,A_{-1}\in R\ge0$ & No Solutions &  \\
&  &  &  &  \\
& $\ne0$ & $\gamma \ge \frac{(A^2_1+A^2_{-1})}{2 A_1A_{-1}}%
[(A_1-A_{-1})^2-\rho]$ & $\gamma \le \frac{(A^2_1+A^2_{-1})}{2 A_1A_{-1}}%
[\rho-(A_1+A_{-1})^2]$ &
$$
\\
&  &  &  &  \\ \hline
&  &  &  &  \\
$
< 0$ & $\ne0$ & $\gamma \le \frac{(A^2_1+A^2_{-1})}{2 A_1A_{-1}}%
[\rho-(A_1-A_{-1})^2]$ & $\gamma \le \frac{(A^2_1+A^2_{-1})}{2 A_1A_{-1}}%
[(A_1+A_{-1})^2-\rho]$ &  \\[1ex] \hline\hline
\end{tabular}
}
\end{table}
\end{center}


\section{Modulational instability (MI)}

For the investigation of MI in the Ps BEC, described by the dynamical
equations (\ref{eq:subeqns3}), we start with addressing small perturbations $%
\delta \phi (z,t)$ added to the CW solutions \cite%
{Li,Robins,Aggarwal,Tasgal,Abad,Ohmi1}:
\begin{equation}
\phi _{j}=[A_{j}+\delta \phi _{j}(z,t)]\exp [i(k_{j}z+\theta _{j}-\omega
_{j}t)]  \label{eq:22}
\end{equation}%
where $j$ defines the spin index $0$,$\pm 1$ and $p$. \newline

Assuming the system size greater than healing length, which determines the
characteristic length scale for MI, we assume the perturbations to be in the
form of plane waves,
\begin{equation}
\delta \phi _{j}(z,t)=\lambda _{j}\ \cos (kz-\omega t)+i\ \eta _{j}\sin
(kz-\omega t),  \label{eq:23}
\end{equation}%
where $\lambda _{j}$ and $\eta _{j}$ are perturbation amplitudes, while $k$
and $\omega $ are the wavenumber and the (generally complex) frequency,
respectively.\newline

The substitution of Eqs. (\ref{eq:22}) and (\ref{eq:23}) in Eq.(\ref%
{eq:subeqns3}), for the four spin indices, gives a set of eight homogeneous
equations, with respect to $\lambda _{j}$ and $\eta _{j}$, in the matrix
form:
\begin{equation}
\mathcal{M}\Psi =\Big(\lbrack -\hbar \omega +\frac{\hbar ^{2}k}{2m}%
(k_{1}+k_{-1})]\check{\mathbf{I}}+\mathbf{X}+\mathbf{Y}+\mathbf{Z}\Big)\Psi
=0  \label{eq:24}
\end{equation}%
where $\Psi =(\eta _{1},\lambda _{1},\eta _{0},\lambda _{0},\eta
_{-1},\lambda _{-1},\eta _{p},\lambda _{p})^{T}$, $\check{\mathbf{I}}$ is
the usual unit matrix and \newline
\begin{subequations}
\label{eq:subeqns28}
\begin{eqnarray}
X=\frac{\hbar^2}{2 m}k\left(
\begin{array}{cccccccc}
\Delta k & k & 0 & 0 & 0 & 0 & 0 & 0 \\
k & \Delta k & 0 & 0 & 0 & 0 & 0 & 0 \\
0 & 0 & 0 & k & 0 & 0 & 0 & 0 \\
0 & 0 & k & 0 & 0 & 0 & 0 & 0 \\
0 & 0 & 0 & 0 & -\Delta k & k & 0 & 0 \\
0 & 0 & 0 & 0 & k & -\Delta k & 0 & 0 \\
0 & 0 & 0 & 0 & 0 & 0 & 0 & k \\
0 & 0 & 0 & 0 & 0 & 0 & k & 0 \\
&  &  &  &  &  &  &
\end{array}
\right)  \label{eq:subeq25a}
\end{eqnarray}%
\bigskip
\begin{eqnarray}
 \resizebox{.90\columnwidth}{!}{$ Y=2\left( \begin{array}{cccccccc} 0 & 0
& 0 & 0 & 0 & 0 & 0 & 0 \\ g_0A^2_1 & 0 & g_0A_0A_1 & 0 &
(g_0+2g_1)A_1A_{-1}& 0 & g_0A_1A_p & 0 \\ 0 & 0 & 0 & g_1 A^2_p & 0 & 0 & 0
& -g_1 A_0A_P \\ g_0A_1A_0 & 0 & (g_0+g_1)A^2_0 & 0 & g_0 A_0A_{-1}& 0 &
(g_0-g_1)A_0A_p & 0 \\ 0 & 0 & 0 & 0 & 0 & 0 & 0 & 0 \\ (g_0+2g_1)A_1A_{-1}&
0 & g_0A_0A_{-1} & 0 & g_0A^2_{-1} & 0 & g_0A_{-1}A_p & 0 \\ 0 & 0 & 0 &
-g_1A_0A_p & 0 & 0 & 0 & g_1A^2_0 \\ g_0 A_1A_p & 0 & (g_0-g_1)A_0A_p& 0 &
g_0A_1A_p & 0 & (g_0+g_1)A^2_p & 0 \\ \end{array} \right)
\label{eq:subeq25b}$}
\end{eqnarray}%
\bigskip
\begin{eqnarray}
 \resizebox{.90\columnwidth}{!}{$ Z=(-1)^n g_{1}\left(
\begin{array}{cccccccc} 0 & \frac{A_{-1}}{A_1}(A^2_0-A^2_p) & 0 &
-2A_0A_{-1} & 0 & (A^2_0-A^2_p) & 0 & 2A_{-1}A_p \\
\frac{A_{-1}}{A_1}(A^2_0-A^2_p) & 0 & -2A_0A_{-1} & 0 & -(A^2_0-A^2_p) & 0 &
2A_{-1}A_p & 0 \\ 0 & -2A_0A_{-1} & 0 & 4A_1A_{-1} & 0 & -2A_0A_1 & 0 & 0 \\
-2A_0A_{-1}& 0 & 0 & 0 & -2A_0A_1 & 0 & 0&0 \\ 0 & (A^2_0-A^2_p) & 0 &
-2A_0A_1 & 0 & \frac{A_1}{A_{-1}}(A^2_0-A^2_p) & 0 & 2A_1A_p \\
-(A^2_0-A^2_p)& 0 & -2A_1A_0 & 0 & \frac{A_1}{A_{-1}}(A^2_0-A^2_p) & 0 &
2A_1A_P & 0 \\ 0 & 2A_{-1}A_p & 0 & 0 & 0 & 2A_1A_p & 0 & -4A_1A_{-1} \\
2A_{-1}A_p & 0 & 0 & 0 & 2A_1A_p & 0 & 0 & 0 \\ \end{array} \right)
\label{eq:subeq25c}$}
\end{eqnarray}
\bigskip 
\newline
Frequency $\omega $ of the perturbations, the average wave number, $%
k_{0}=k_{p}=\frac{(k_{1}+k_{-1})}{2}$, and the wavenumber difference, $%
\Delta k$, are diagonal entries of the stability matrix. In the most general
case, the wavenumber difference assumes a nonzero value, $\Delta k\neq 0$,
while $\Delta k=0$ corresponds to the limit case, in which all the
wavenumbers are equal. The off diagonal elements depend upon various CW
parameters either explicitly or implicitly via $A_{0}$ [see Eq. (\ref{eq:9}%
)] and $A_{p}$ [see Eq. (\ref{eq:15})].\newline

Equation (\ref{eq:24}) represents an eigenvalue problem for matrix $\mathcal{%
M}$. The solution to the problem aims to identify all possible eigenvalues $%
\hbar \omega $ and the corresponding eigenvectors $\big(\eta _{j},\lambda
_{j};\ j=1,2,\cdots ,8\big)$. In this study, we produce solutions for
eigenvalues $\hbar \omega $ for both cases of $\Delta k=0$ and $\Delta k\neq
0$. Complex eigenvalues $\hbar \omega $, if obtained by solving the
dispersion equation, $\mathrm{det}(\mathcal{M})=0$, for some positive values
of $k^{2}$, render the spinor condensate modulationally unstable. In the
present case this is the eighth-order equation with respect to $\hbar \omega
$, with the dependence on the CW parameters implied in the coefficients, $%
C_{k^{\alpha },\omega ^{\beta }}^{n=0,1}$, while $\alpha $ and $\beta $ are
integer numbers:\newline
\end{subequations}
\begin{eqnarray}
(\hbar \omega )^{8}+(\hbar \omega )^{7}C_{k,\omega ^{7}}^{n=0,1}k+(\hbar
\omega )^{6}\sum_{j=1}^{3}C_{k^{2j-2},\omega ^{6}}^{n=0,1}k^{2j-2}  \nonumber
\\
+(\hbar \omega )^{5}\sum_{j=1}^{3}C_{k^{2j-1},\omega
^{5}}^{n=0,1}k^{2j-1}+(\hbar \omega )^{4}\sum_{j=1}^{4}C_{k^{2j},\omega
^{4}}^{n=0,1}k^{2j}  \nonumber \\
+(\hbar \omega )^{3}\sum_{j=1}^{4}C_{k^{2j+1},\omega
^{3}}^{n=0,1}k^{2j+1}+(\hbar \omega )^{2}\sum_{j=1}^{5}C_{k^{2j+2},\omega
^{2}}^{n=0,1}k^{2j+2}  \nonumber \\
+(\hbar \omega )\sum_{j=1}^{5}C_{k^{2j+3},\omega
^{3}}^{n=0,1}k^{2j+3}+\sum_{j=1}^{6}C_{k^{2j+4},\omega
^{0}}^{n=0,1}k^{2j+4}=0  \label{eq:26}
\end{eqnarray}%
The coefficients $C_{k^{\alpha },\omega ^{\beta }}^{n=0,1}=\frac{1}{\alpha
!\,\beta !}\frac{\partial ^{\alpha }}{\partial k^{\alpha }}\frac{\partial
^{\beta }}{\partial \,\omega ^{\beta }}\mid \mathcal{M}\mid $ represent a
blend of nonlinearity coefficients and various CW parameters, being too
cumbersome to be included in the main text. They are explicitly displayed in
Supplement 1. \newline
Equation (\ref{eq:26}) can be solved analytically only in a few limit cases,
as shown below. \newline

For the ortho Ps condensate $(\psi _{p}=0)$ for which only the in-phase CW
solutions (with $n=0$) exist, the sixth-order dispersion equation is%
\begin{eqnarray}
(\hbar \omega )^{6}+(\hbar \omega )^{5}C_{k,\omega ^{5}}^{n=0}k+(\hbar
\omega )^{4}\sum_{j=1}^{3}C_{k^{2j-2},\omega ^{4}}^{n=0}k^{2j-2}  \nonumber
\\
+(\hbar \omega )^{3}\sum_{j=1}^{3}C_{k^{2j-1},\omega
^{3}}^{n=0}k^{2j-1}+(\hbar \omega )^{2}\sum_{j=1}^{4}C_{k^{2j},\omega
^{2}}^{n=0}k^{2j}  \nonumber \\
+(\hbar \omega )\sum_{j=1}^{4}C_{k^{2j+1},\omega
}^{n=0}k^{2j+1}+\sum_{j=1}^{5}C_{k^{2j+2},\omega ^{0}}^{n=0}k^{2j+2}=0.
\label{eq:27}
\end{eqnarray}%
Expressions for coefficients, $C_{k^{\alpha },\omega ^{\beta }}^{n=0}$ are
again too cumbersome for the main text, being displayed in Supplement 2.
\newline

In the case of zero fields with $M=\pm 1$, the dynamics of the system is
governed by a pair of coupled GP equations, and the resulting characteristic
polynomial of the fourth order is analytically soluble:
\begin{eqnarray}
(\hbar \omega )^{4}+(\hbar \omega )^{3}C_{k,\omega ^{3}}k+(\hbar \omega
)^{2}\sum_{j=1}^{3}C_{k^{2j-2},\omega ^{2}}k^{2j-2}  \nonumber \\
+(\hbar \omega )\sum_{j=1}^{3}C_{k^{2j-1},\omega
}k^{2j-1}+\sum_{j=1}^{4}C_{k^{2j},\omega ^{0}}k^{2j}=0.  \label{eq:28}
\end{eqnarray}%
Here coefficients $C_{k^{\alpha },\omega ^{\beta }}$ being functions of the
nonlinearity coefficients $g_{0}$ and $g_{1}$ are included in Supplement 3.
These coupled GP equations have been extensively studied in\ terms of optics
\cite{Yu, Yang} and BEC \cite{Kasamatsu1, Ishfaq}.

\section{Results and Discussions}

 Below, we discuss cases for which the dispersion
relations of different orders obtained above are solved for the
perturbation frequency in the limiting case of equal wavenumbers, $\Delta k = 0$,
when simple analytical solutions are obtained, as well as in the general case
when wavenumbers of different components are not equal, and
related solutions are not available. In the latter case, results are plotted
for difference $\Delta k = 1$ between scaled wavenumbers of the spin components
with $M = \pm 1$.

\subsection{Limit cases with simple analytic solutions}

The simplest and the most generic case corresponds to the single-component
model with three of the four spin components equal to zero. In this case,
the CW frequency (chemical potential) is
\begin{equation}
\hbar \omega _{j}=\frac{\hbar
^{2}k_{j}^{2}}{2m}+\epsilon+gA_{j}^{2},
\nonumber
\end{equation}
and the eigenfrequency of the perturbations is given by
\begin{equation}
\hbar \omega =\frac{\hbar ^{2}k}{m}k_{j}\pm \sqrt{\frac{\hbar ^{2}k^{2}}{2m}%
\Big(\frac{\hbar ^{2}k^{2}}{2m}+2gA_{j}^{2}\Big)},  \label{eq:29}
\end{equation}%
with $g=g_{0}$ for spins $j=\pm 1$, $g=g_{0}+g_{1}$ for spins $j=0,p$, and $%
\epsilon =0$ for the para component. Equation (\ref{eq:29}) is the
Bogoliubov dispersion relation \cite{Bogoliubov} for the propagation of
small perturbations (sound waves) on top of CW solutions (\ref{eq:6}). The
results obtained here are found to be in complete agreement with the
previously reported ones \cite{Aggarwal1,Theocharis,Salasnich}. Namely, for
the attractive nonlinearity ( $g<0$ ), there is MI against the perturbations
in the wavenumber range $0<k<2\sqrt{|g|m}A_{j}/\hbar$, characterized
by the MI gain, $\xi = $Im$(\omega)$. The maximum MI gain,
\begin{equation}
\xi _{\max }=|g|A_{j}/\hbar,
\label{xi_max}
\end{equation}
is attained at
\begin{equation}
k_{\max }=\sqrt{2|g|m}A_{j}/\hbar.
\label{k_max}
\end{equation}

\begin{figure}[!tbp]
\begin{center}
\subfigure[]{\includegraphics[width=0.4\textwidth]{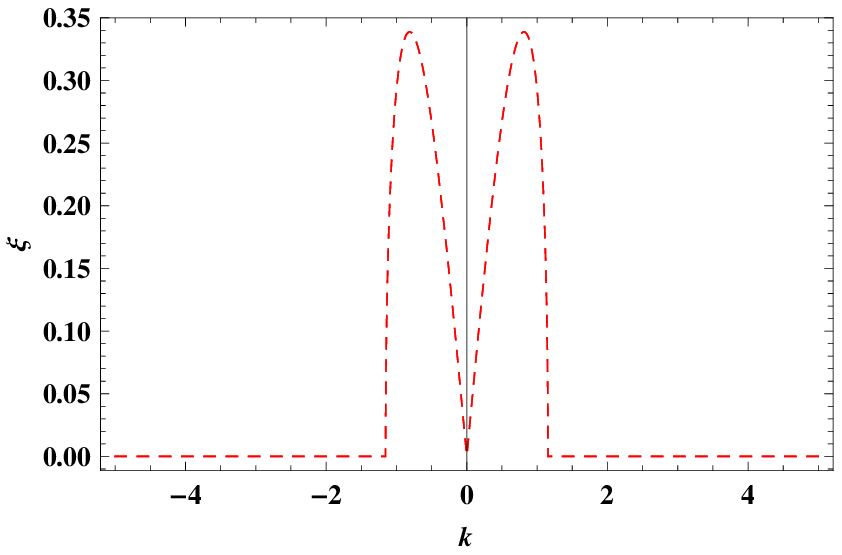}} %
\subfigure[]{\includegraphics[width=0.4\textwidth]{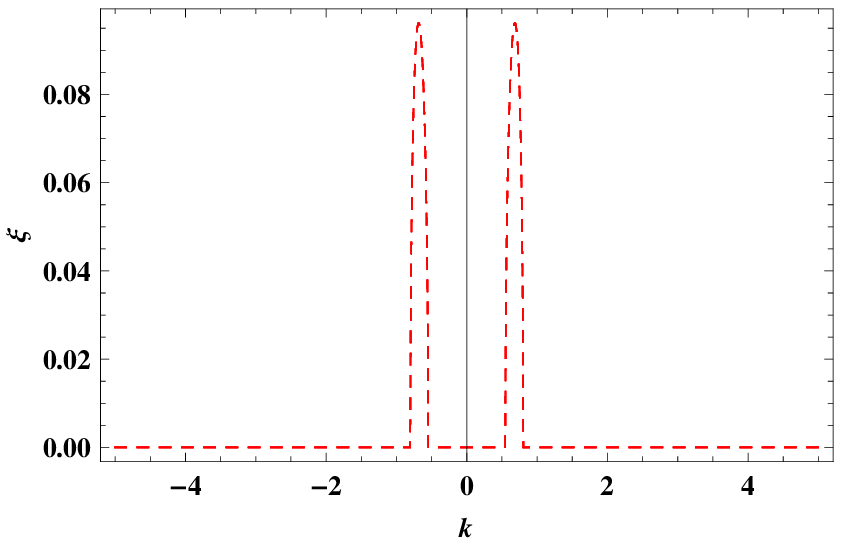}}
\end{center}
\caption{(Color online) The MI gain for the CW states in the two-component
positronium for $A_{0}=\frac{1}{3}$ and $A_{p}=\frac{1}{2}$. The CW
background is modulationally unstable for a proper choice of the amplitudes
of the spin components if either of the nonlinearities are repulsive.\newline
\textbf{(a)}: The MI gain for $g_{0}^{\prime }=-1$ and $g_{1}^{\prime }=1$
\newline
\textbf{(b)}: The MI gain for $g_{0}^{\prime }=1$ and $g_{1}^{\prime }=-1$ }
\label{Fig.3}
\end{figure}

\begin{figure}[tbp]
\begin{center}
\subfigure[]{\includegraphics[width=0.33\textwidth]{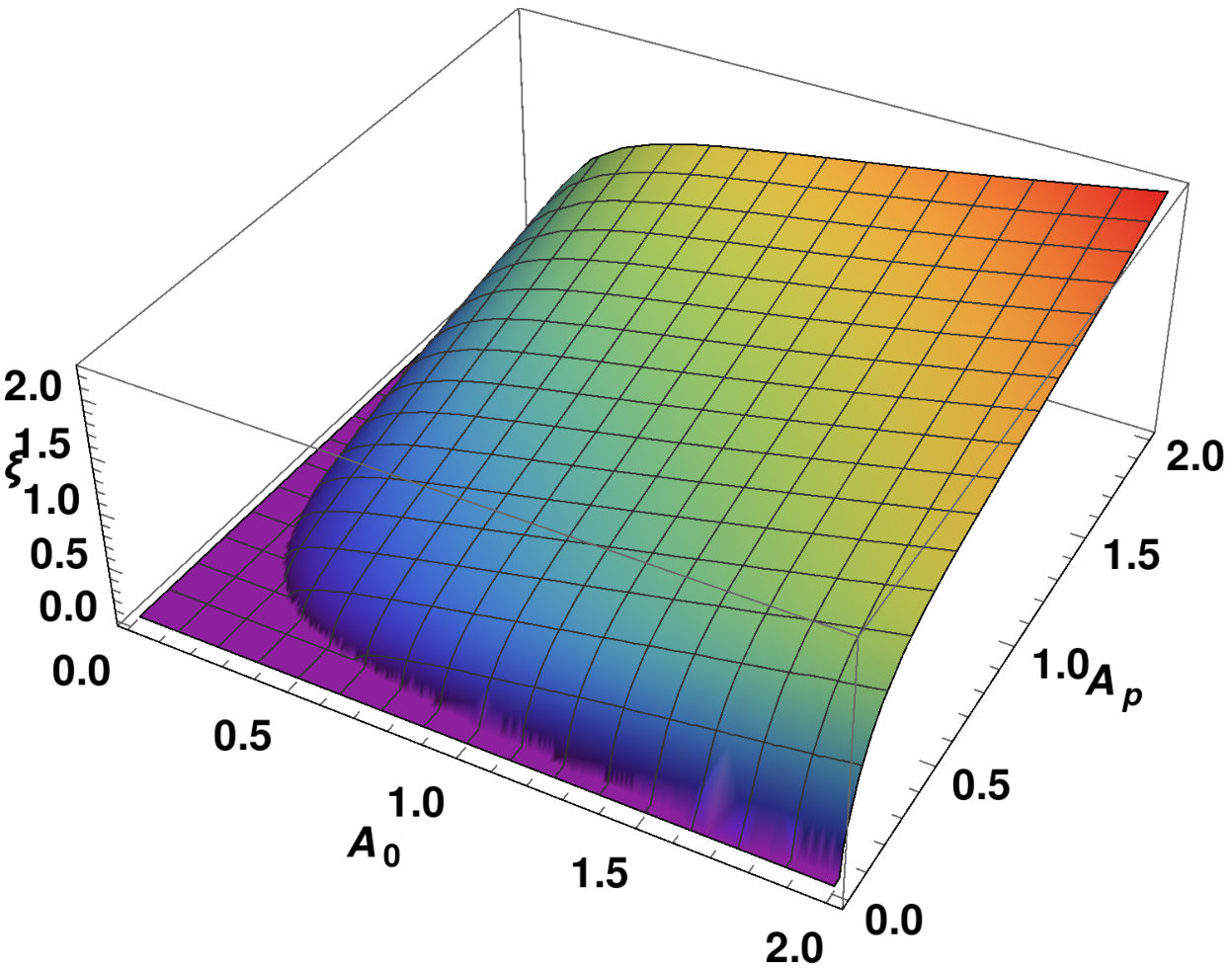}} %
\includegraphics[width=0.4 cm, height=4 cm]{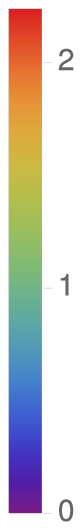} %
\subfigure[]{\includegraphics[width=0.33\textwidth]{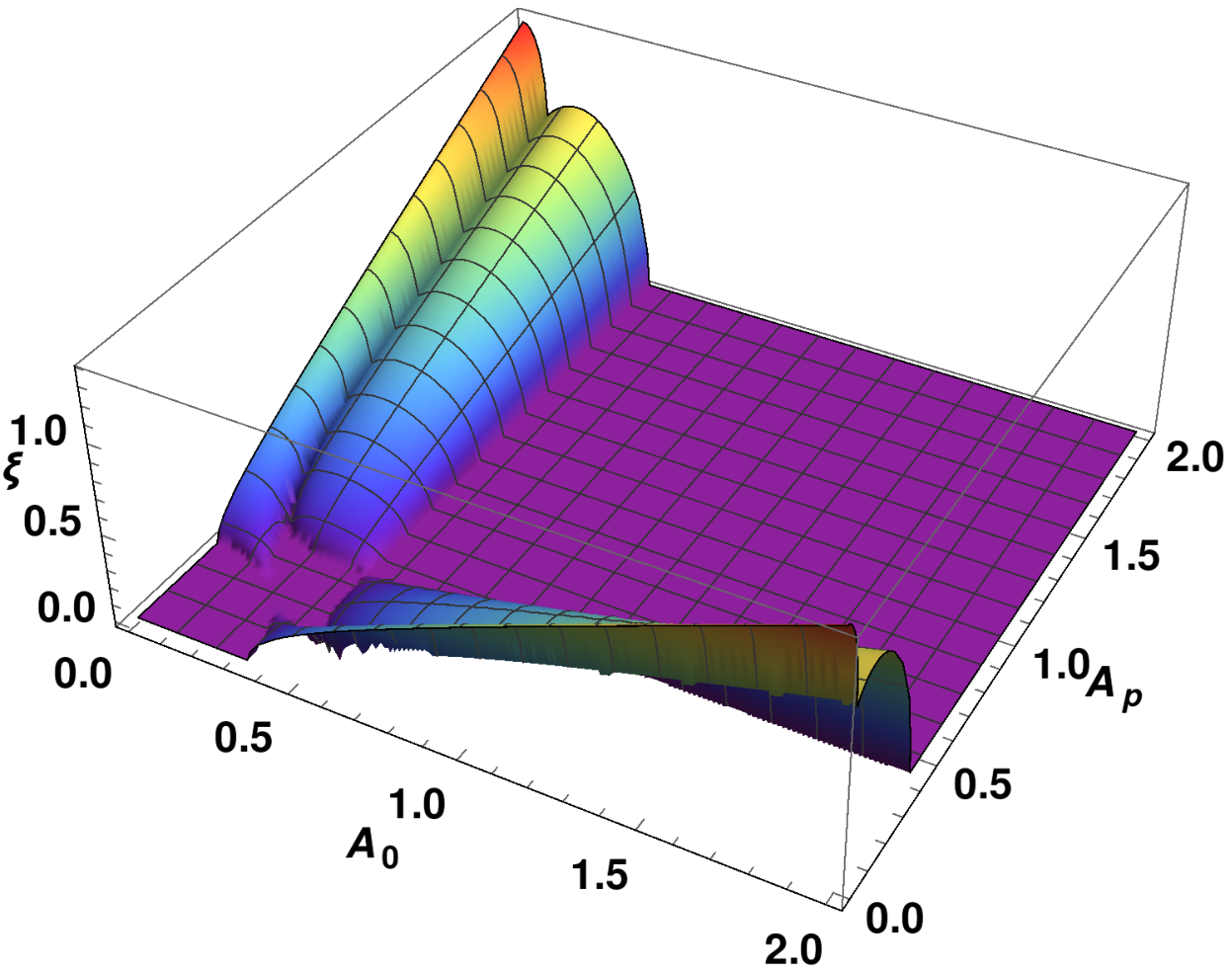}} %
\includegraphics[width=0.6 cm, height=4 cm]{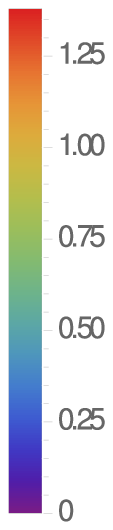}
\end{center}
\caption{(Color online) The dependenece of the largest MI gain on the
amplitudes of the two CW components with identical wavenumbers, $k_{p}=k_{0}$%
. \newline
\textbf{(a)}: The dependence of the maximum gain on $A_{0}$ and $A_{p}$ for $%
g_{0}^{\prime }=-1$ and $g_{1}^{\prime }=1$.\newline
\textbf{(b)}: The dependence of the maximum gain on $A_{0}$ and $A_{p}$ for $%
g_{0}^{\prime }=1$ and $g_{1}^{\prime }=-1$.}
\label{Fig.4}
\end{figure}

\begin{figure}[]
\begin{center}
\subfigure[]{\includegraphics[width=0.4\textwidth]{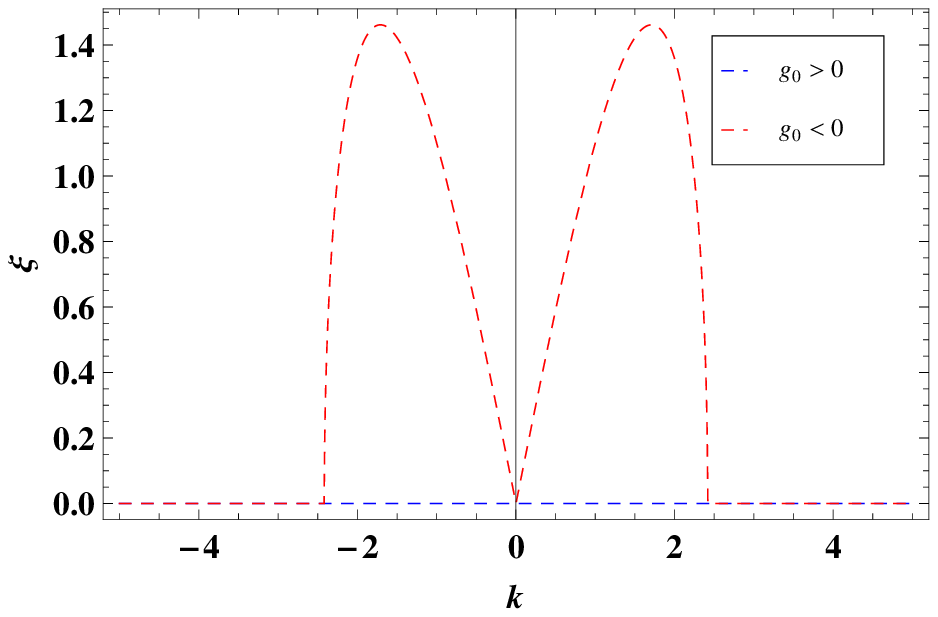}} %
\subfigure[]{\includegraphics[width=0.4\textwidth]{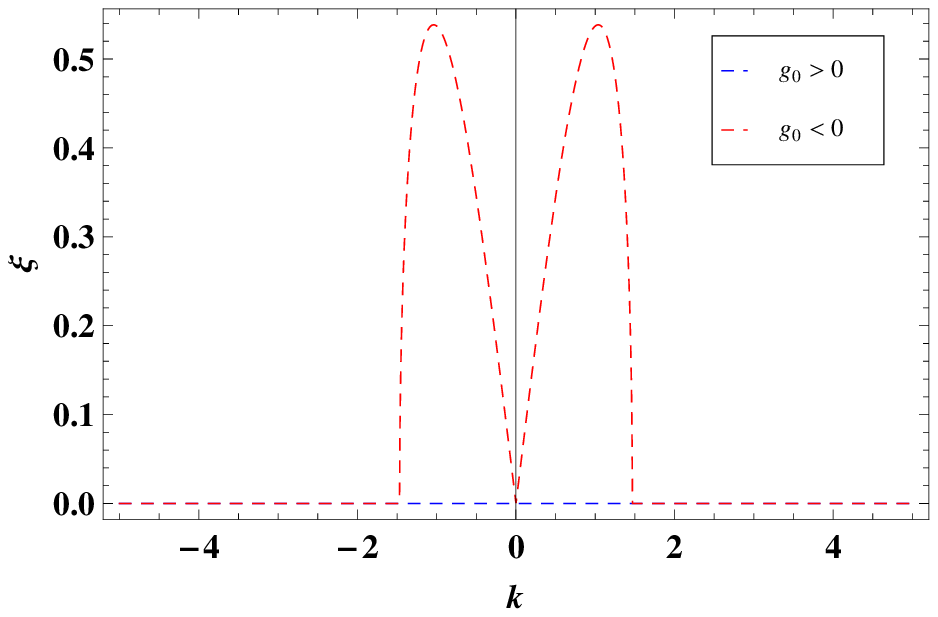}}
\end{center}
\caption{(Color online) The MI gain for the CW solutions with $n=0,1$, for $%
\protect\rho =1$, $A_{1}=\frac{1}{2}$ and $A_{-1}=\frac{1}{3}$. The CW
solutions with both $n=0$ and $n=1$ are modulationally unstable for $%
g_{0}^{\prime }<0$.\newline
\textbf{(a)}: The MI gain for $n=0$, $g_{1}^{\prime }=1$ or $n=1$, $%
g_{1}^{\prime }=-1$.\newline
\textbf{(b)}: The MI gain for $n=0$, $g_{1}^{\prime }=-1$ or $n=1$, $%
g_{1}^{\prime }=1$.}
\label{Fig.5}
\end{figure}

In case of zero components with $M=\pm 1$, the BEC consists of a mixture of $%
\ket{1,0}$ and $\ket{0,0}$ spin components with vanishing interspecific
interaction. The frequency of the perturbations are in fact solutions of Eq.(%
\ref{eq:28}) for the same wavenumbers ($k_{0}=k_{p}=k$) and the Bogoliubov
dispersion relation with $\rho _{2}=A_{0}^{2}+A_{p}^{2}$ and $\alpha
=8A_{0}^{2}A_{p}^{2}g_{1}^{2}$ modifies as 
\begin{eqnarray}
\hbar \omega = \frac{\hbar^2 k^2}{m}\pm \ \Bigg[\frac{\hbar^2 k^2}{2 m}\Big(%
\frac{\hbar^2 k^2}{2 m} + \rho_2(g_0 + 2g_1)\Big) + \alpha \pm  \nonumber \\
\left\{\frac{\hbar^2 k^2}{2 m}\left(\frac{\hbar^2 k^2}{2 m}\bigg(\rho_2
g^2_0 + 2\alpha(1-\frac{g_0}{g_1})\bigg)+ 2\rho_2\alpha(2-\frac{g_0}{g_1}%
)\right) + \alpha^2 \right\}^{\frac{1}{2}}\Bigg]^{\frac{1}{2}}  \label{eq:30}
\end{eqnarray}
From the above equation Eq. (\ref{eq:30}) it is found that MI in the system
for a proper choice of $A_{0}$ and $A_{p}$, as shown in Fig. \ref{Fig.3}, is
possible if either of the nonlinearities, $g_{0}$ or $g_{1}$, is attractive (%
$g_{0}<0$ or $g_{1}<0$), which in fact does not hold for the Ps BEC, hence
the system is modulationally stable. The computation of relations for $%
k_{\max }$ and maximum growth rate $\xi $ in terms of the nonlinearities and
the respective wavenumbers is too complex to be performed analytically.
\newline
In accordance with the above result, and aiming to properly choose
$A_{0}$ and $A_{p}$, we studied the variation of the largest MI gain for
a range of values of $A_{0}$ and $A_{p}$. In the limit case of equal
wavenumbers, Fig.\ref{Fig.4} shows the peak MI gain versus amplitudes of
the two spin components for $g_{0}<0$ and $g_{0}>0$, respectively. In the
latter case, the system is modulationally unstable if and only if condition $%
g_{1}<0$ holds, and amplitudes of the spin components are different, as
seen in Fig. \ref{Fig.4}(b). 
The analytic solutions for the perturbation frequency around the $n=0$ CW
background in case of the ortho Ps condensate, i.e., the three-component
system with identical wavenumbers are
\begin{subequations}
\label{eq:subeqns31}
\begin{equation}
\hbar \omega =(2\pm 1)\frac{\hbar ^{2}k^{2}}{2m},
\end{equation}%
\begin{equation}
\hbar \omega =\Bigg[(2\pm 1)\frac{\hbar ^{2}k^{2}}{2m}\pm
2g_{1}(A_{1}+A_{-1})^{2}\Bigg].
\end{equation}%
\begin{equation}
\hbar \omega =\frac{\hbar ^{2}k^{2}}{m}\pm \sqrt{\frac{\hbar ^{2}k^{2}}{2m}%
\Bigg[\frac{\hbar ^{2}k^{2}}{2m}+2g_{0}(A_{1}+A_{-1})^{2}\Bigg].}
\end{equation}%
The results given by Eq. (\ref{eq:subeqns31}) are in exact agreement with
the previously known results for the ferromagnetic $F=1$ spinor condensates
\cite{Tasgal}. Thus, the $n=0$ CW background of the ferromagnetic ortho Ps
with $g_{0}>0$ is stable against the exponential growth of perturbations
when the wavenumbers are equal, and it is modulationally unstable if $%
g_{0}<0 $, so that the largest MI gain, $\xi _{\max }=-g_{0}\frac{%
(A_{1}+A_{-1})^{2}}{\hbar }$, occurs at $k_{\max }=\sqrt{-2mg_{0}}\frac{%
(A_{1}+A_{-1})^{2}}{\hbar }$. \newline

Analytical solutions of Eq. (\ref{eq:26}) in the form of Bogoliubov
dispersion relations are obtained for the perturbations propagating on top
of the CWs with $n=0$ and $1$ CW, in case of identical wavenumbers:
\end{subequations}
\begin{subequations}
\label{eq:subeqns32}
\begin{equation}
\hbar \omega =(2\pm 1)\frac{\hbar ^{2}k^{2}}{2m},
\end{equation}%
\begin{equation}
\hbar \omega =(2\pm 1)\frac{\hbar ^{2}k^{2}}{2m},
\end{equation}%
\begin{equation}
\hbar \omega =(2\pm 1)\frac{\hbar ^{2}k^{2}}{2m}\pm 2\rho g_{1}\pm (-1)^{n}%
\frac{2\epsilon A_{1}A_{-1}}{(A_{1}^{2}+A_{-1}^{2})},  \label{eq:subeq32c}
\end{equation}%
\begin{equation}
\hbar \omega =\frac{\hbar ^{2}k^{2}}{m}\pm \sqrt{\frac{\hbar ^{2}k^{2}}{2m}%
\Bigg[\frac{\hbar ^{2}k^{2}}{2m}+2g_{0}\Bigg(\rho +\frac{(-1)^{n}\epsilon
A_{1}A_{-1}}{g_{1}(A_{1}^{2}+A_{-1}^{2})}\Bigg)\Bigg]}.  \label{eq:subeq32d}
\end{equation}%
\end{subequations}%
From the above equations it is clear that, in the limit case of identical
wavenumbers, the CW backgrounds with $n=0$ and $1$ give rise to MI and,
thereby, formation of solitons if and only if condition $g_{0}<0$ holds,
with the largest growth rate, $\xi _{\max }=-\frac{g_{0}}{\hbar }\Big(\rho
+(-1)^{n}\frac{\epsilon A_{1}A_{-1}}{g_{1}(A_{1}^{2}+A_{-1}^{2})}\Big)$,
corresponding to $k_{\max }=\frac{1}{\hbar }\sqrt{-2mg_{0}\Big(\rho +(-1)^{n}%
\frac{\epsilon A_{1}A_{-1}}{g_{1}(A_{1}^{2}+A_{-1}^{2})}\Big)}$. Thus, the
Ps condensate with repulsive nonlinearities is modulationally stable in such
a case. The total particle density, $\rho $, together with the densities of
the components with $M=\pm 1$, have a significant impact on the stability of
the CW solutions, while the order $(n)$ of the CW solution does not affect
the stability. However, the structure of the CW solution and the
nonlinearity coefficient $g_{1}$ affect the instability strength for $g_{0}<0
$, with a larger MI\ gain for $n=0$, $g_{1}>0$ and $n=1$, $g_{1}<0$, as
shown in Fig. \ref{Fig.5}. 

\begin{figure}[tbp]
\begin{center}
\subfigure[]{\includegraphics[width=0.33\textwidth]{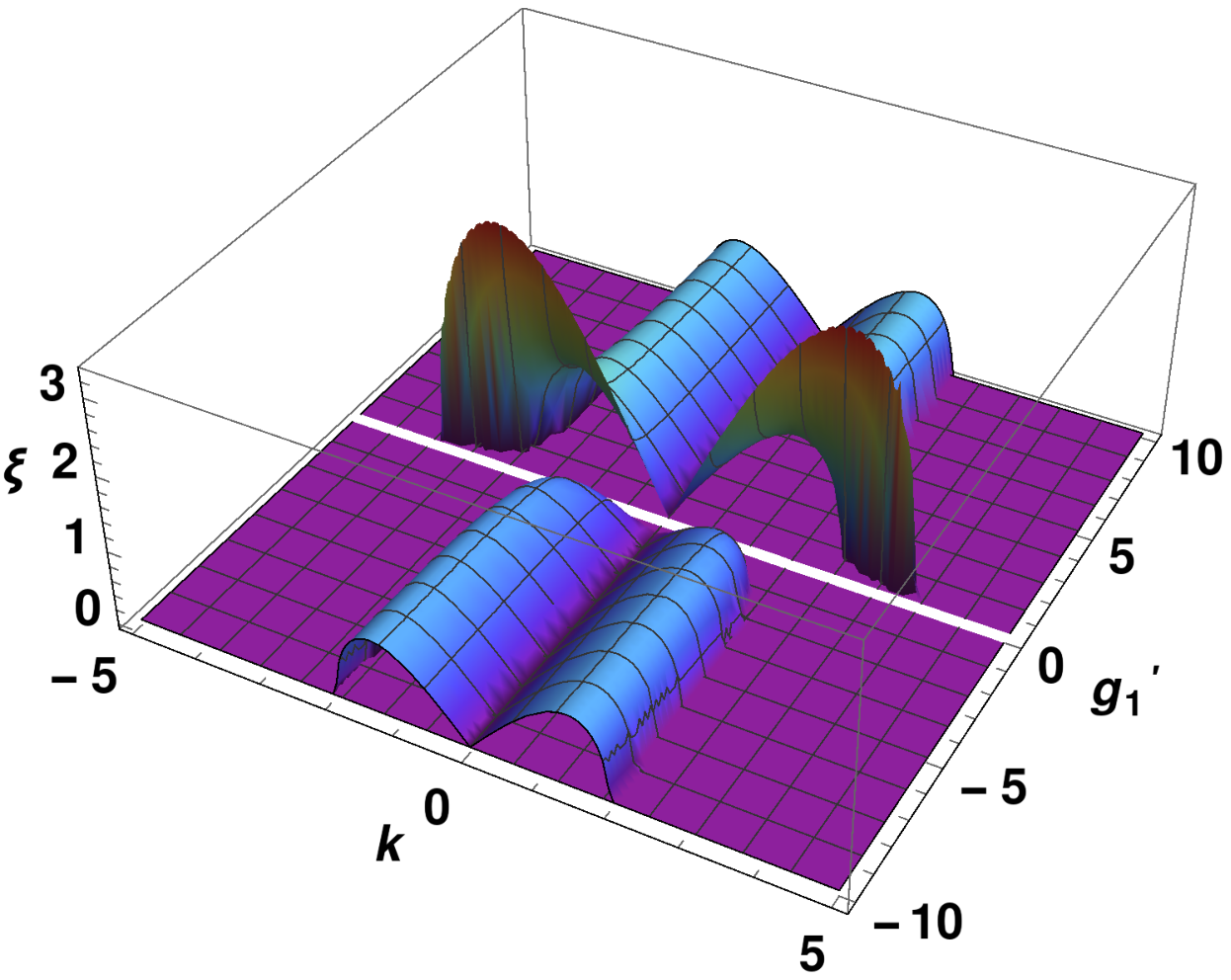}} %
\includegraphics[width=0.4 cm, height=4 cm]{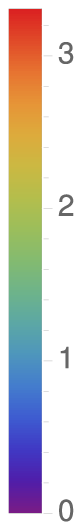} %
\subfigure[]{\includegraphics[width=0.33\textwidth]{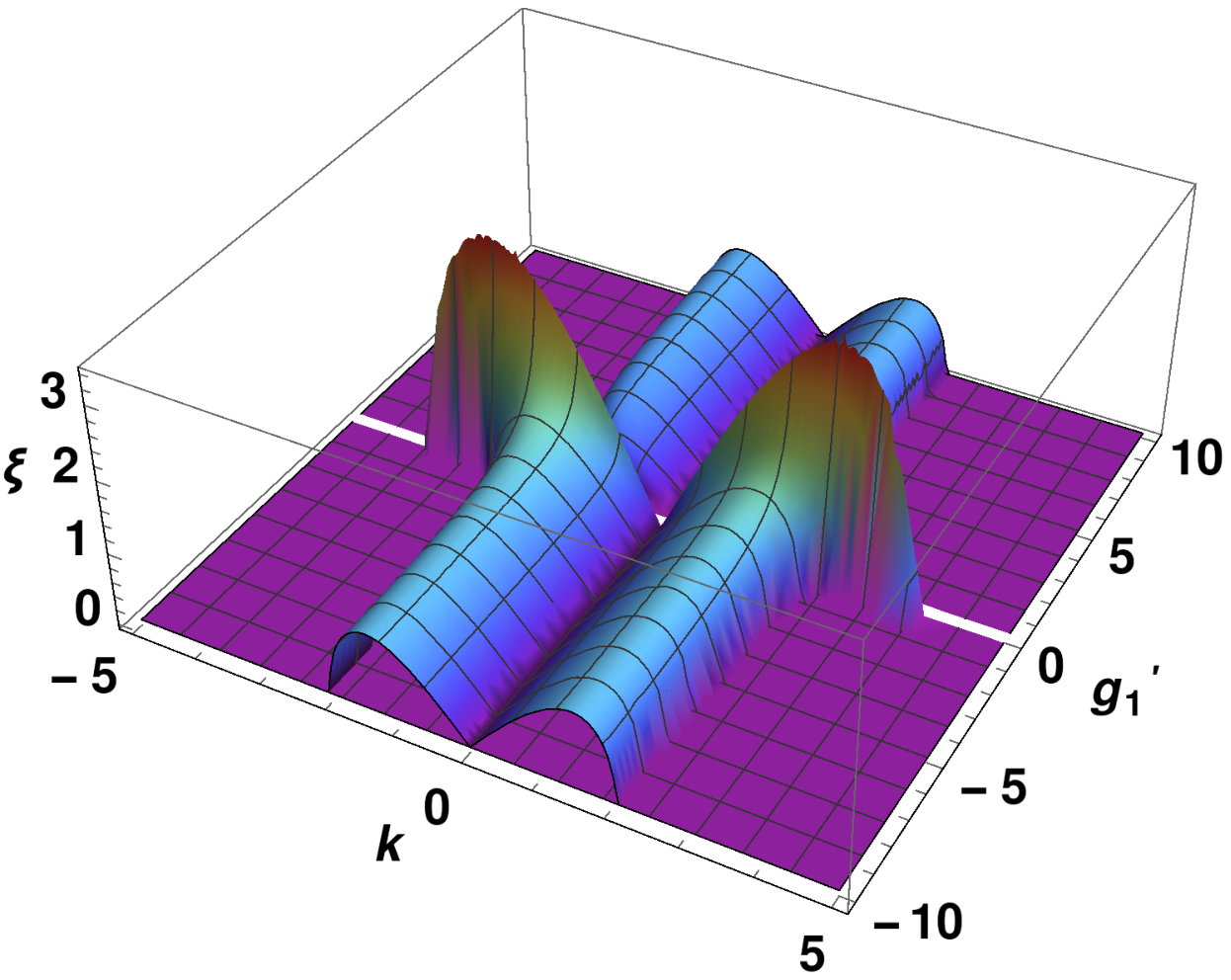}} %
\includegraphics[width=0.4 cm, height=4 cm]{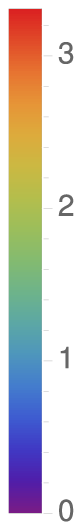}
\end{center}
\caption{(Color online) The dependence of the MI gain on $g_{1}$ for the CW
solutions with $n=0\ $and $1$ for $\protect\rho =1$, $A_{1}=\frac{1}{2}$, $%
A_{-1}=\frac{1}{3}$ and $k_{1}-k_{-1}=0$. Both CW solutions with $n=0$ and $%
1 $ are modulationally unstable for $g_{0}^{\prime }<0$.\newline
\textbf{(a)}: The dependence of the MI gain on $g_{1}^{\prime }$ for $n=0$
and $g_{0}^{\prime }=-1$.\newline
\textbf{(b)}: The dependence of the MI gain on $g_{1}^{\prime }$ for $n=1$
and $g_{0}^{\prime }=-1$.}
\label{Fig.6}
\end{figure}

For $g_{0}<0$, Fig. \ref{Fig.6} shows the dependence of the gain on the $%
g_{1}$ nonlinearity coefficient in the limit case of $k_{1}=k_{-1}$. It is
evident that the CW backgrounds with both $n=0$ and $1$ exhibit similar MI
regions for opposite signs of nonlinearity $g_{1}$. The MI gain is larger at
$g_{1}>0$ for the CW solutions with $n=0$, and at $g_{1}<0$ for the CW with $%
n=1$, respectively, for a fixed attractive $g_{0}$ nonlinearity. Note that,
on either side of $g_{1}=0$ at fixed $g_{0}$, the gain attains constant
value after an initial change. 

\subsection{General case with no simple analytic solutions.}

We have derived the dispersion relations for different cases in the limit
case of identical wavenumbers in different spin components. However, in the
general case, when the wavenumbers of the components with $M=\pm 1$ are
different, so simple analytic results cannot be obtained. Note that even if $%
k_{1}$ and $k_{-1}$ assume different values, wavenumbers $k_{p}$ and $k_{0}$
are constrained to be equal, according to Eq.(\ref{eq:subeq8a}), unless the
field components with $M=\pm 1$ vanish. The effect of the CW order $n$, and
of the nonlinearity coefficients $g_{0}$ and $g_{1}$, on the condensate's
MI, for both zero and nonzero differences in the wavenumbers of the $M=\pm 1$
components, were analyzed by plotting the gain, $\xi =\mathrm{Im}(\omega )$,
versus the perturbation wavenumber $k$ for suitable values of the
parameters, including $A_{1}$ and $A_{-1}$. The possibility of having
attractive nonlinearities, achievable via the Feshbach resonance, and their
effect on the stability of the condensate, are also discussed in the
respective cases. \newline

We chose $\rho =1,A_{1}=\frac{1}{2}$ and $A_{-1}=\frac{1}{3}$ for the case
when the wavenumbers are identical, and $\rho =1$ and $A_{1}=A_{-1}=\frac{1}{%
2}$, as the total particle density and the dimensionless amplitudes of the
field components with $M=\pm 1$, respectively, in the case of the wavenumber
difference $\Delta k=1$. The case of $\Delta k\neq 0$ with different
amplitudes $A_{1}\neq A_{-1}$ is not analytically solvable. \newline

\begin{figure}[tbp]
\begin{center}
\subfigure[]{\includegraphics[width=0.33\textwidth]{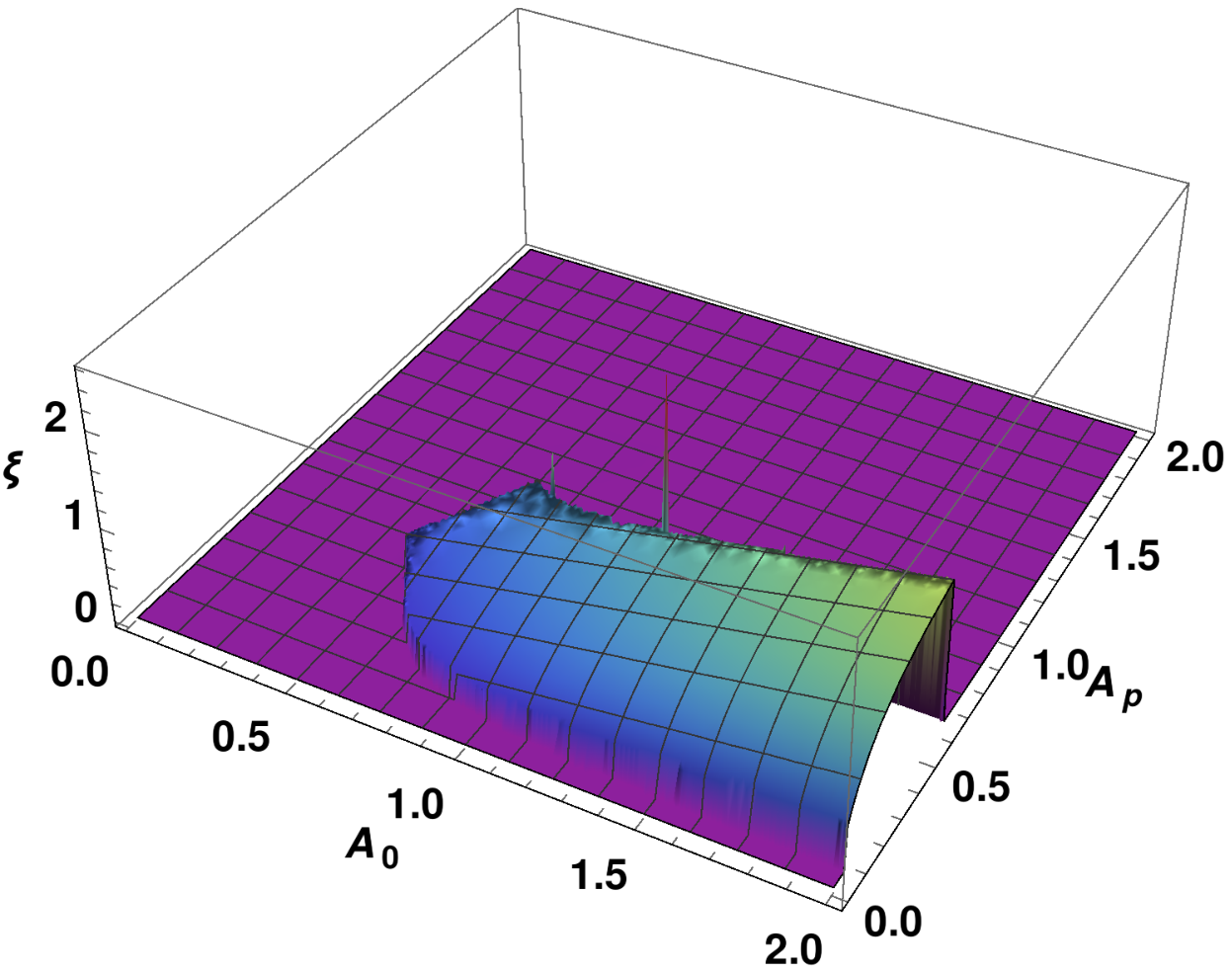}} %
\includegraphics[width=0.4 cm, height=4 cm]{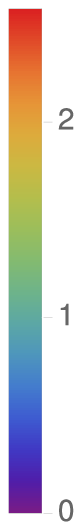} %
\subfigure[]{\includegraphics[width=0.33\textwidth]{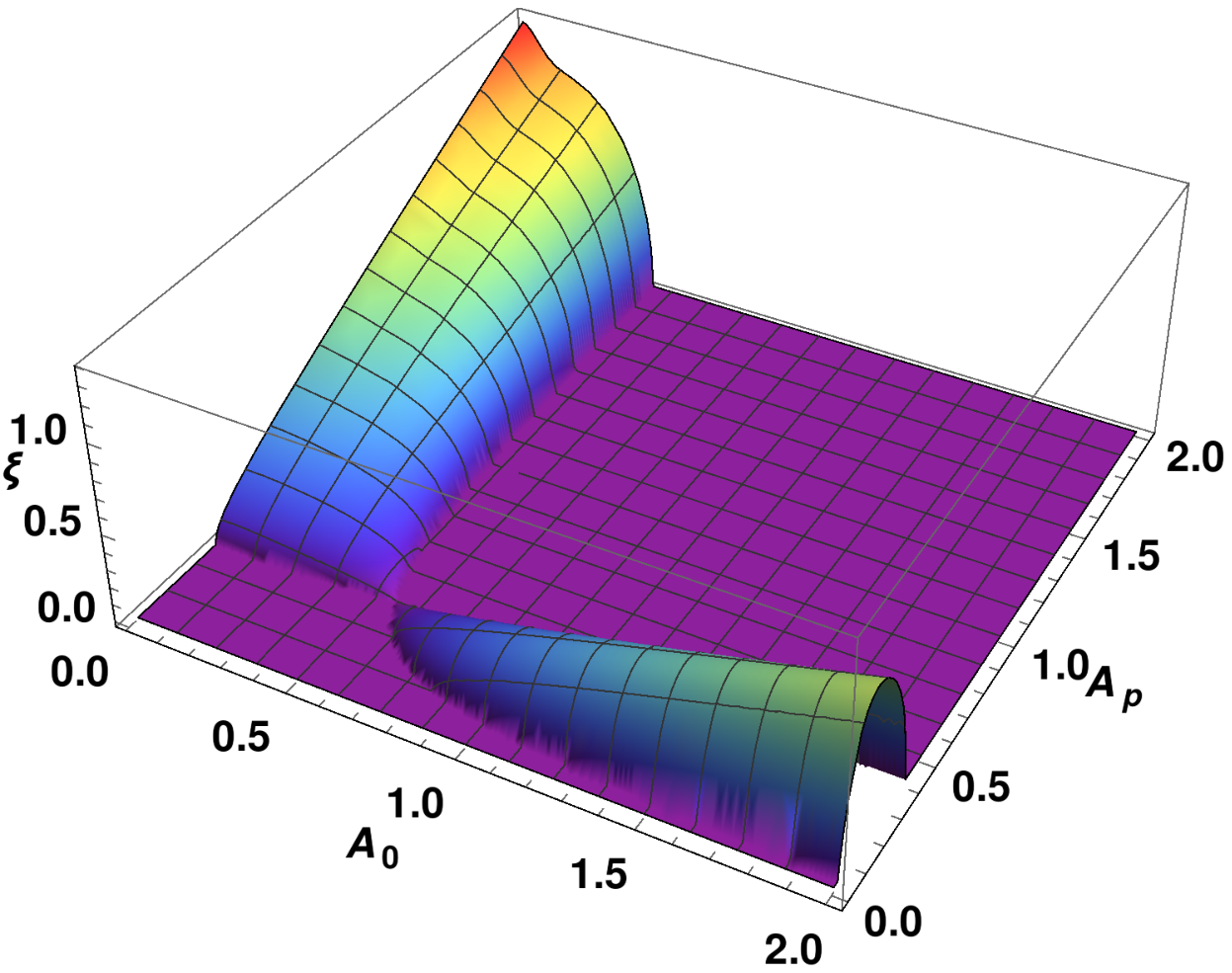}} %
\includegraphics[width=0.5 cm, height=4 cm]{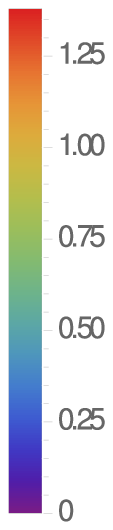}
\end{center}
\caption{(Color online) The dependence of the maximum MI gain on amplitudes
of two-component condensate with the wavenumbers difference $k_{0}-k_{p}=1$.
\newline
\textbf{(a)}: The dependence of maximum MI gain on $A_{0}$ and $A_{p}$ for $%
g_{0}^{\prime }=-1$ and $g_{1}^{\prime }=1$.\newline
\textbf{(b)}: The dependence of maximum MI gain on $A_{0}$ and $A_{p}$ for $%
g_{0}^{\prime }=1$ and $g_{1}^{\prime }=-1$.}
\label{Fig.7}
\end{figure}

\begin{figure}[tbp]
\begin{center}
\subfigure[]{\includegraphics[width=0.33\textwidth]{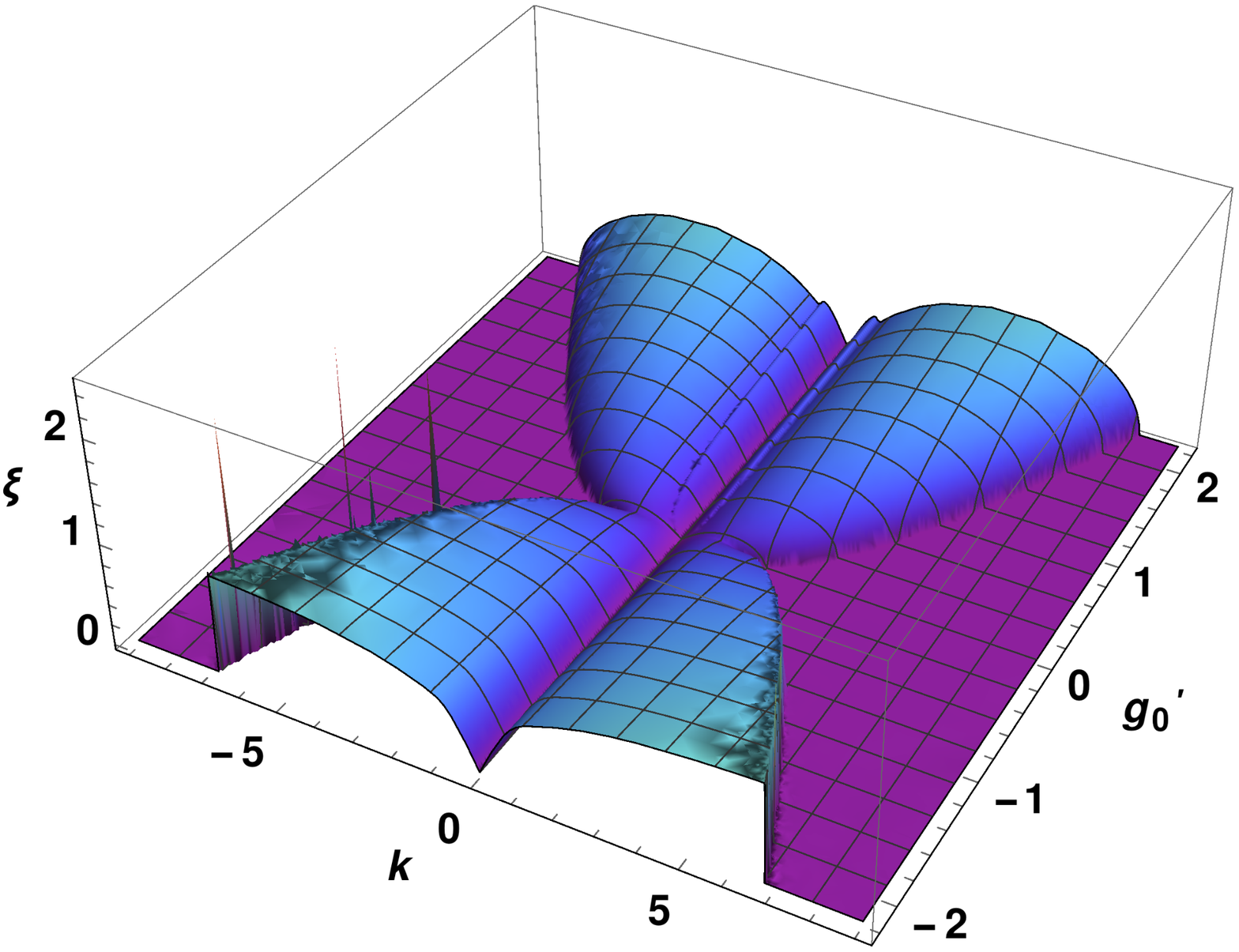}} %
\includegraphics[width=0.45 cm, height=4 cm]{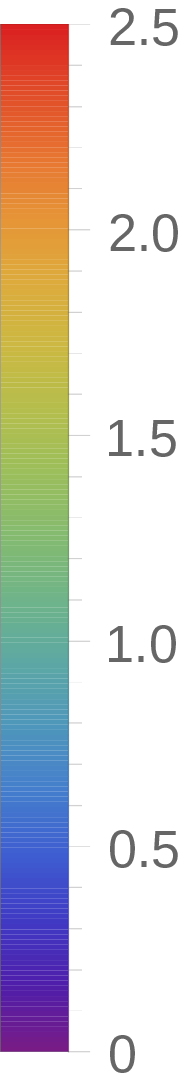} %
\subfigure[]{\includegraphics[width=0.33\textwidth]{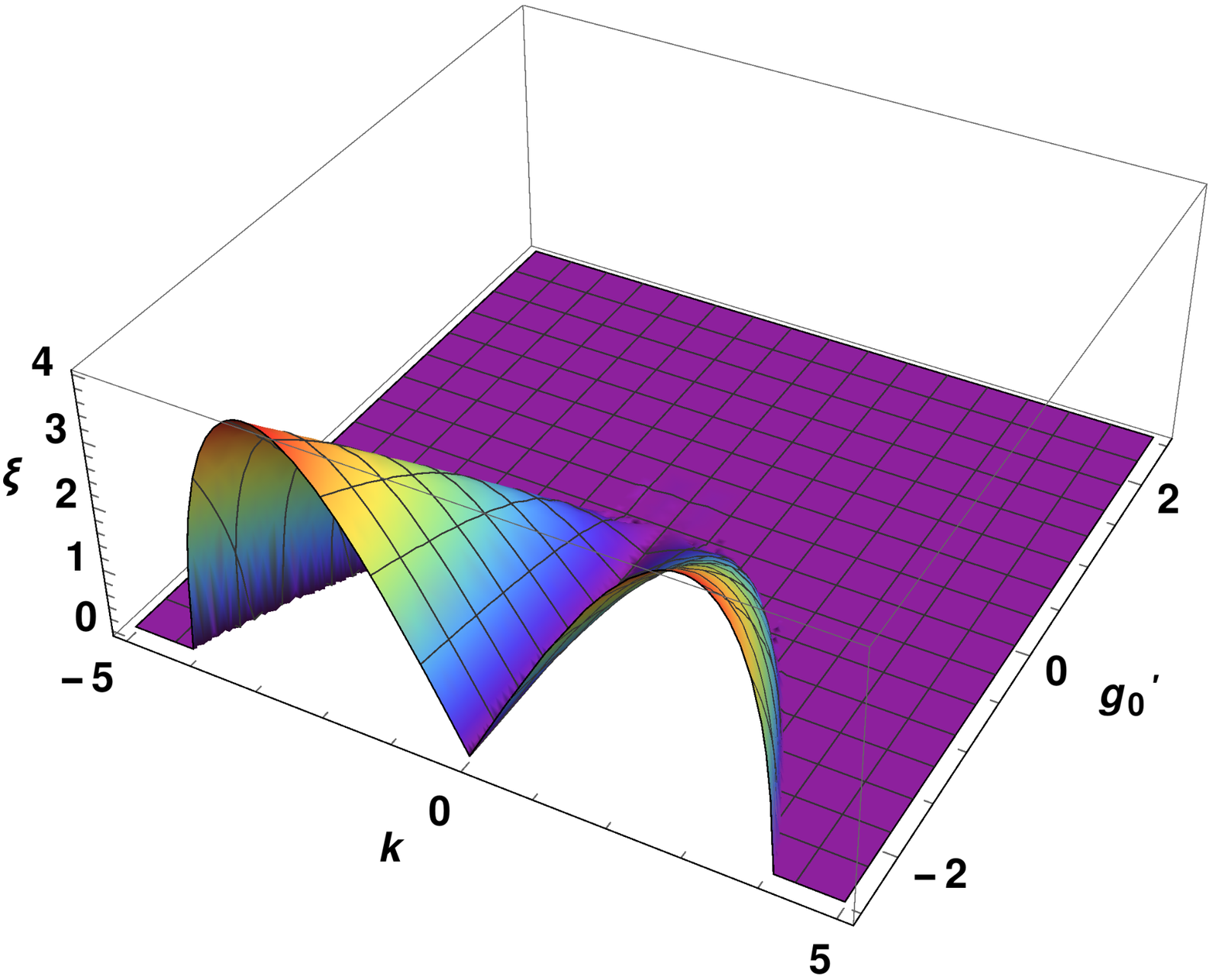}} %
\includegraphics[width=0.4 cm, height=4 cm]{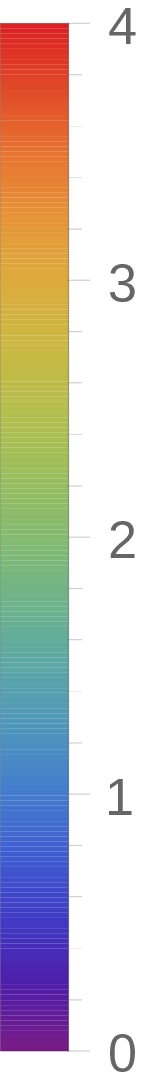}
\end{center}
\caption{(Color online) The dependence of the MI gain on $g_{0}$ in the
out-of-phase CW solution for $\protect\rho =1$, $A_{1}=A_{-1}=\frac{1}{2}$
and $\Delta k=1$. The CW solutions with $n=1$ are modulationally stable if
the nonlinearities $g_{0}^{\prime }$ and $g_{1}^{\prime }$ have opposite
signs and $g_{1}^{\prime }<0$. \newline
\textbf{(a)}: The variation of the MI gain for $g_{1}^{\prime }=1$. \newline
\textbf{(b)}: The variation of the MI gain for $g_{1}^{\prime }=-1$.}
\label{Fig.8}
\end{figure}

In accordance with the above set parameters, for a CW solution with zero $%
M=\pm 1$ components, we chose $A_{p}=\frac{1}{2}$ and $A_{0}=\frac{1}{3}$ in
the case when the wavenumbers are identical, the MI sets in for $g_{1}<0$ if
and only if the amplitudes of the spin components $\ket{1,0}$ and $\ket{0,0}$
are different. However, when the wavenumbers of the two spin components are
different, MI occurs even if $A_{0}$ and $A_{p}$ share the same values for
repulsive $g_{0}$ and attractive $g_{1}$ nonlinearities, as shown in Fig. %
\ref{Fig.7}. \newline
Comparison of Figs. \ref{Fig.4}(a) and \ref{Fig.7}(a) makes it obvious that,
for $g_{0}<0$, the CW states with the zero $M=\pm 1$ component and nonzero
difference between the wavenumbers of the other components are less
vulnerable to the MI in a larger domain of values of the $A_{0}$ and $A_{p}$
amplitudes, than in the limit case of identical wavenumbers. \newline
 As demonstrated in the previous subsection, the CW states of
spinor Ps with both $n=0$ and $1$, $g_{0}>0$, and identical wavenumbers are
modulationally stable. Nevertheless, when the
wavenumbers of the $M=\pm 1$ components are different,
the $n=1$ CW solution is modulationally unstable for the natural repulsive
signs of the $g_{0}$ and $g_{1}$ nonlinearities, as shown in Fig. \ref{Fig.8}%
. The gain is found to increase with the increase in magnitude of $g_{0}$
for fixed $g_{1}>0$. 

\begin{figure}[tbp]
\begin{center}
\subfigure[]{\includegraphics[width=0.33\textwidth]{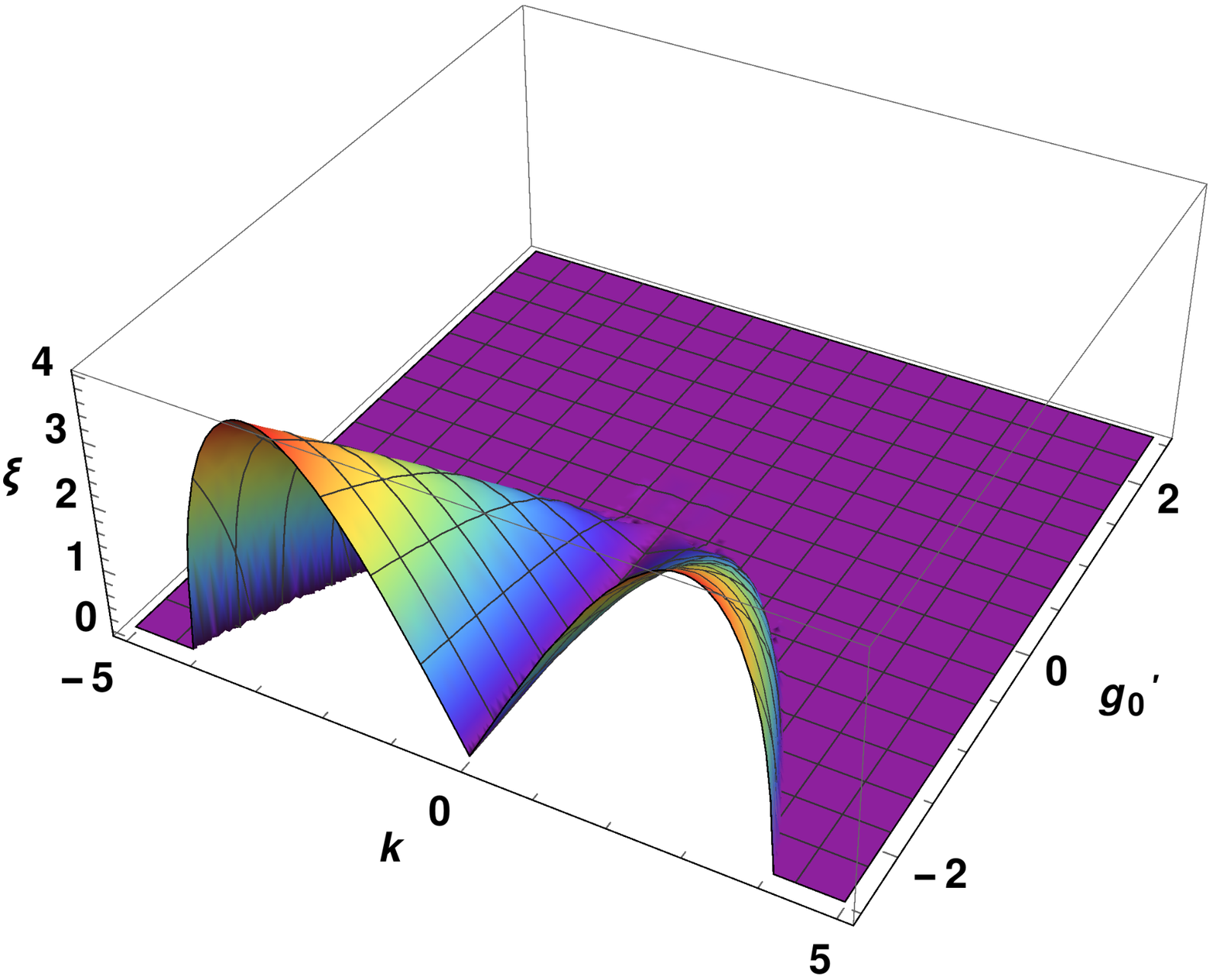}} %
\includegraphics[width=0.4cm, height=4 cm]{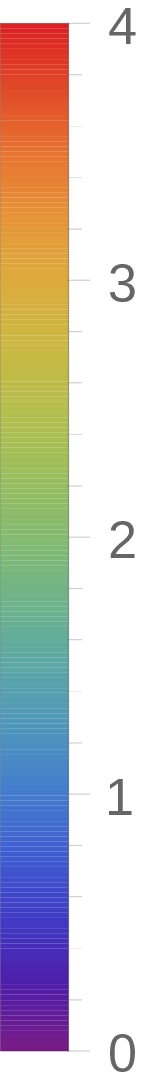}\subfigure[]{%
\includegraphics[width=0.33\textwidth]{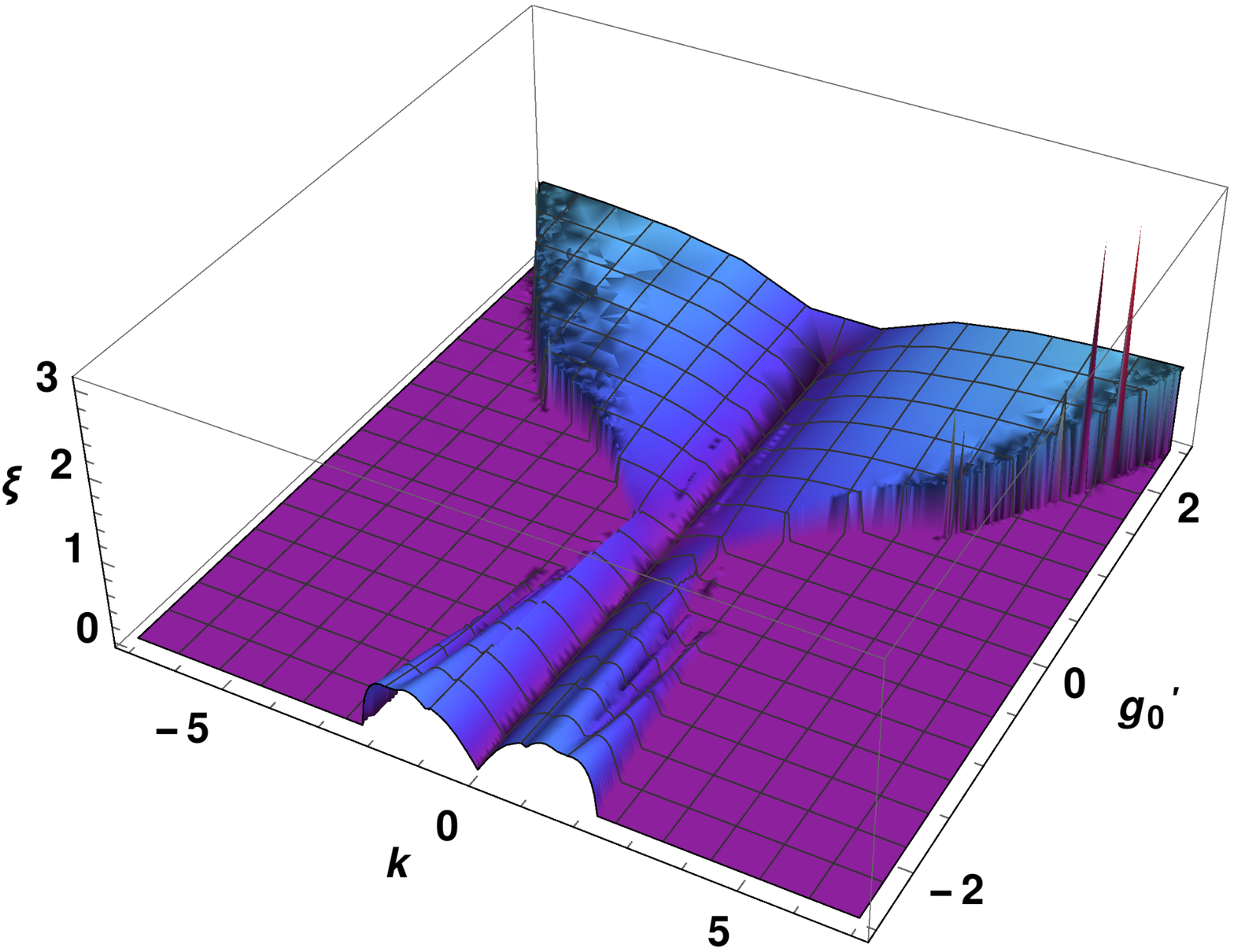}} %
\includegraphics[width=0.4cm, height=4 cm]{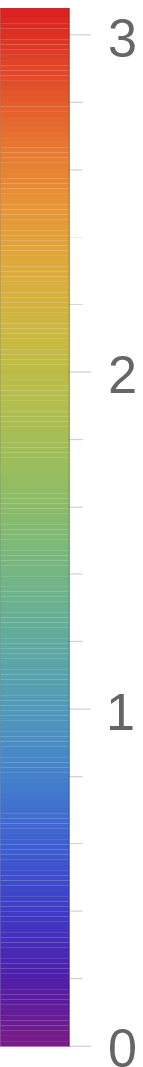}
\end{center}
\caption{(Color online) The dependence of the MI gain on $g_{0}$ for CW
solutions with $n=0$ for $\protect\rho =1$, $A_{1}=A_{-1}=\frac{1}{2}$ and $%
\Delta k=1$. These solutions are modulationally stable if the nonlinearities
$g_{0}^{\prime }$ and $g_{1}^{\prime }$ have identical signs and $%
g_{1}^{\prime }>0$.\newline
\textbf{(a)}: The variation of the MI gain for $g_{1}^{\prime }=1$. \newline
\textbf{(b)}: The variation of MI gain for $g_{1}^{\prime }=-1$. }
\label{Fig.9}
\end{figure}

The effect of the CW parameter $n$ and the nature of the nonlinearity on the
stability of the Ps condensate can be understood by comparing the subplots
(a) and (b) in Figs. \ref{Fig.8} and \ref{Fig.9}, respectively.
The CW background with $n=0$ and the attractive sign of $g_{1}$ is
modulationally unstable for all values of $g_{0}$, and for the CW solutions
with $n=1$, the same is true for the repulsive sign of $g_{1}$. For
small $\Delta k$, even CW solutions $(n=0)$ are found to be stable for
identical signs of $g_{0}$ and $g_{1}$ nonlinearities and $g_{1}>0$, while
odd CW backgrounds $(n=1)$ are stable for opposite signs of the
nonlinearities and $g_{1}<0$. Large differences in the wavenumbers
of the spin components with $M=\pm1$ give rise to the MI even in such
backgrounds, which might be expected to be stable. The final conclusion in the
case of the nonzero wavenumber difference
between the spin components is that the nonlinearities and
the CW parameter, $n$, in Eq. (\ref{eq:subeq32d}) tend to increase the
susceptibility of the CW backgrounds to MI. \newline

 Throughout the paper, we have studied CW solutions and their MI
in the condensate of positronium, which is a
mixture of $F=1$ and $F=0$ spinor components with vanishing
interaction between them. \newline
The $n=0, 1$ CW solutions in the case of $F=1$ spinor condensates (CW
solutions with $n=1$ do not exist for the ferromagnetic case) were found to be
completely stable, with the existence ranges independent of the density of the
condensate. However, for the spinor positronium condensate,
the existence ranges of the $n=0, 1$ CW solutions is
found to be limited by the density of the condensate. In addition, such CW
solutions can be stable or unstable depending on the relative densities of
the different constituents and the order ($n$) of the CW solution. \newline
The absence of the inter-species interaction between the $F=1$ and $F=0$
spinor components invalidates the familiar MI condition $(g^2_{12}>g_{1}g_{2}$, where
$g_{1}$, $g_{2}$ and $g_{12}$ are, respectively, strengths of the intra-species
and inter-species interactions) for the case of vanishing fields with $M=\pm1$.
Note that spin-orbit coupling is also found to alter the MI condition \cite%
{Ishfaq}. MI is possible if either of the nonlinearities, $g_{0}$ or $g_{1}$,
are attractive. In the case of the attractive spin-mixing interaction, $g_{1}<0$,
binary condensates of $\ket{1,0}$ and $\ket{0,0}$ spin components are
stable for equal amplitudes of the components and $g_{0}>0$.
\newline
The $n=0, 1$ CW states in the spinor positronium are stable against MI
similar to the stability of $%
n=0$ CW states in the condensate with $F=1$ for repulsive nonlinearities. Nevertheless,
the instability gain in the $n=0$ CW states in case of $F=1$ spinor
condensates depends solely on the interaction strength, while for the spinor condensate
of positronium, the order $n$ of the CW state, in the combination with the nonlinear
spin-exchange interaction, $g_{1}$, modifies the instability gain for $%
g_{0}<0$. Wavenumber differences between the different spin components are
found to have significant destabilizing effects. \newline
The summary of the results of the MI analysis for the CW solutions with $n=0$
and $1$ for zero and nonzero values of $k_{1}\neq k_{-1}$ is presented in
Table II. 
\FloatBarrier

\begin{center}
\begin{table}[tph]
\caption{Summary of the results of the MI analysis for different
combinations of the parameters.}
\label{tab2}{\footnotesize \centering
\begin{tabular}{p{2.5cm}p{1cm}p{1cm}p{1cm}p{7.5cm}c}
\hline\hline
$\Delta k=(k_{1}-k_{-1})$ & $n$ & $g_{1}$ & $g_{0}$ & Inference &  \\%
[0.5ex] \hline
\multirow{8}{*}{$0$} & \multirow{4}{*}{$0$} & \multirow{2}{*}{$+$} & $+$ & %
\multirow{8}{*}{\shortstack{The $n=0, 1$ CW backgrounds are modulationally
\\ unstable for the attractive $g_0$ nonlinearity $(g_{0}<0)$.\\ It is
independent of the nature of $g_{1}$.}} &  \\
&  &  & $-$ &  &  \\
&  & \multirow{2}{*}{$-$} & $+$ &  &  \\
&  &  & $-$ &  &  \\
& \multirow{4}{*}{$1$} & \multirow{2}{*}{$+$} & $+$ &  &  \\
&  &  & $-$ &  &  \\
&  & \multirow{2}{*}{$-$} & $+$ &  &  \\
&  &  & $-$ &  &  \\
&  &  &  &  &  \\ \hline
\multirow{8}{*}{$\ne 0$} & \multirow{4}{*}{$0$} & \multirow{2}{*}{$+$} & $+$
& stable for small values of $\Delta k$ &  \\
&  &  & $-$ & unstable &  \\
&  & \multirow{2}{*}{$-$} & $+$ & unstable &  \\
&  &  & $-$ & unstable &  \\
& \multirow{4}{*}{$1$} & \multirow{2}{*}{$+$} & $+$ & unstable for natural
values of nonlinearities. &  \\
&  &  & $-$ & unstable &  \\
&  & \multirow{2}{*}{$-$} & $+$ & stable for small values of $\Delta k$ &
\\
&  &  & $-$ & unstable &  \\[1ex] \hline\hline
\end{tabular}
}
\end{table}
\end{center}


\section{Conclusion}

We have obtained the CW solutions in spinor BEC of positronium, composed of
para $(F=0)$ and ortho $(F=1)$ spin fields in the absence of external
magnetic fields. For the condensate without the para component, $\psi _{p}=0$%
, the BEC is tantamount to a spinor $F=1$ ferromagnetic condensate with CW
solutions existing only for even $n$, which determines the phase shift, $\pi
m$, between the different components (i.e., for the in-phase components).
The CW solutions in such a case are stable.\newline

In the presence of the para component $(\psi _{p}\neq 0)$, there exist CW
solutions with both $n=0$ and $n=1$, whose existence ranges are limited by
the total density, $\rho $. The ground state of positronium for $g_{1} \gg
\epsilon$ is found to have equal densities of ortho and para components for $%
n=1$ CW solutions, provided that the amplitudes of the $M=\pm 1$ components
are equal, and the condensate is a polar one. Since the ortho-to-para
interconversion is minimal, if measured in units of the scaled energy
difference, for $g_{1}\ll \epsilon $ and can be neglected. In such a case,
the CW solutions are stable if the density of the para-Ps obeys either
condition $A_{p}^{2}<A_{0}^{2}$, or the conditions given by Eqs. (\ref{eq:18}%
) and (\ref{eq:19}). \newline

The obtained CW solutions with $n=0$ and $1$ in spinor Ps were subsequently
examined for the MI (modulational instability), using the linear stability
analysis for the small perturbations. In the case of zero fields with $M=\pm
1$, the stability of the CW background is found to depend on the amplitudes
of the respective components, and on the nonlinearity coefficients $g_{0}$
and $g_{1}$. For $g_{0}>0$ and $g_{1}<0$, the CW solutions with identical
wavenumbers are modulationally unstable only for different amplitudes of the
components. In the limit case of identical wavenumbers of all the spinor
components, MI for both CW backgrounds with $n=0$ and $n=1$ depends on the
sign of $g_{0}$ alone. The result is that, both for $n=0$ and $1$, the CW
backgrounds are stable for $g_{0}>0$. The total particle density, $\rho $,
with the help of individual densities of the $M=\pm 1$ components,
suppresses the influence of the CW parity, $n$, on the MI. For $g_{0}<0$,
the gain attains a constant value after a rapid initial change with the
variation of the $g_{1}$ nonlinearity coefficient. \newline

In the general case of the nonzero differences between the wavenumbers of
the $M=\pm 1$ spin components, the $n=1$ (out-of-phase) CW background is
unstable for the natural repulsive signs of $g_{0}$ and $g_{1}$, with the
gain maximum large for larger values of $g_{0}$ at a fixed value of $g_{1}$.
The wavenumber difference $\Delta k\equiv k_{1}-k_{-1}\neq 0$, thereby makes
the CW solutions with $n=0$ and $1$ much more vulnerable to MI. \newline

 Finally, it is relevant to briefly discuss peculiarities of
possible experimental realization of the MI in the positronium BEC. The
MI directly applies if its characteristic growth time, $\sim 1/\xi_{\max}$
(see Eq. (\ref{xi_max}))
does not exceed the (ortho) positronium lifetime, $\approx 140$ ns \cite{lifetime}.
However, it is possible to realize the MI in a less extreme form if the positronium
condensate is permanently replenished from an external source. Then, an essential
difference of the expected experimental situation from that typical for usual atomic
condensates \cite{Randy,experiment-2} is the fact that the scattering length
for the positronium is smaller by a factor $\sim 10$ \cite{Adhikari1}-\cite{Morandi},
and, most essentially, its mass is smaller than the atomic mass of $^7$Li  by a large
factor $\approx 6300$. For this reason, Eq. (\ref{k_max}) demonstrates that the same
range of spatial scales of the MI as in the experiments with atomic gases ($\sim 0.1$ mm)
may be achieved for the positronium density exceeding the atomic one by $\sim 5$ orders of magnitude,
which may be achieved in the experiment. Then, Eq. (\ref{xi_max}) suggests that,
in this region of the densities, the characteristic growth time of the MI may be
$\sim 4$ orders of magnitude smaller than in the atomic BEC, i.e., roughly,
on the microsecond scale, which is closer to the above-mentioned positronium lifetime.

\section{Acknowledgements}

K.P. thanks Department of Science and Technology (DST), Council of Scientific
and Industrial Research (CSIR), National Board for Higher Mathematics
(NBHM), IndoFrench Centre for the Promotion of Advanced Research (IFCPAR),
and Government of India, for the financial support through major projects.
We are grateful to Dr. R. S. Tasgal for fruitful discussions. I. A. Bhat thanks
University Grants Commission (UGC), Government of India, for financial
support via BSR fellowships in Sciences. T.M. acknowledges the grant
IBS-R024-D1 for the support. The work of B.A.M. is supported, in part,
by a joint program of the National Science Foundation and Binational
(US-Israel) Science Foundation through grant No. 2015616, and by
the Israel Science Foundation (grant No. 1287/17).

\clearpage
\newpage
\begin{center}
\textbf{\large Supplemental Material for: 'Continuous-wave solutions and modulational instability in spinor
condensates of positronium'}
\end{center}
\setcounter{equation}{0}
\setcounter{figure}{0}
\setcounter{table}{0}
\setcounter{page}{1}
\makeatletter
\renewcommand{\theequation}{\arabic{equation}}
\renewcommand{\thefigure}{\arabic{figure}}
\renewcommand{\bibnumfmt}[1]{[#1]}
\renewcommand{\citenumfont}[1]{#1}
\section{Supplement 1}
The eighth-order characteristic polynomial in the case of spinor Positronium is
\begin{eqnarray}
0= (\hbar \omega )^{8}+(\hbar \omega )^{7}C_{k,\omega
^{7}}^{n=0,1}k+(\hbar \omega )^{6}\sum_{j=1}^{3}C_{k^{2j-2},\omega
^{6}}^{n=0,1}k^{2j-2} +(\hbar \omega )^{5}\sum_{j=1}^{3}C_{k^{2j-1},\omega
^{5}}^{n=0,1}k^{2j-1}+ \nonumber \\
(\hbar \omega )^{4}\sum_{j=1}^{4}C_{k^{2j},\omega
^{4}}^{n=0,1}k^{2j}+(\hbar \omega )^{3}\sum_{j=1}^{4}C_{k^{2j+1},\omega
^{3}}^{n=0,1}k^{2j+1}+(\hbar \omega )^{2}\sum_{j=1}^{5}C_{k^{2j+2},\omega
^{2}}^{n=0,1}k^{2j+2}+\nonumber\\
(\hbar \omega )\sum_{j=1}^{5}C_{k^{2j+3},\omega
^{3}}^{n=0,1}k^{2j+3}+\sum_{j=1}^{6}C_{k^{2j+4},\omega ^{0}}^{n=0,1}k^{2j+4}
\label{eq:1}
\end{eqnarray}%
The coefficients $C^{n=0,1}_{k^\alpha, \omega^\beta}$ shown explicitly below
contain the variables like continuous wave (CW) parameter $n$, the nonlinearities $g_0$ and $g_1$,
and the amplitudes and the wavenumbers of the of the $M=\pm1$ components that along with the other
spin components form the condensate of density $\rho$. $\hbar$ is the reduced Planck constant,
$m$ is the mass of the Positronium atom and
$\epsilon$ is the internal energy of the ortho states of the Positronium.
\begin{eqnarray}
C^{n=0,1}_{k, \omega^7}=-\frac{4 \hbar^9}{m}(k_{1}+k_{-1})
\label{eq:2}
\end{eqnarray}%
\begin{eqnarray}
C^{n=0,1}_{k^0, \omega^6}= -\frac{\hbar^6}{4 m^2(A^2_1+A^2_{-1})^2}(16 m^2 \rho^2 g^2_{1} ( A^4_{1}+A^4_{-1})+
8 (-1)^n m \rho A_{1}A_{-1} g_{1}(A^2_{1}+A^2_{-1})\nonumber \\
(4 m \epsilon +\hbar^2(k^2_{-1}-k^2_{1})^2)+ A^2_{1}A^2_{-1}\big(32 m^2 \rho^2 g^2_1 +
(4 m \epsilon +\hbar^2 (k^2_{-1}-k^2_{1})^2)^2))
\label{eq:3}
\end{eqnarray}%
\begin{eqnarray}
C^{n=0,1}_{k^2, \omega^6}= -\frac{\hbar^8}{4 m^2(A^2_1+A^2_{-1})g_{1}}(4 m g_{1} (A^2_{1}+A^2_{-1})
(2 g_{1}A^2_{1}-g_{0}A^2_{-1}) + (-1)^n A_{1}A_{-1}(g_{0}+ \nonumber\\
2 g_{1})(4 m \epsilon +\hbar^2(k^2_{-1}-k^2_{1})^2)+\ 2 g_{1}(A^2_{1}+A^2_{-1})(2 m g_{0}
(\rho + A^2_{-1})+ \nonumber \\
4 m g_{1}(\rho-A^2_{1})-\hbar^2(13 k^2_{-1}+30 k_{1}k_{-1}+13 k^2_{1})))
\label{eq:4}
\end{eqnarray}%
\begin{eqnarray}
C^{n=0,1}_{k^4, \omega^6}=-\frac{\hbar^{10}}{m^2}
\label{eq:5}
\end{eqnarray}%
\begin{eqnarray}
C^{n=0,1}_{k^2, \omega^5}= -\frac{\hbar^7}{4 m^3 (A^2_1+A^2_{-1})^2}(16 m^2 \rho g^2_{1} A^6_{-1}(k_{-1}-k_{1})+
16 m^2  \rho g^2_{1} A^4_{1}((3\rho + A^2_{1})k_{-1}+\nonumber \\
(3 \rho-A^2_{1})k_{1})+  16 m^2 \rho g^2_{1} A^4_{-1}((3\rho + A^2_{1})k_{-1}+(3 \rho + A^2_{1})k_{1})- \nonumber \\
4 m A^5_{-1}A_{1} g_{1}(k_{-1}-k_{1})(4m \epsilon + \hbar^2(k^2_{-1}-k^2_{1})^2)+ 24 m A^3_{-1}A_{1}((-1)^n \rho -\nonumber \\
(1+(-1)^n)A^2_{1}))g_{1}(k_{1}+k_{-1}(4 m \epsilon + \hbar^2(k^2_{-1}-k^2_{1})^2)+ 4 m A_{-1}A^3_{1}g_{1}(6(-1)^n \nonumber \\
(\rho-A^2_{1})(k_{1}+k_{-1})-A^2_{1}(5 k_{-1}+7 k_{1}))(4 m \epsilon +\hbar^2(k^2_{-1}-k^2_{1})^2)+
A^2_{1}A^2_{-1}\nonumber \\
(3\hbar^4 k^9_{-1}+ 3\hbar^4 k^8_{-1}k_{1} - 12\hbar^4 k^7_{-1}k^2_{1}-12 \hbar^4 k^6_{-1}k^3_{1}+
6 k^5_{-1}(4 m \epsilon \hbar^2 + 3\hbar^4 k^4_{1})+ \nonumber \\
6 k^4_{-1}(4 m \epsilon \hbar^2 k_{1}+3\hbar^4 k^5_{1})- 12 k^3_{-1}(4 m \epsilon \hbar^2 k^2_{1}+\hbar^4 k^6_{1})- 12 k^2_{-1}((4 m \epsilon \hbar^2 k^3_{1}+ \nonumber \\
\hbar^4 k^7_{1})+ k_{-1}(16 m^2 \rho (6 \rho - A^2_{1})g^2_{1}+3(4 m \epsilon + \hbar^2 k^4_{1})^2)+
k_{1}(16 m^2 \rho (6 \rho + \nonumber \\
A^2_{1})g^2_{1}+3(4 m \epsilon + \hbar^2 k^4_{1})^2))
\label{eq:6}
\end{eqnarray}%
\begin{eqnarray}
C^{n=0,1}_{k^3, \omega^5}= -\frac{\hbar^9}{4 m^3 (A^2_1+A^2_{-1})g_{1}}(4 m g_{1}A^4_{-1} (g_{0}-2 g_{1})(k_{1}-k_{-1})+
3 (-1)^n A_{1}A_{-1}(g_{0}+ 2 g_{1}) \nonumber\\
(k_{1}+k_{-1}) (4 m \epsilon +\hbar^2 k^4_{-1}-2 \hbar^2 k^2_{-1}k^2_{1}+\hbar^2 k^4_{1})+ 2 A^2_{-1}g_{1}(k_{1}+
k_{-1})(6 m \rho g_{0}+ \nonumber \\
12 m \rho g_{1}-\hbar^2(11 k^2_{-1}+ 34 k_{1}k_{-1}+11k^2_{1}))+ 2 A^2_{1}g_{1}(2 m g_{0}((3 \rho +A^2_{1})k_{1}+\nonumber \\
(3 \rho-A^2_{1})k_{1})+ 4 m gh_{1}((3 \rho-A^2_{1})k_{-1}+(3 \rho+A^2_{1})k_{1})-\hbar^2(11 k^3_{-1}+ \nonumber \\
45 k^2_{-1}k_{1}+45 K_{-1}k^2_{1}+11 k^3_{1})))
\label{eq:7}
\end{eqnarray}%
\begin{eqnarray}
C^{n=0,1}_{k^5, \omega^5}=-\frac{3 \hbar^{11}}{m^3}(k_{1}+k_{-1})
\label{eq:8}
\end{eqnarray}%
\begin{eqnarray}
C^{n=0,1}_{k^2, \omega^4}= -\frac{\hbar^6}{16 m^4 (A^2_1+A^2_{-1})^3 g_{1}}(16 m^2\hbar^2 A^{10}_{-1}g^3_{1}(k_{1}-
k_{-1})^2-16 m^2\hbar^2 A^{8}_{-1}g^3_{1}(k_{1}-k_{-1})\nonumber \\
((12 \rho+A^2_{1})k_{-1}+(8\rho - A^2_{1})k_{1})- 16 m A^7_{-1}A_{1}g^2_{1}(2(1+(-1)^n)m \nonumber \\
(\rho-A^2_{1})g_{0}-\hbar^2(k_{1}-k_{-1})(3 k_{-1}+2 k_{1}))(4 m \epsilon+\hbar^2 (k_{1}-k_{-1})^2)- \nonumber \\
16 m^2 A^6_{1}g^3_{1} (4 m \rho^3 g_{0}+\hbar^2(-(13\rho^2-8\rho A^2_{1}+A^4_{1})k^2_{-1} + 2(-17\rho^2+2\rho A^2_{1}+ \nonumber \\
A^4_{1})k_{1}k_{-1} - (13\rho^2 + 12\rho A^2_{1}+ A^4_{1})k^2_{1}))-16 m^2 A^6_{-1}g^3_{1}(4 m \rho^3 g_{0}+\hbar^2 \nonumber \\
((-13\rho^2 + 28\rho A^2_{1}+ 2 A^4_{1})k^2_{-1}- 2(17 \rho^2- 8\rho A^2_{1}+ 2 A^4_{1})k_{1}k_{-1} + (-13\rho^2 + \nonumber \\
12\rho A^2_{1}+ 2 A^4_{1})k^2_{1}))-8 m A^5_{-1}A_{1}g^2_{1}(4 m \epsilon+\hbar^2(k^2_{-1}-k^2_{1})^2)(6 m((-1)^n\rho^2- \nonumber \\
(1+(-1)^n)A^4_{1})g_{0} +\hbar^2(A^2_{1}(21 k^2_{-1}+28 k_{1}k_{-1} + 11 k^2_{1}) + (-1)^n(A^2_{1}-\rho) \nonumber \\
(13 k^2_{-1}+34 k_{1}k_{-1}+13 k^2_{1})))- 8 m A_{-1}A^5_{1}g^2_{1}(4 m \epsilon +\hbar^2(k^2_{-1}-k^2_{1})^2) \nonumber \\
((-1)^n (A^2_{1}-\rho)(2 m(A^2_{1}-3\rho)g_{0}+ \hbar^2 (13 k^2_{-1}+34 k_{1}k_{-1}+13 k^2_{1}))+ A^2_{1} \nonumber \\
(2 m (A^2_{1}-4\rho)g_{0}+\hbar^2(9 k^2_{-1}+32 k_{1}k_{-1}+19 k^2_{1})))+ A^3_{1}A^3_{-1}(4 m \epsilon + \hbar^2 \nonumber \\
(k^2_{-1}-k^2_{1})^2)(16 m \hbar^2 g^2_{1} ((-1)^n (\rho - A^2_{1})(13 k^2_{-1}+34 k_{1}k_{-1}+13 k^2_{1})-A^2_{1} \nonumber \\
(12 k^2_{-1} + 31 k_{1}k_{-1} + 17 k^2_{1}) + g_{0}(-96 m^2 \rho ((-1)^n \rho -(1+(-1)^n)A^2)g^2_{1}+ \nonumber \\
( 4 m \epsilon +\hbar^2 (k^2_{-1}-k^2_{1})^2)^2)) + A^2_{-1}A^4_{1} g_{1}(-12 m \rho g_{0}(16 m^2 \rho^2 g^2_{1}
+(4 m \epsilon + \nonumber \\
\hbar^2(k^2_{-1}-k^2_{1})^2)^2)+ \hbar^2( 13 \hbar^4 k^{10}_{-1} + 34 \hbar^4 k^9_{-1}k_{1} -
39 \hbar^4 k^8_{-1}k^2_{1} - 136 \hbar^4 k^7_{-1}k^3_{1} + \nonumber \\
26\hbar^4 k^4_{-1}k^2_{1}(\hbar^2 k^2_{1}-4 m \epsilon) + 26 k^6_{-1}(4 m \epsilon \hbar^2 +
\hbar^4 k^4_{1})+68 k^5_{-1}(4 m \epsilon \hbar^2 k_{1} + \nonumber \\
3\hbar^4 k^5_{1})- 136 k^3_{-1}(4 m \epsilon \hbar^2 k^3_{1}+ \hbar^4 k^7_{1}) + k^2_{-1}(16 m^2(39 \rho^2 -
12\rho A^2_{1}+A^4_{1}) g^2_{1}+ \nonumber \\
13 (4 m \epsilon -3 \hbar^2 k^5_{1})(4 m \epsilon + \hbar^2 k^4_{1}) + k^2_{1}(16 m^2 (39 \rho^2 -
28\rho A^2_{1}+A^4_{1}) g^2_{1}+ \nonumber \\
13 (4 m \epsilon + \hbar^2 k^4_{1})^2) + 2 k_{-1}k_{1}(16 m^2(51 \rho^2 - 8\rho A^2_{1} -
A^4_{1})g^2_{1} + 17(4 m \epsilon +\nonumber \\
\hbar^2 k^4_{1})^2)))+ A^4_{-1}A^2_{1} g_{1}(-12 m \rho g_{0}(16 m^2 \rho^2 g^2_{1}+(4 m \epsilon +
\hbar^2(k^2_{-1}-k^2_{1})^2)^2) +\nonumber \\
\hbar^2( 13 \hbar^4 k^{10}_{-1} + 34 \hbar^4 k^9_{-1}k_{1} - 39 \hbar^4 k^8_{-1}k^2_{1} - 136 \hbar^4 k^7_{-1}k^3_{1} + 26\hbar^4 k^4_{-1}k^2_{1}(\hbar^2 k^4_{1}+\nonumber \\
4 m \epsilon) + 26 k^6_{-1}(4 m \epsilon \hbar^2 + \hbar^4 k^4_{1})+ 68 k^5_{-1}(4 m \epsilon \hbar^2 k_{1} +
3\hbar^4 k^5_{1})- 136 k^3_{-1}\nonumber \\
(4 m \epsilon \hbar^2 k^3_{1}+ \hbar^4 k^7_{1}) + k^2_{-1}(16 m^2(39 \rho^2 + 12\rho A^2_{1} -
2 A^4_{1}) g^2_{1} + 13 (4 m \epsilon -\nonumber \\
3 \hbar^2 k^4_{1})(4 m \epsilon + \hbar^2 k^4_{1}) + k^2_{1}(16 m^2 (39 \rho^2 - 12\rho A^2_{1} -
2 A^4_{1}) g^2_{1}+ 13 (4 m \epsilon +\nonumber \\
\hbar^2 k^4_{1})^2)+ 2 k_{-1}k_{1}(16 m^2(51 \rho^2 - 12\rho A^2_{1}+ 2 A^4_{1})g^2_{1}+ 17(4 m \epsilon +
\hbar^2 k^4_{1})^2))))
\label{eq:9}
\end{eqnarray}%
\begin{eqnarray}
C^{n=0,1}_{k^4, \omega^4}= -\frac{\hbar^{6}}{16 m^4(A^2_1+A^2_{-1})^2 g_{1}}(4 m \hbar^2 A^6_{-1}g_{1}(k_{-1}-k_{1})
(-2 g_{1}(13 k_{-1} + 7 k_{1}) + \nonumber \\
g_{0}(7 k_{-1} + 13 k_{1}))+ 8(1+(-1)^n)m A^5_{-1}A_{1}g_{0} g_{1}(4 m \epsilon + \hbar^2(k^2_{-1}-k^2_{1})^2) + \nonumber \\
A^3_{-1}A_{1}(4 m \epsilon +\hbar^2(k^2_{-1}-k^2_{1})^2)(24 m ((-1)^n \rho - (1+(-1)^n)A^2_{1})g^2_{1}-2(-1)^n \nonumber \\
\hbar^2 g_{1}(13 k^2_{-1}+ 34 k_{-1}k_{-1}+12 k^2_{1})+ (-1)^n g_{0}(16 m \rho g_{1}-\hbar^2(13 k^2_{-1}+ \nonumber \\
34 k_{-1}k_{-1}+12 k^2_{1}))) + A^4_{1}g_{1}(48 m^2 \rho^2 g^2_{1}-8 m \hbar^2 g_{1}((13 \rho - 7 A^2_{1})k^2_{-1}+ \nonumber \\
2 (17 \rho - 3 A^2_{1})k_{-1}k_{1}+13(\rho + A^2_{1})k^2_{1})+\hbar^4 (41 k^4_{-1}+276 k^3_{-1}k_{1}+486 k^2_{-1}k^2_{1}+ \nonumber \\
276 k^3_{1}k_{-1}+41 k^4_{1})+ 4 m g_{0}(8 m \rho^2 g_{1} + \hbar^2(-13(\rho + A^2_{1})k^2_{-1}+2(-17\rho + 3 A^2_{1}) \nonumber \\
k_{-1}k_{1}+(7 A^2_{1}-13 \rho )k^2_{1})) + A^4_{1}g_{1}(48 m^2 \rho^2 g^2_{1}- 8 m \hbar^2 g_{1}((13 \rho +19 A^2_{1})k^2_{-1}+ \nonumber\\
2 (17 \rho - 9 A^2_{1})k_{-1}k_{1}+ 13(\rho - A^2_{1})k^2_{1})+\hbar^4 (41 k^4_{-1}+276 k^3_{-1}k_{1}+486 k^2_{-1}k^2_{1}+ \nonumber \\
276 k^3_{1}k_{-1}+ 41 k^4_{1})+ 4 m g_{0}(8 m \rho^2 g_{1} + \hbar^2((A^2_{1}-13\rho )k^2_{-1}+2(9 A^2_{1} -17\rho )\nonumber \\
k_{-1}k_{1}- (13 \rho + 19 A^2_{1})k^2_{1}))) + A_{-1}A^3_{1}(4 m \epsilon + \hbar^2(k^2_{-1}-k^2_{1})^2)(- 8 m A^2_{1}g_{1}\nonumber \\
(g_{0}+3 g_{1})- (-1)^n(g_{0}(8 m(-2 \rho+A^2_{1})g_{1}+ \hbar^2(13 k^2_{-1} + 34 k_{1}k_{-1} + 13 k^2_{1})) + \nonumber \\
2 g_{1}(12 m(A^2_{1}-\rho) g_{1}+\hbar^2 (13 k^2_{-1} + 34 k_{1}k_{-1} + 13 k^2_{1})))) + A^2_{1}A^2_{-1}(2 g_{0} \nonumber \\
(32 m^2 \rho^2 g^2_{1}-2 m \hbar^2 g_{1}((26 \rho + 19 A^2_{1})k^2_{-1} + 2(34 \rho - 9 A^2_{1}) k_{1} k_{-1} + (26 \rho - \nonumber \\
A^2_{1}) k^2_{1}) +(4 m \epsilon + \hbar^2(k^2_{-1}-k^2_{1})^2) + g_{1}(48 m^2 \epsilon ^2 +96 m^2 \rho^2 g^2_{1}-8 m \hbar^2 g_{1} \nonumber \\
(( 26 \rho-A^2_{1})k^2_{1}+2(34\rho-9 A^2_{1})k_{1}k_{1}+(26\rho +19 A^2_{1}) k^2_{1})+ \hbar^2(3 \hbar^2 k^8_{-1}+ \nonumber \\
552\hbar^2 k^3_{-1}k_{1}-12 \hbar^2 k^6_{-1}k^2_{1}+552 \hbar^2 k_{-1}k^3_{1}- 12 k^2_{1}k^2_{1}(4 m \epsilon-81 \hbar^2 +
\hbar^2 k^4_{1}) \nonumber \\
+k^4_{1}(24 m \epsilon +82 \hbar^2 +3 \hbar^2 k^4_{1}) + 2k^4_{-1}(12 m \epsilon + 41 \hbar^2 + 9 \hbar^2 k^4_{1})))))
\label{eq:10}
\end{eqnarray}%
\begin{eqnarray}
C^{n=0,1}_{k^6, \omega^4}= -\frac{\hbar^{10}}{16 m^4 g_{1} (A^2_{1}+A^2_{-1})}(3 (-1)^n A_{1}A_{-1} (g_{0}+2 g_{1})(4 m \epsilon +\hbar^2(k^2_{-1}-k^2_{1})^2)+ \nonumber \\
2 (A^2_{1}+A^2_{-1}) g_{1}   (6 m \rho (g_{0} + 2 g_{1}) - \hbar^2 (29 k^2_{-1} + 62 k_{1}k_{-1}+29 k^2_{1})))
\label{eq:11}
\end{eqnarray}%
\begin{eqnarray}
C^{n=0,1}_{k^8, \omega^4}=& -\frac{\hbar^{12}}{8 m^4}
\label{eq:12}
\end{eqnarray}%
\begin{eqnarray}
C^{n=0,1}_{k^8, \omega^4}= -\frac{\hbar^{12}}{8 m^5 (A^2_{1}+A^2_{-1})^3 g_{1}} (
8 m^2 \hbar^2 A^{10}_{1}g^3_{1}(k_{-1}-k_{1})^2 (3 k_{-1}+k_{1})+8 m^2 \hbar^2 A^8_{-1}
g^3_{1}\nonumber \\
(k_{-1}-k_{1})((13 \rho + 5 A^2_{1})k^2_{-1} + 2(11 \rho -3A^2_{1})k_{1}k_{-1} +
(5 \rho + A^2_{1})k^2_{1})-\nonumber \\
2 m A^7_{-1}A_{1}g^2_{1}(4 m \epsilon + \hbar^2(k^2_{-1}-k^2_{1})^2)(8(1+(-1)^n)
m (\rho - A^2_{1})g_{0}\nonumber \\
 (k_{-1}+3 k_{1})+ \hbar^2 (k_{-1}-k_{1})(13 k^2_{-1}+22 k_{-1}k_{1}
+5 k^2_{1}))- 16 m^2 A^6_{-1} g^3_{1}\nonumber \\
(4 m \rho^3 g_{0}(k_{-1}+k_{1}) +
 \hbar^2 ((-3 \rho^2 - 17 \rho A^2_{1}+ A^4_{1})k^3_{-1} + (A^4_{1} - 17 \rho^2 -\nonumber \\
5 \rho A^2_{1}) k^2_{-1}k_{1} + (21 \rho A^2_{1}-17 \rho^2- 5 A^4_{1})k_{-1}k^2_{1} +
 (3 A^4_{1} + \rho A^2_{1}-3 \rho^2)k^3_{1}))-\nonumber \\
8 m^2 A^6_{1}g^3_{1}( 8 m \rho^3 g_{0}(k_{-1}+k_{1})- \hbar^2((6 \rho^2-5 \rho A^2_{1}+
A^4_{1})k^3_{-1} + (34 \rho^2 - \nonumber \\
 17\rho A^4_{1}) k^2_{-1}k_{1} + (34 \rho^2 + 9 \rho A^2_{1} - 5 A^4_{1}) k_{-1} k^2_{1} +
(6 \rho^2 - 13 \rho A^2_{1} + 3 A^4_{1}) k^3_{1}))\nonumber \\
-2 m  A^5_{-1}A_{1}g^2_{1}( 4 m \epsilon + \hbar^2 (k^2_{-1}-k^2_{1})^2)(8 m g_{0}
((3(-1)^n \rho^2-(1+(-1)^n)\nonumber \\
A^2_{1}(\rho+2 A^2_{1}))k_{-1}+(3(-1)^n \rho^2 +(1+(-1)^n )A^2_{1}(\rho - 4 A^2_{1}))
 k_{1})+ \nonumber \\
\hbar^2(-4(-1)^n(\rho -A^2_{1})(k_{-1}+k_{1})(3 k^2_{-1}+14 k_{-1}k_{1} + 3 k^2_{1}) +
 A^2_{1}(33 k^3_{-1}+\nonumber \\
69 k^2_{-1}k_{1} + 43 k_{-1}k^2_{1} + 15 k^3_{1})))- 2 m A_{-1}A^5_{1}g^2_{1}(4 m \epsilon +
\hbar^2 (k^2_{-1}-k^2_{1})^2) \nonumber \\
(4 (-1)^n (A^2_{1}-\rho)(-2 m g_{0}(3 \rho k_{-1} + (3 \rho - 2 A^2_{1})k_{1}) +
\hbar^2 (k_{-1}+k_{1})(k^2_{-1}+ \nonumber \\
14 k_{1}k_{-1} + 3 k^2_{1}) + A^2_{1}(-8 m g_{0}(3 \rho k_{-1}+(5 \rho - 2 A^2_{1})k_{1})+
\hbar^2(7 k^3_{-1} + \nonumber \\
 51 k_{1}k_{-1}+77 k_{-1}k^2_{1}+25 k^3_{1})))+ A^3_{-1} A^3_{1}(4 m \epsilon +
\hbar^2 (k^2_{-1}-k^2_{1})^2)( 2 m \hbar^2 g^2_{1}\nonumber\\
(8(-1)^n(\rho-A^2_{1})(k_{-1}+k_{1}) (3 k^2_{-1} + 14 k_{-1}k_{1}+3 k^2_{1})-
A^2_{1}(3 k_{-1}+5 k_{1})\nonumber\\
(9 k^2_{-1} + 22 k_{1}k_{-1} + 9 k^2_{1})) + g_{0}(\hbar^4 k^9_{-1} +
\hbar^4 k^8_{-1}k_{1}- 4\hbar^4 k^7_{-1}k^2_{1}-4 \hbar^4 k^6_{-1}k^3_{1}+\nonumber \\
 k^5_{-1}(8 m \epsilon \hbar^2 +6 \hbar^4 k^4_{1}) + k^4_{-1}(8 m \epsilon \hbar^2 k_{1} +
6 \hbar^4 k^5_{1})-4 k^3_{-1}(4 m \epsilon \hbar^2 k^2_{1} + \hbar^4 k^6_{1}) - \nonumber \\
 4 k^2_{-1}(4 m \epsilon \hbar^2 k^3_{1} + \hbar^4 k^7_{1})+ k_{1}(-16 m^2(6(-1)^n \rho^2
+(1+(-1)^n)A^2_{1}(A^2_{1}-7 \rho))\nonumber \\
g^2_{1}+(4 m \epsilon + \hbar^2 k^4_{1})^2)+ k_{-1}(-16 m^2(6(-1)^n \rho^2-(1+(-1)^n)A^2_{1}(A^2_{1}+
5 \rho))\nonumber \\
g^2_{1} + (4 m \epsilon + \hbar^2 k^4_{1})^2))) + A^4_{-1}A^2_{1} g_{1} (-12 m \rho g_{0}(k_{-1}+k_{1})
(16 m^2 \rho^2 g^2_{1}+(4 m \epsilon  \nonumber \\
+\hbar^2(k^2_{-1}-k^2_{1})^2)^2) + \hbar^2(3 \hbar^4 k^{11}_{-1} + 17 \hbar^4 k^{10}_{-1}k_{1}+
 5\hbar^4 k^9_{-1}k^2_{1} -65 \hbar^4 k^8_{-1}k^3_{1}+ \nonumber \\
k^7_{-1}(24 m \epsilon \hbar^2 -50 \hbar^4 k^4_{1}) + 2 k^6_{-1}
(68 m \epsilon \hbar^2 k_{1} + 45 \hbar^4 k^5_{1}) + k^5_{-1}(88 m \epsilon \hbar^2 k^2_{1} +\nonumber \\
90 \hbar^4 k^6_{1}) - 2 k^4_{-1}(124 m \epsilon \hbar^2 k^3_{1} + 25 \hbar^4 k^7_{1}) + k^3_{-1}( 48 m^2(3 \rho^2 +
4 \rho A^2_{1}-A^4_{1})g^2_{1}+ \nonumber \\
(12 m \epsilon - 65 \hbar^2 k^4_{1})(4 m \epsilon + \hbar^2 k^4_{1}))+k^3_{1}( 16 m^2(9 \rho^2 +
12 \rho A^2_{1}-A^4_{1})g^2_{1}+ \nonumber \\
3 (4 m \epsilon + \hbar^2 k^4_{1})^2)+ k_{-1}k^2_{1}(16 m^2(9 \rho^2 -
12 \rho A^2_{1}-A^4_{1})g^2_{1} + 17 (4 m \epsilon + \hbar^2 k^4_{1})^2)+ \nonumber \\
 k^2_{-1}k_{1}( 16 m^2(51 \rho^2 - 12 \rho A^2_{1} +
5 A^4_{1})g^2_{1} + (4 m \epsilon + \hbar^2 k^4_{1})(68 m \epsilon + 5 \hbar^2 k^4_{1}))))+ \nonumber \\
 A^2_{-1}A^4_{1} g_{1} (-12 m \rho g_{0}(k_{-1}+k_{1})(16 m^2 \rho^2 g^2_{1}+(4 m \epsilon +
\hbar^2(k^2_{-1}-k^2_{1})^2)^2)+ \nonumber \\
\hbar^2(3 \hbar^4 k^{11}_{-1} + 17 \hbar^4 k^{10}_{-1}k_{1}+
 5\hbar^4 k^9_{-1}k^2_{1} - 65 \hbar^4 k^8_{-1}k^3_{1}+ k^7_{-1}(24 m \epsilon \hbar^2 -
50 \hbar^4 k^4_{1}) +\nonumber \\
 2 k^6_{-1} (68 m \epsilon \hbar^2 k_{1} + 45 \hbar^4 k^5_{1}) + k^5_{-1}
(88 m \epsilon \hbar^2 k^2_{1} + 90 \hbar^4 k^6_{1}) - 2 k^4_{-1}(124 m \epsilon \hbar^2 k^3_{1} +\nonumber \\
 25 \hbar^4 k^7_{1}) + k^3_{-1}(8 m^2(18 \rho^2 -
2 \rho A^2_{1}-A^4_{1})g^2_{1}+(12 m \epsilon - 65 \hbar^2 k^4_{1})(4 m \epsilon + \hbar^2 k^4_{1}))\nonumber \\
+ k^3_{1}( 8 m^2(18 \rho^2 + 34 \rho A^2_{1} + 5 A^4_{1})g^2_{1} + 3 (4 m \epsilon
+ \hbar^2 k^4_{1})^2)+ k_{-1}k^2_{1}(8 m^2(102 \rho^2 +  \nonumber \\
10 \rho A^2_{1}- 11 A^4_{1})g^2_{1} + 17 (4 m \epsilon + \hbar^2 k^4_{1})^2)+ k^2_{-1}k_{1}( 8 m^2(102 \rho^2 -
42 \rho A^2_{1} +  \nonumber \\
7 A^4_{1})g^2_{1} + (4 m \epsilon + \hbar^2 k^4_{1})(68 m \epsilon +5 \hbar^2 k^4_{1})))))
\label{eq:13}
\end{eqnarray}%
\begin{eqnarray}
C^{n=0,1}_{k^5, \omega^3}= -\frac{\hbar^9}{8 m^5 g_{1}(A^2_{-1}+A^2_{1})^2}(-8 m A^6_{-1}g_{1}
(k_{-1}-k_{1})(2 m \rho g^2_{1} - 2 \hbar^2 g_{1} (k_{-1}+k_{1})\nonumber \\
(4 k_{-1}+k_{1}) + \hbar^2 g_{0}(k_{-1}+k_{1})(k_{-1}+4 k_{1})-4 m A^5_{-1}A_{1}g_{1}
(g_{1}(k_{1}-k_{-1}) + \nonumber \\
(1+(-1)^n)g_{0}(k_{-1}+3 k_{1}))(4 m \epsilon + \hbar^2 (k^2_{-1}-k^2_{1})^2)+ A^4_{1}g_{1}
(-16 m^2 \rho g^2_{1} \nonumber \\
((3 \rho-A^2_{1}) k_{-1}+(3 \rho + A^2_{1}) k_{1})+ 8 m \hbar^2 g_{1}(k_{-1}+k_{1})
((3 \rho-2 A^2_{1}) k^2_{-1}+\nonumber \\
2(7\rho-3 A^2_{1})k_{1}k_{-1}+ (3 \rho + 8 A^2_{1}) k^2_{1})- \hbar^4 (k_{1}+k_{-1}(5 k^4_{-1}+
52 k^3_{-1}k_{1}+  \nonumber \\
110 k^2_{-1}k^2_{1} + 52 k_{-1}k^3_{1} + 5 k^4_{1})- 4 m g_{0} (k_{-1}+k_{1})(8 m \rho^2 g_{1} +
\hbar^2(-(3 \rho +\nonumber \\
8 A^2_{1})k^2_{-1} + 2(-7\rho + 3 A^2_{1})k_{1}k_{-1}+ (-3 \rho + 2 A^2_{1}) k^2_{1})))-
A^4_{-1}g_{1}(16 m^2 \rho g^2_{1}\nonumber \\
((3\rho+A^2_{1})k_{-1} + (3\rho-A^2_{1})k_{1})-8 m \hbar^2 g_{1}(k_{1}+k_{-1})((3 \rho +
14 A^2_{1})k^2_{-1} + \nonumber \\
2(7\rho-9 A^2_{1})k_{1}k_{-1}+ (3 \rho + 4 A^2_{1}) k^2_{1}) +
\hbar^4 (k_{1}+k_{-1}(5 k^4_{-1}+ 52 k^3_{-1}k_{1}+ \nonumber \\
110 k^2_{-1}k^2_{1} + 52 k_{-1}k^3_{1} + 5 k^4_{1})-
 4 m g_{0} (k_{-1}+k_{1})(-8 m \rho^2 g_{1} + \hbar^2((3 \rho + \nonumber \\
4 A^2_{1})k^2_{-1} + 2(7\rho - 9 A^2_{1})k_{1}k_{-1}+ (3 \rho + 14 A^2_{1}) k^2_{1})))+
 A^3_{1}A_{1}(k_{1}+k_{-1})\nonumber \\
(4 m \epsilon + \hbar^2(k^2_{-1}-k^2_{1})^2)(24 m A^2_{1}g^2_{1}+(-1)^n
(g_{0}(-16 m \rho g_{1} + \hbar^2(3 k^2_{-1}+ \nonumber \\
14 k_{1}k_{-1} + 3 k^2_{1}))+ 2 g_{1}(12 m(A^2_{1}-\rho) g_{1}+
\hbar^2(3 k^2_{-1}+14 k_{1}k_{-1} +3 k^2_{1})))) +\nonumber \\
A_{-1}A^3_{1}(4 m \epsilon + \hbar^2 (k^2_{-1}-k^2_{1})^2)
(-4 m A^2_{1}g_{1}((g_{0} + 5 g_{1}) k_{-1} + (3 g_{0} +
 7 g_{1}) k_{1})\nonumber \\
-(-1)^n (2 g_{1}(k_{1}+k_{-1})(12 m(A^2_{1}-\rho) g_{1}+\hbar^2(3 k^2_{-1}+
 14 k_{1}k_{-1} +3 k^2_{1})) +\nonumber \\
g_{0}(4 m g_{1}((A^2_{1}-4\rho)k_{1}+(3A^2_{1}-4\rho) k_{1}) +
\hbar^2(k_{1} + k_{-1})(3 k^2_{-1}+14 k_{1}k_{-1} + \nonumber \\
 3 k^2_{1})))) - A^2_{-1}A^2_{1}(2 g_{0}(k_{-1}+k_{1})(32 m^2 \rho^2 g^2_{1}
- 4 m \hbar^2 g_{1}((3 \rho + 7A^2_{1}) k^2_{-1} + \nonumber \\
(14\rho- 9A^2_{1}) k_{1}k_{-1} + (3\rho+2A^2_{1})k^2_{1}) +(4 m \epsilon +
\hbar^2(k^2_{-1}-k^2_{1})^2)+ \nonumber \\
g_{1}(3 \hbar^4k^9_{-1} + 16 m^2(3\epsilon^2+\rho(6\rho+A^2_{1})g^2_{1})k_{1} +3\hbar^4 k^8_{-1}k_{1}+
 12 \hbar^4 k^7_{-1}k^2_{1}-\nonumber \\
16 m \hbar^2(3\rho+7 A^2_{1})g_{1}k^3_{1}+2 \hbar^2(12 m \epsilon+5 \hbar^2)k^5_{1}+3\hbar^4 k^9_{1} +
6\hbar^4 k^4_{-1}k_{1} ( 4 m \epsilon+\nonumber\\
 19\hbar^2 + 3\hbar^2 k^4_{1} + 2 k^5_{-1}(12 m\epsilon \hbar^2 +5 \hbar^4 +9\hbar^4 k^4_{1}) +
k_{-1}(16 m^2(3\epsilon^2+\rho(6\rho- \nonumber\\
A^2_{1})g^2_{1})+ 16 m \hbar^2(2A^2_{1}-17\rho)g_{1}k^2_{1}+ 6\hbar^2(4 m \epsilon +
19\hbar^2)k^4_{1}+3 \hbar^4 k^8_{1})- \nonumber \\
4\hbar^2 k^2_{-1}k_{1}(4 m(17\rho-7A^2_{1})
g_{1}+ 3 k^2_{1}(4 m \epsilon -27\hbar^2+\hbar^2 k^4_{1}))- 4\hbar^2 k^3_{-1}\nonumber \\
(4 m(3\rho+2A^2_{1}) g_{1}+ 3 k^2_{1}(4 m \epsilon -27\hbar^2+\hbar^2 k^4_{1})))))
\label{eq:14}
\end{eqnarray}%
\begin{eqnarray}
C^{n=0,1}_{k^7, \omega^3}= -\frac{\hbar^{11}}{8 m^5 g_{1}(A^2_{-1}+A^2_{1})}(-4 m A^4_{-1}
(g_{0}-2 g_{1})g_{1}(k_{-1}-k_{1})+3(-1)^n A_{1}A_{-1}(g_{0}+\nonumber \\
2g_{1})(k_{1}+k_{-1}) (4 m \epsilon + \hbar^2 (k^2_{-1}-k^2_{1})^2)+ 2 A^2_{-1}g_{1}(k_{1}+
k_{-1})( 6 m \rho g_{0}+\nonumber \\
12 m \rho g_{1}-\hbar^2(9k^2_{1}+22k_{1}k_{-1}+9k^2_{1}))+ 2 A^2_{1} g_{1}(2 m g_{0}
((3\rho+A^2_{1})k_{-1}+\nonumber \\
(3\rho - A^2_{1})k_{1})+4 m g_{1}((3\rho -A^2_{1})k_{1}-\hbar^2(9k^2_{-1}k_{1}+31 k_{-1}k^2_{1}+9 k^3_{1})))
\label{eq:15}
\end{eqnarray}%
\begin{eqnarray}
C^{n=0,1}_{k^9, \omega^3}=& -\frac{3\hbar^{13}(k_{1}+k_{-1})}{4 m^5}
\label{eq:16}
\end{eqnarray}%
\begin{eqnarray}
C^{n=0,1}_{k^4, \omega^2}= -\frac{\hbar^6}{16 m^6 g_{1}(A^2_{-1}+A^2_{1})^3} ( -12 m^2 \hbar^2
A^9_{-1} A_{1}g_{0}g^2_{1}(k_{-1}-k_{1})^2(4 m \epsilon +\hbar^2(k^2_{-1}- \nonumber \\
k^2_{1})^2)+  4 m^2 \hbar^2 A^{10}_{-1}g^3_{1}(k_{-1}-k_{1})^2(12 m \rho g_{0}+\hbar^2(13 k^2_{-1}+
10 k_{1}k_{-1}+\nonumber \\
k^2_{1}))+4 m^2 \hbar^2 A^{8}_{-1}g^2_{1}(k_{-1}-k_{1})(4 g_{0}
(m \rho(3 A^2_{1}-5\rho)g_{1} + \hbar^2 A^2_{1}(k_{-1}- \nonumber \\
k_{1})^2)(k_{-1}-k_{1})+ \hbar^2 g_{1}((24\rho+29 A^2_{1})k^3_{-1}+
3(28\rho-9A^2_{1})k^2_{-1}k_{1}+ \nonumber \\
3(16\rho-3A^2_{1})k^2_{1}k_{-1}+(4\rho+ 7 A^2_{1})k^3_{1}))+4 m^2 \hbar^2 A^6_{1}g^3_{1}
(-4 m\rho g_{0}((4\rho^2 + \nonumber \\
5\rho A^2_{1}-3 A^4_{1})k^2_{-1}+2(8\rho^2- 5\rho A^2_{1}+3A^4_{1}) k_{-1}k_{1} +
(4 \rho^2+5\rho A^2_{1}-3 A^4_{1})\nonumber \\
k^2_{1})+\hbar^2((A^2_{1}-2 \rho)^2 k^4_{-1}+4(14\rho^2-11\rho A^2_{1}+2A^4_{1})
 k^3_{-1}k_{1} + 6(20\rho^2-\nonumber \\
6\rho A^2_{1}-A^4_{1})k^2_{-1}k^2_{1}+4(14\rho^2+
15\rho A^2_{1}-4A^4_{1})k_{-1}k^3_{1}+(4\rho^2+24\rho A^2_{1}+\nonumber \\
13 A^4_{1})k^4_{1})) - 4 m \hbar^2 A_{-1}A^5_{1}g^2_{1}(4 m \epsilon+\hbar^2(k^2_{-1}-k^2_{1})^2)
(m g_{0}(2(-1)^n(\rho-\nonumber \\
A^2_{1})((6\rho + 7 A^2_{1})k^2_{-1}+2(12\rho - 7 A^2_{1})k_{-1}k_{1}
 +(6\rho-5A^2_{1}) k^2_{1})+A^2_{1}\nonumber \\
(-(8\rho+11 A^2_{1})k^2_{-1}+2(11A^2_{1} -28\rho)k_{-1}k_{1}+
(13 A^2_{1}-32\rho)k^2_{1}))+\nonumber \\
 \hbar^2(A^2_{1}(k_{1}+k_{-1})(k^3_{-1}+16 k^2_{-1}k_{1}+35k^2_{1}k_{-1}+
8 k^3_{1})+2(-1)^n (A^2_{1}-\rho)\nonumber \\
(k^4_{-1}+14 k^3_{-1}k_{1}+30 k^2_{-1}k^2_{1}+14 k_{-1}k^3_{1}+k^4_{1})))-
4 m A^7_{-1}A_{1}g^2_{1}(4 m \epsilon +\nonumber \\
\hbar^2(k^2_{-1}-k^2_{1})^2)(-32(1+(-1)^n) m^2 A^4_{1} g_{1}(g_{0}+g_{1})+
\hbar^2(2 m \rho g_{0}((-4+ \nonumber \\
(-1)^n)k^2_{-1}+10(2+(-1)^n)k_{-1}k_{1}+(8+13(-1)^n)k^2_{1}) +
\hbar^2(k_{-1}-k_{1})\nonumber \\
(6 k^3_{-1}+ 21 k^2_{-1}k_{1}+12 k^2_{1}k_{-1}+k^3_{1}))+2(1+(-1)^n) mA^2_{1}
(16 m \rho g^2_{1}+ \nonumber \\
g_{0}(16 m \rho g_{1}- \hbar^2(k^2_{-1}+10 k_{1}k_{-1}+
 13 k^2_{1}))))-4 m A^5_{-1}A_{1}g^2_{1}(4 m \epsilon+\hbar^2 \nonumber \\
(k^2_{-1}-k^2_{1})^2)(-64(1+(-1)^n) m^2 A^6_{1}g_{1}(g_{0}+g_{1})+
2(-1)^n \rho \hbar^2 (6 m \rho g_{0} \nonumber \\
(k^2_{-1}+4 k_{-1}k_{1}+k^2_{1})- \hbar^2(k^4_{-1}+
14 k^3_{-1} k_{1}+30 k^2_{-1}k^2_{1} + 14 k_{-1}k^3_{1}+k^4_{1}))+\nonumber \\
\hbar^2 A^2_{1}(6 m \rho g_{0}(k_{-1}-k_{1})((-4+(-1)^n)k_{-1} - 5(-1)^n k_{1}+
\hbar^2((13 + 2 (-1)^n)\nonumber \\
k^4_{-1} + (47 + 28(-1)^n) k^3_{-1}k_{1}+ 3(11+20(-1)^n) k^2_{-1}k^2_{1} +
7(3+4(-1)^n) \nonumber \\
k_{-1}k^3_{1}+2(3+(-1)^n)k^4_{1}))+2 m A^4_{1}(32(1+(-1)^n) m \rho g^2_{1}+
 g_{0}(32(1+\nonumber \\
(-1)^n) m \rho g_{1}-3 \hbar^2((4+3(-1)^n)k^2_{-1}+
2(-1)^n k_{-1}k_{1}+(8+7(-1)^n)\nonumber \\
k^2_{1}))))+m \hbar^2 A^6_{-1}g_{1}
 (-8 m A^4_{1}g_{1}(k_{-1}-k_{1})^2(6 g_{0}(2 m\rho g_{1}-\hbar^2
(k_{-1}-k_{1})^2)+\nonumber \\
\hbar^2 g_{1}(-5 k^2_{-1}+22 k_{-1}k_{1}+7 k^2_{1}))-16 m \rho^2 g^2_{1}
 (4 m \rho g_{0}(k^2_{-1}+4 k_{-1}k_{1}+k^2_{1})- \nonumber \\
\hbar^2(k^4_{-1}+14 k^3_{-1}k_{1} + 30 k^2_{-1}k^2_{1}+14 k_{-1}k^3_{1} +
k^4_{1}))- A^2_{1}(k_{-1}-k_{1})\nonumber \\
 (-16 m \rho \hbar^2 g^2_{1}(17 k^3_{-1}+51 k^2_{-1}k_{1}+15 k_{-1} k^2_{1}
-3 k^3_{1})+ 5 g_{0}(k^2_{-1}-k^2_{1}) \nonumber \\
(64 m^2 \rho^2 g^2_{1} + (4 m \epsilon+\hbar^2(k^2_{-1}-
 k^2_{1})^2)^2)))- A^3_{-1}A^3_{1}(4 m \epsilon+\hbar^2(k^2_{-1}-k^2_{1})^2)\nonumber \\
(-128 (1+(-1)^n) m^3 A^6_{1}g^3_{1}(g_{0}+g_{1})+ 4 m \hbar^2 A^2_{1}g^2_{1}(6 m \rho g_{0}
 ((-4+(-1)^n) \nonumber \\
k^2_{-1}-2(6+11(-1)^n)k_{-1}k_{1}-(8+3(-1)^n)k^2_{1})+
\hbar^2(4(2+(-1)^n)k^4_{-1} + \nonumber \\
7(7+8(-1)^n) k^3_{-1}k_{1} + 3(31+40(-1)^n)k^2_{-1}k^2_{1}+(75+56(-1)^n)k_{-1}k^3_{1}\nonumber \\
+(15+4(-1)^n)k^4_{1})) + 8(1+(-1)^n) m^2
 A^4_{1} g^2_{1}(16 m \rho g^2_{1}+g_{0}(16 m \rho g_{1}-3 \hbar^2 \nonumber \\
(5k^2_{-1}-6 k_{-1}k_{1}+k^2_{1})))-
\hbar^2(16(-1)^n m \rho \hbar^2 g^2_{1}(k^4_{-1}+14 k^3_{-1}k_{1}+30 k^2_{-1}k^2_{1}+ \nonumber \\
14 k_{-1}k^3_{1}+k^4_{1})+ g_{0}(k^2_{-1}+4 k_{-1}k_{1}+k^2_{1})
(-96 (-1)^n m^2 \rho^2 g^2_{1} + (4 m \epsilon + \nonumber \\
\hbar^2(k^2_{-1}-k^2_{1})^2)^2))) + \hbar^2 A^4_{-1}A^2_{1}g_{1}(\hbar^2
(\hbar^4k^{12}_{-1}+14\hbar^4k^{11}_{-1}k_{1}+
26\hbar^4k^{10}_{-1}k^2_{1}- \nonumber\\
42\hbar^4k^{9}_{-1}k^3_{1}+28\hbar^2k^{5}_{-1}k^3_{1}(\hbar^2 k_{1}-4 m \epsilon) + k^8_{-1}
(8 m \epsilon \hbar^2 -113 \hbar^4 k^4_{1})+28 k^7_{-1}\nonumber\\
(4 m \epsilon \hbar^2 k_{1} + \hbar^4 k^5_{1})
+4 k^6_{-1}(56 m \epsilon \hbar^2 k^2_{1} + 43 \hbar^4 k^6_{1})+k^4_{-1}
(16 m^2 \epsilon^2+8 m^2 \nonumber \\
(6 \rho^2 + 30 \rho A^2_{1}-7A^4_{1}) g^2_{1}- 464 m \epsilon \hbar^2 k^4_{1}-
113 \hbar^4 k^8_{1})+2k^3_{-1}k_{1}(16 m^2(21 \rho^2 +\nonumber \\
6 \rho A^2_{1}-2 A^4_{1}) g^2_{1} +7 (4 m \epsilon-3 \hbar^2 k^4_{1})
(4 m \epsilon + \hbar^2 k^4_{1}))+ k^4_{1}(8 m^2(6 \rho^2 + 30 \rho A^2_{1} + \nonumber \\
5 A^4_{1}) g^2_{1} + (4 m \epsilon + \hbar^2 k^4_{1})^2)+
2 k_{-1}k^3_{1}(16 m^2 (21 \rho^2 + 6 \rho A^2_{1}- 8 A^4_{1}) g^2_{1}+  \nonumber \\
7 (4 m \epsilon + \hbar^2 k^4_{1})^2)+2 k^2_{-1}k^2_{1}(24 m^2 (30 \rho^2 -
18 \rho A^2_{1} + 7 A^4_{1}) g^2_{1}+(4 m \epsilon +\hbar^2 k^4_{1}) \nonumber \\
 (60 m \epsilon + 13 \hbar^2 k^4_{1})))-2 m g_{0}(\hbar^4 (6\rho + 5A^2_{1})
k^{10}_{-1}+2\hbar^4(12\rho - 5A^2_{1})k^9_{-1}k_{1}-\nonumber \\
 3\hbar^4 (6\rho + 5 A^2_{1}) k^8_{-1}k^2_{1}+8\hbar^4(-12\rho + 5A^2_{1})k^7_{-1}k^3_{1}+ 2\hbar^2
(6\rho + 5A^2_{1})k^6_{-1}(4 m \epsilon +\nonumber \\
\hbar^2 k^4_{1})+4\hbar^2(12\rho - 5A^2_{1})k^5_{-1}k_{1}(4 m \epsilon + 3 \hbar^2 k^4_{1})+
 8 \hbar^2 k^3_{-1}k_{1}(12 m A^4_{1}g_{1}+4 m \epsilon \nonumber \\
(5 A^2_{1}-12\rho)k^2_{1}+\hbar^2(5 A^2_{1}-12\rho)k^6_{1})
2 \hbar^2 k^4_{-1}-12 m A^4_{1} g_{1}- 4 m \epsilon(6\rho+
5\rho A^2_{1})\nonumber \\
k^2_{1}+\hbar^2 (6\rho+ 5\rho A^2_{1})k^6_{1})+k^2_{-1}
(16 m^2(\epsilon^2(6\rho+ 5\rho A^2_{1})+3\rho(2\rho^2+5\rho A^2_{1}+\nonumber \\
A^4_{1})g^2_{1})- 144 m \hbar^2 A^4_{1} g_{1} k^2_{1}-
8 m \epsilon \hbar^2(6\rho+
 5 A^2_{1})k^4_{1}3\hbar^4(6\rho+ 5\rho A^2_{1})k^8_{1})+\nonumber\\
2 k_{-1}k_{1}(48 m A^4_{1}g_{1}(\hbar^2 k^2_{1}-m \rho g_{1})+
12\rho(m^2 \rho^2 g^2_{1}+(4 m \epsilon+\hbar^2 k^4_{1})^2)- \nonumber\\
5 A^2_{1}(48 m^2 \rho^2 g^2_{1}+(4 m \epsilon+\hbar^2 k^4_{1})^2))+
k^2_{1}(24 m A^4_{1} g_{1}(2 m \rho g_{1}-\hbar^2k^2_{1})+ 6\rho \nonumber\\
(16 m^2 \epsilon^2 \rho^2 g^2_{1}+ (4 m \epsilon -\hbar^2k^4_{1})^2)+
5 A^2_{1}(16 m^2 \epsilon^2 \rho^2 g^2_{1}+(4 m \epsilon -\hbar^2k^4_{1})^2))))+\nonumber\\
\hbar^2 A^2_{-1}A^2_{1}g_{1}(\hbar^2(\hbar^4k^{12}_{-1}+14\hbar^4 k^{11}_{-1}k_{1}+
 26\hbar^4 k^{10}_{-1}k^2_{1}- 42\hbar^4 k^9_{-1}k^3_{1}+ \nonumber\\
28\hbar^2 k^5_{-1}k^3_{1}(\hbar^2k^4_{1}-4 m \epsilon)+
k^8_{-1}(4 m \epsilon \hbar^2 - 113\hbar^4k^4_{1})+28 k^7_{-1}(4 m \epsilon \hbar^2 k_{1}+ \nonumber\\
 \hbar^4k^5_{1})+ 4k^6_{-1}(56 m \epsilon \hbar^2 k^2_{1}+ 43\hbar^4k^6_{1})+ k^4_{-1}(16 m^2 \epsilon^2 +
(12\rho^2 + 12\rho A^2_{1} - \nonumber\\
7 A^2_{1})g^2_{1}-464 m \epsilon \hbar^2 k^4_{1}-
113 \hbar^4 k^8_{1})+ 2k^3_{-1}k_{1}(16 m^2(21\rho^2 - 9\rho A^2_{1}+2 A^2_{1})\nonumber\\
g^2_{1}+ 7(4 m \epsilon
-3 \hbar^2 k^4_{1})(4 m \epsilon+\hbar^2 k^4_{1})^2) + k^4_{1}(4 m^2(12\rho^2 +
 68\rho A^2_{1}+ 29 A^4_{1}) \nonumber\\
g^2_{1}+(4 m \epsilon+\hbar^2 k^4_{1})^2)+ 2 k_{-1}k^3_{1}
(16 m^2(21\rho^2 + 17\rho A^2_{1}-7 A^4_{1})g^2_{1}+ 7(4 m \epsilon +\nonumber\\
\hbar^2 k^4_{1})^2)+
 2 k^2_{-1}k^2_{1}(36 m^2(20\rho^2 - 8\rho A^2_{1}+ A^4_{1})g^2_{1}+ (4 m \epsilon + \hbar^2 k^4_{1})
(60 m \epsilon + \nonumber\\
13 \hbar^2 k^4_{1})))- m g_{0}(\hbar^4(12\rho + 5A^2_{1}) k^{10}_{-1}+
 2 \hbar^4(24\rho - 5A^2_{1}) k^{9}_{-1}k_{1}- 3 \hbar^4(12\rho + \nonumber\\
5A^2_{1}) k^{8}_{-1}k^2_{1}+8 \hbar^4 (-24\rho +5A^2_{1})k^{7}_{-1}k^3_{1}+
2\hbar^2(12\rho + 5A^2_{1}) k^{6}_{-1}(4 m \epsilon+\hbar^2 k^4_{1}) +\nonumber\\
 4\hbar^2 (24\rho - 5A^2_{1})k^{5}_{-1}k_{1}(4 m \epsilon+3\hbar^2 k^4_{1})+
8 \hbar^2 k^3_{-1}k_{1}(8 m A^4_{1}g_{1}+4 m \epsilon(-24\rho + \nonumber\\
5A^2_{1})k^{2}_{1}+ \hbar^2(-24\rho + 5A^2_{1})k^{6}_{1})+ 2 \hbar^2 k^4_{-1}
(-8 m A^4_{1}g_{1}-4 m \epsilon(12\rho + 5A^2_{1}) k^{6}_{1})+\nonumber\\
 k^2_{-1}(192 m^2 \epsilon^2 \rho + 16 m^2(5\epsilon^2 A^2_{1}+
\rho(12\rho^2+20\rho A^2_{1}-3 A^4_{1}) g^2_{1})- 96 m \hbar^2A^4_{1}) \nonumber\\
g_{1}k^2_{1}-8 m\epsilon \hbar^2 (12\rho + 5 A^2_{1})k^{8}_{1})+ 2 k_{-1}k_{1}
(16 m A^4_{1}g_{1}(3 m\rho g_{1} + 2\hbar^2k^2_{1})+24 \rho  \nonumber\\
(16 m^2\rho^2 g^2_{1}+(4 m \epsilon+\hbar^2 k^4_{1})^2)-
5 A^2_{1}(16 m^2\rho^2 g^2_{1}+(4 m \epsilon+\hbar^2 k^4_{1})^2))+  \nonumber\\
k^2_{1}(-16 m A^4_{1} g_{1}(3 m \rho g_{1}+\hbar^2 k^2_{1})+12\rho(16 m^2\rho^2 g^2_{1}+
(4 m \epsilon+\hbar^2 k^4_{1})^2)+ \nonumber\\
5 A^2_{1}(64 m^2\rho^2 g^2_{1}+(4 m \epsilon+\hbar^2 k^4_{1})^2)))))
\label{eq:17}
\end{eqnarray}%
\begin{eqnarray}
C^{n=0,1}_{k^6, \omega^2}= -\frac{\hbar^8}{32 m^6 g_{1}(A^2_{-1}+A^2_{1})^3}
(-16 m^2 \hbar^2 A^{10}_{-1}(4 g_{0}-g_{1})g^2_{1}(k_{-1}-k_{1})^2-4 m A^7_{-1}A_{1}g_{1} \nonumber\\
(4 m \epsilon+\hbar^2 (k^2_{-1}-k^2_{1})^2)(-2 g_{1}(8(1+(-1)^n)m A^2_{1} g_{1}+
3\hbar^2(-k^2_{1}-k^2_{1}))+  \nonumber\\
g_{0}(8(1+(-1)^n) m (\rho-4 A^2_{1}g_{1}- \hbar^2(((-1)^n - 5)k^2_{-1}+
2(11+5(-1)^n)  \nonumber\\
k_{1}k_{-1}+(7+13(-1)^n)k^2_{1})))+ 2 m \hbar^2 A^8_{-1}g_{1}(k_{-1}-k_{1})
(2 g_{1}(4 m g_{1}((6\rho+  \nonumber\\
A^2_{1})k_{1}+(6\rho- A^2_{1})k_{1})-\hbar^2(37k^3_{-1}+81k^2_{-1}k_{1}+39k_{-1}k^2_{1}+3k^3_{1}))+
g_{0} \nonumber\\
(16 m(3\rho-2A^2_{1})g_{1}(k_{-1}-k_{1})+\hbar^2(3k^3_{-1}+
 39k^2_{-1}k_{1}+81k_{-1}k^2_{1}+37k^3_{1})))  \nonumber\\
- 2A^6_{-1}g_{1}(-\hbar^2(8 m62 g^2_{1}
((9\rho^2+12\rho A^2_{1}-2 A^4_{1})k^2_{-1}+2(9\rho^2+2 A^4_{1})k_{1}k_{-1}+  \nonumber\\
(9\rho^2-2(6\rho A^2_{1}+A^4_{1}))k^2_{1})+ \hbar^4(k_{-1}+k_{1})^2(k^4_{-1}+22 k^3_{-1}k_{1}+
66 k^2_{1}k^2_{1}+  \nonumber\\
22k_{-1}k^3_{1}+k^4_{1})-8 m \hbar^2 g_{1}((\rho+ 27A^2_{1})k^4_{-1}+
2(7\rho+12 A^2_{1})k^3_{-1}k_{1}+6(5\rho-  \nonumber\\
7 A^2_{1})k^2_{-1}k^2_{1} + 2(7\rho-8A^2_{1})k_{-1}k^3_{1}+
(\rho+7A^2_{1})k^4_{1}))+4 m g_{0} (8 m^2 \rho^3 g^2_{1}-  \nonumber\\
8 m\hbar^2 g_{1}((\rho^2+ 6\rho A^2_{1}+2 A^4_{1})k^2_{-1}+
4(\rho^2-3\rho A^2_{1}-A^4_{1})k_{-1}k_{1}+(\rho^2+  \nonumber\\
6\rho A^2_{1}+2 A^4_{1})k^2_{-1}+ 4(\rho^2-3\rho A^2_{1}-A^4_{1})k_{-1}k_{1}+(\rho^2+6\rho A^2_{1}
+2 A^4_{-1})k^2_{1})+  \nonumber\\
\hbar^4 ((\rho+7A^2_{1})k^4_{-1}+2(7\rho-8A^2_{1})k^3_{-1}k_{1}+
 6(5\rho-7 A^2_{1})k^2_{-1}k^2_{2}+2 (\rho+12 A^2_{1})  \nonumber\\
k_{-1}k^3_{1}+(\rho +2 A^2_{1})k^4_{1})))+
2A^6_{1}g_{1}(-m g_{0}(32 m^2 \rho^3 g^2_{1}- 16 m \hbar^2 g_{1}
 ((2\rho -A^2_{1})  \nonumber\\
(\rho+2 A^2_{1})k^2_{-1}+2(4 \rho^2-3\rho A^2_{1}+2 A^4_{1})k_{-1}k_{1}+
(2\rho-A^2_{1})(\rho + 2A^2_{1})k^2_{1})+\hbar^4  \nonumber\\
(( 4\rho+3 A^2_{1})k^4_{-1}+ 4(14\rho+11A^2_{1})k^3_{-1}k_{1}+ 6(20\rho-
7A^2_{1})k^2_{-1}k^2_{1}+4(14\rho-  \nonumber\\
9 A^2_{1})k_{-1}k^3_{1}+(4\rho-3A^2_{1})k^4_{1}))+\hbar^2 (8 m^2 g^2_{1}
((3\rho - A^2_{1})k_{-1}+ (3\rho+A^2_{1})k_{1})^2+ \nonumber\\
\hbar^4(k_{-1}+k_{1})^2(k^4_{-1}+22 k^3_{-1}+66 k^2_{-1}k^2_{1}
+22 k_{-1}k^3_{1}+ k^4_{1})-2 m \hbar^2 g_{1}((4\rho- \nonumber\\
3 A^2_{1})k^4_{-1}+4(14\rho-9 A^2_{1})k^3_{1}k_{1}+6(20\rho-
7 A^2_{1})k^2_{-1}k^2_{1}+4(14\rho+11 A^2_{1}) \nonumber\\
k_{-1}k^3_{1}+(4\rho +37A^2_{1})k^4_{1})))-
 2 A_{-1}A^5_{1}(4 m \epsilon+\hbar^2 (k^2_{-1}-k^2_{1})^2)(g_{0}(8 m^2  \nonumber\\
 (3(-1)^n \rho^2+(1+(-1)^n) A^2_{1}(A^2_{1}-4 \rho))g^2_{1}- 2 m \hbar^2 g_{1}(( 8(-1)^n\rho +  \nonumber\\
(-1+5(-1)^n)A^2_{1})k^{2}_{-1}+2(16(-1)^n\rho -(5+11(-1)^n)A^2_{1})k_{-1}k_{1}+ \nonumber\\
(8(-1)^n\rho -(13+7(-1)^n)A^2_{1})k^2_{1})+
(-1)^n\hbar^4(k^4_{-1}+14k^3_{-1}k_{1}+30 k^2_{-1}k^2_{1}+ \nonumber\\
14 k_{-1}k^3_{1}+k^4_{1}))+2\hbar^2 g_{1}((-1)^n\hbar^2(k^4_{-1}+14k^3_{-1}k_{1}+
30 k^2_{-1}k^2_{1} + 14 k_{-1}k^3_{1} + \nonumber\\
k^4_{1})+6 m g_{1}(k_{-1}+k_{1})(-3(-1)^n(\rho-A^2_{1})k_{-1}(k_{-1}+k_{1})+
2A^2_{1}(k_{-1}+2k_{1})))) \nonumber\\
+A^3_{-1}A^3_{1}(4 m \epsilon+\hbar^2 (k^2_{-1}-k^2_{1})^2)
(g_{0}(96 m^2(-(-1)^n \rho^2+(1+(-1)^n)A^2_{1}(\rho+ \nonumber\\
A^2_{1}))g^2_{1}+ 4 m \hbar^2 g_{1}(( 16(-1)^n\rho +
(-7 +11(-1)^n)A^2_{1})k^2_{-1}+2(A^2_{1}+(-1)^n \nonumber\\
(32\rho- 17 A62_{1}))k_{-1}k_{1}+(16(-1)^n\rho - (19+(-1)^n)A^2_{1})k^2_{1})-
4(-1)^n \hbar^4 \nonumber\\
(k^4_{-1}+14 k^3_{-1}k_{1}+30 k^2_{-1}k^2_{1}+14 k_{-1}k^3_{1}+
k^4_{1}+(4 m \epsilon+\hbar^2 (k^2_{-1}-k^2_{1})^2)^2)+ \nonumber\\
8 g_{1}(8(1+(-1)^n )m^2 A^4_{1}g^2_{1}- 3 m \hbar^2 A^2_{1} g_{1}
(k_{-1}+k_{1})((5+6(-1)^n )k_{-1}+ \nonumber\\
(7+6(-1)^n)k_{1})-(-1)^n \hbar^2(-18 m\rho g_{1}(k_{-1}+k_{1})^2+
 \hbar^2(k^4_{-1}(k^4_{-1}+14k^3_{-1}k_{1}+ \nonumber\\
30 k^2_{-1}k^2_{1} + 14 k_{-1}k^3_{1}+k^4_{1}))))- 2A^5_{-1}A_{1}(4 m \epsilon+
\hbar^2 (k^2_{-1}-k^2_{1})^2)(g_{0}(24 m^2\nonumber\\
 ((-1)^n \rho^2 - 5(1+(-1)^n)A^4_{1}) g^2_{1}- 2 m \hbar^2 g_{1}((8(-1)^n\rho +
(-11+7(-1)^n)A^2_{1}) \nonumber\\
k^2_{-1}+ 2(16 (-1)^n \rho^2 -(-17+(-1)^n)A^2_{1})k_{1}k_{-1}+
(8(-1)^n \rho+(1+19 (-1)^n) \nonumber\\
A^2_{1})k^2_{1})+(-1)^n \hbar^4(k^4_{-1}+14 k^3_{-1}k_{1}+30 k^2_{-1}k^2_{1}+
14 k_{-1}k^3_{1}+ k^4_{1}))+2 g_{1}\nonumber\\
(-32(1+(-1)^n) m^2 A^4_{1}g^2_{1}+6 m \hbar^2 A^2_{1}g_{1}(k_{-1}+k_{1})
((4 + 3(-1)^n)k_{-1}+ \nonumber\\
(2+ 3(-1)^n)k_{1})+(-1)^n \hbar^2(-18 m\rho g_{1}(k_{-1}+k_{1})^2+
\hbar^2(k^4_{-1}+14 k^3_{-1}k_{1}+ \nonumber\\
30 k^2_{-1}k^2_{1}+ 14 k_{-1}k^3_{1}+k^4_{1}))))+A^2_{-1} A^4_{1}(\hbar^2 g_{1}
 (9\hbar^4 k^{10}_{-1}+18\hbar^4 k^9_{-1}k_{1}+16 m^2  \nonumber\\
(9 \epsilon^2+(3\rho+A^2_{1})((9\rho+A^2_{1})g^2_{1})k^2_{1})-27 \hbar^4 k^8_{-1}k^2_{1}-
72\hbar^4k^7_{-1}k^3_{1}-  \nonumber\\
48 m \hbar^2(\rho+9 A^2_{1})g_{1}k^4_{1}+6(12 m \epsilon \hbar^2+\hbar^4)k^6_{1}+9 \hbar^4k^{10}_{1}+
36 \hbar^2 k^5_{-1}k_{1}(4(m \epsilon+  \nonumber\\
\hbar^2)+3\hbar^2 k^4_{1})+6 k^6_{-1}(12 m \epsilon \hbar^2+
 \hbar^4+ 3\hbar^4 k^4_{1}) + k^2_{-1}(16 m^2(9 \epsilon^2 + (27 \rho^2 - \nonumber\\
12\rho A^2_{1})A^4_{1})g^2_{1})+ 96 m \hbar^2 (7 A^2_{1}-15 \rho) g_{1} k^2_{1} + 18 \hbar^2(37 \hbar^2-
 4 m \epsilon) k^4_{1} - 27 \hbar^4 k^8_{1})  \nonumber\\
+2 k_{1}k_{-1}(16 m^2(9 \epsilon^2 + (27 \rho^2 - A^2_{1})g^2_{1})-
48 m \hbar^2(7\rho + 4A^2_{1})g_{1} k^2_{1} + \nonumber\\
72 \hbar^2(m \epsilon + \hbar^2) k^4_{1} + 9\hbar^4 k^8_{1}) + 2 \hbar^2 k^4_{-1}
(-8 m(3\rho+7A^2_{1}) g_{1}+9 k^2_{1} (-4 m\epsilon + \nonumber \\
37 \hbar^2+ 3 \hbar^2k^4_{1}))-8 \hbar^2 k^3_{-1}k_{1}(4 m(21 \rho-
8 A^2_{1}) g_{1}+ 3 k^2_{1}(12 m \epsilon-44 \hbar^2 +\nonumber \\
3 \hbar^2 k^4_{1})))-4 g_{0}(48 m^3 \rho^3 g^3_{1} + 16 m^2 \hbar^2 g^2_{1}((A^4_{1}-
3\rho^2-6\rho A^2_{1})k^2_{-1} -2(6 \rho^2 -\nonumber \\
6\rho A^2_{1}+ A^4_{1}) k_{-1}k_{1}+ (A^4_{1}-3\rho^2 - 6\rho A^2_{1}) k^2_{-1})-( k^2_{-1}+4 k_{1}k_{-1}+k^2_{1})
(4 m \epsilon \hbar \nonumber \\
+ \hbar^3(k^2_{1}-k^2_{-1})^2)^2+ m g_{1}(48 m^2 \epsilon^2 \rho +
 \hbar^2(3 \rho \hbar^2 k^8_{-1}+12 \hbar^2( 7\rho+4A^2_{1})k^3_{-1}k_{1} \nonumber \\
- 12 \rho \hbar^2 k^6_{-1} k^2_{1} + 4 \hbar^2(21\rho - 8 A^2_{1}) k_{-1}k^3_{1} -
12 k^2_{-1}k^2_{1}(4 m \epsilon \rho-15\rho \hbar^2+ 7 \hbar^2 A^2_{1}\nonumber \\
+\rho \hbar^2 k^4_{1}) + 6 k^4_{-1}(\rho(4 m\epsilon + \hbar^2)+3 \hbar^2(3 A^2_{1}+\rho k^4_{1}))+
k^4_{1}(6\rho(4 m \epsilon + \hbar^2) + \nonumber \\
\hbar^2(14 A^2_{1}+ 3\rho k^4_{1}))))))+ A^4_{-1} A^2_{1}(\hbar^2 g_{1}
(9\hbar^4 k^{10}_{-1}+18\hbar^4 k^9_{-1}k_{1}+16 m^2(9 \epsilon^2+\nonumber \\
(27 \rho^2-2A^4_{1})g^2_{1})k^2_{1}-27\hbar^4 k^8_{-1}k^2_{1}- 72 \hbar^4 k^7_{-1}k^3_{1}-
 24 m \hbar^2(2\rho + 17A^2_{1})g_{1}k^4_{1}+ \nonumber \\
6(12 m \epsilon \hbar^2 + \hbar^4)k^6_{1}+9 \hbar^4
k^{10}_{1}+36 \hbar^2 k^5_{-1}k_{1}(4( m \epsilon + \hbar^2)+3 \hbar^2 k^4_{1})+6 k^6_{-1}\nonumber \\
(12 m \epsilon \hbar^2 + \hbar^4+ 3\hbar^4 k^4_{1}) + k^2_{-1}(16 m^2(9\epsilon^2 +
(27 \rho^2 -2 A^4_{1}) g^2_{1}) + 144 m \hbar^2 \nonumber \\
(7 A^2_{1}-10\rho)g_{1}k^2_{1}+18 \hbar^2
(37\hbar^2-4 m \epsilon) k^4_{1}- 27 \hbar^4 k^8_{1})+ 2k_{1}k_{-1}(16 m^2 \nonumber \\
(9\epsilon^2 + (27 \rho^2+2 A^4_{1})g^2_{1})-48 m \hbar^2(7\rho + A^2_{1}) g_{1} k^2_{1}+72\hbar^2
(m \epsilon+\hbar^2)k^4_{1}+ \nonumber \\
9 \hbar^4 k^8_{1})+6 \hbar^2 k^4_{-1}(-4 m(2\rho+17 A^2_{1})g_{1} +
3 k^2_{1}(-4 m \epsilon + 37 \hbar^2 + \hbar^2 k^4_{1}))- \nonumber \\
24 \hbar^2 k^3_{-1}k_{1}(4 m (7\rho + A^2_{1})g_{1} + k^2_{1}(12 m \epsilon -
44 \hbar^2+3\hbar^2 k^4_{1}))) - 4 g_{0} \nonumber \\
(48 m^3 \rho^3 g^3_{1}-16 m^2 \hbar^2 g^2_{1}((3\rho^2+
9 \rho A^2_{1}+2 A^4_{1})k^2_{-1} + 2(6\rho^2 - 9\rho A^2_{1}- 2A^4_{1})\nonumber \\
k_{-1}k_{1}+ (3 \rho^2 + 9\rho A^2_{1} + 2A^4_{1})k^2_{-1})-(k^2_{-1} + 4 k_{-1}k_{1} + k^2_{1})
(4 m \epsilon \hbar + \nonumber \\
\hbar^3(k^2_{-1}-k^2_{1})^2)^2 + 3 m g_{1}(16 m^2 \epsilon^2 \rho+
 \hbar^2 (\rho \hbar^2 k^8_{-1} + 4 \hbar^2(7\rho+ A^2_{1})k^3_{-1}k_{1} - \nonumber \\
4 \rho \hbar^2 k^6_{-1} k^2_{1} + 4 \hbar^2(7\rho+ A^2_{1})k^3_{1}k_{-1} + k^4_{1}(2\rho(4 m \epsilon+ \hbar^2)+
 \hbar^2 (17 A^2_{1} + \rho k^4_{1}))\nonumber \\
- 2 k^2_{-1}k^2_{1}(8 m \epsilon \rho - 30 \rho \hbar^2
\hbar^2(21 A^2_{1}+ 2\rho k^4_{1})) + k^4_{-1}(17 \hbar^2 A^4_{1} + 2 \rho (4m \epsilon + \nonumber \\
 \hbar^2 + 3 \hbar^2 k^4_{1})))))))
\label{eq:18}
\end{eqnarray}%
\begin{eqnarray}
C^{n=0,1}_{k^8, \omega^2}= -\frac{\hbar^{10}}{64 m^6 g_{1}(A^2_{-1}+A^2_{1})^2}(- 8 m \hbar^2 A^6_{-1}g_{1}
(k_{-1}-k_{1})(-2g_{1}(7k_{-1}+5k_{1})+ g_{0}(5k_{-1}+ \nonumber \\
7k_{1}))-16(1+ (-1)^n) m A^5_{-1} A_{1}g_{0}g_{1}(4 m \epsilon +\hbar^2(k^2_{-1}-k^2_{1})^2)-
2 A^3_{-1}A_{1}\nonumber \\
(4 m \epsilon +\hbar^2(k^2_{-1}-k^2_{1})^2)((-1)^n g_{0}(16 m \rho g_{1}-
 9 \hbar^2(k_{-1}+k_{1})^2)+6 g_{1}(2 m \nonumber \\
((-1)^n \rho - (1+(-1)^n) A^2_{1})g_{1} - 3(-1)^n \hbar^2(k_{-1}+k_{1})^2))- 2A^4_{-1}g_{1}\nonumber \\
(24 m^2 \rho^2 g^2_{1}-24 m \hbar^2 g_{1}(3(\rho+A^2_{1})k^2_{-1}+2(3\rho-A^2_{1})k_{-1}k_{1}+(3\rho-A^2_{1})\nonumber \\
k^2_{1})+\hbar^4(25 k^4_{-1}+116 k^3_{-1}k_{1}+ 198 k^2_{-1}k^2_{1}+116 k_{-1}k^3_{1}+  25k^4_{1})+4 m g_{0}\nonumber \\
(8 m \rho^2 g_{1} + 3\hbar^2((A^2_{1}-3\rho)k^2_{-1}+2(A^2_{1}-3\rho)k_{1}k_{-1} - 3(A^2_{1}+\rho)k^2_{1})))-\nonumber \\
2 A^2_{1}g_{1}(24 m^2 \rho^2 g^2_{1}- 8 m \hbar^2 g_{1}((9\rho-5A^2_{1})k^2_{-1}+2(9\rho-A^2_{1})k_{-1}k_{1}+\nonumber \\
(9\rho+7A^2_{1})k^2_{1})+\hbar^4 (25 k^4_{-1}+116 k^3_{-1}k_{1}+ 198 k^2_{-1}k^2_{1}+116 k_{-1}k^3_{1}+  25k^4_{1})+\nonumber \\
4 m g_{0}(8 m \rho^2 g_{1}+\hbar^2(-(7A^2_{1}+9\rho)k^2_{-1}+2 (A^2_{1}-9\rho)k_{-1}k_{1}+(5A^2_{1}-9\rho)k^2_{1})))\nonumber \\
-2 A_{-1}A^3_{1}(4 m \epsilon +\hbar^2(k^2_{-1}-k^2_{1})^2)(-4 m g_{1} A^2_{1}(2 g_{0}+ 3g_{1})- (-1)^n(6 g_{1}\nonumber \\
 (2 m (A^2_{1}-\rho)g_{1}+3\hbar^2(k_{-1}+k_{1})^2) + g_{0}(8 m(A^2_{1}-2\rho)g_{1}+9\hbar^2(k_{-1}+k_{1})^2)))\nonumber \\
- A^2_{1}A^2_{-1}(4 g_{0}(32 m^2 \rho^2 g^2_{1}-6 m \hbar^2 g_{1}(3(A^2_{1}+2\rho)k^2_{-1}+2(6\rho-A^2_{1})k_{-1}k_{1} +\nonumber \\
(6\rho-A^2_{1})k^2_{1}) + (4 m \epsilon +\hbar^2(k^2_{-1}-k^2_{1})^2)^2)+ g_{1}(48 m^2 \epsilon^2 + 96 m^2 \rho^2 g^2_{1}- \nonumber \\
48 m \hbar^2 g_{1}((6\rho-A^2_{1})k^2_{-1}+2(6\rho-A^2_{1})k_{1}k_{-1}+3(2\rho+A^2_{1})k^2_{1})+ \hbar^2 (3\hbar^2 k^8_{-1}\nonumber \\
+464 \hbar^2 k^3_{-1}k_{1}-12 \hbar^2 k^6_{-1}k^2_{1}+464\hbar^2 k_{-1}k^3_{1}-12 k^2_{-1}k^2_{1}(4 m \epsilon - 66 \hbar^2 + \nonumber \\
\hbar^2 k^4_{1})+ k^4_{1}(24 m \epsilon - 100 \hbar^2 + 3\hbar^2 k^4_{1})+ 2k^4_{-1}(12 m \epsilon + 50 \hbar^2 + 9\hbar^2 k^4_{1})))))
\label{eq:19}
\end{eqnarray}%
\begin{eqnarray}
C^{n=0,1}_{k^7, \omega^3}= -\frac{\hbar^{12}}{64 m^6 g_{1}(A^2_{-1}+A^2_{1})}(-3 (-1)^n A_{1}A_{-1}(g_{0}+2 g_{1})
(4 m \epsilon +\hbar^2(k^2_{-1}-k^2_{1})^2)+ \nonumber \\
2 (A^2_{-1}+A^2_{1})g_{1}(-6 m \rho(g_{0}+2 g_{1}) + \hbar^2(19 k^2_{-1} + 34 k_{-1}k_{1} + 19 k^2_{1})))
\label{eq:20}
\end{eqnarray}%
\begin{eqnarray}
C^{n=0,1}_{k^9, \omega^3}=& -\frac{\hbar^{14}}{16 m^6}
\label{eq:21}
\end{eqnarray}%
\begin{eqnarray}
C^{n=0,1}_{k^5, \omega}= \frac{\hbar^{7}}{8 m^7 g_{1}(A^2_{-1}+A^2_{1})^3}(8 m^3 \hbar^2 A^{12}_{-1}
g_{0}g^3_{1}(k_{-1}-k_{1})^3+ 4 m^2 \hbar^2 A^{10}_{-1}g^3_{1}(k_{-1}-k_{1})^2\nonumber \\
(\hbar^2 k_{-1}(k_{-1}+k_{1})(3 k_{-1}+k_{1})+2 m \rho g_{0}(k_{-1}+5 k_{1}))-2 m^2 \hbar^2 A^{9}_{-1}A_{1}g_{0}g^2_{1}\nonumber \\
(k_{-1}-k_{1})^2(k_{1}+5 k_{1})(4 m \epsilon + \hbar^2(k^2_{-1}-k^2_{1})^2)+ 4 m^2 \hbar^2 A^8_{-1}g^2_{1}(k_{-1}-k_{1})\nonumber \\
(\hbar^2 g_{1}(k_{-1}+k_{1})(2 (4 A^2_{1}+\rho)k^3_{-1}+(14\rho-11A^2_{1})k^2_{-1}k_{1}+2(2\rho+A^2_{1})k_{-1}k^2_{1}\nonumber \\
+A^2_{2}+A^2_{1}k^3_{1})-2g_{0}(k_{-1}-k_{1})(-\hbar^2 A^2_{1}(k_{-1}-k_{1})^2(k_{1}+k_{-1}) + m g_{1}\nonumber \\
((4\rho^2 + 3 A^2_{1}(\rho+A^2_{1}))k_{-1}+3(2\rho^2-3\rho A^2_{1}-A^4_{1})k_{1})))- 4 m^2 \hbar^2 A^6_{1}g^3_{1}\nonumber \\
((2\rho-A^2_{1})k_{-1}+ A^2_{1}k_{1})(-2 m A^4_{1} g_{0}(k_{-1}-k_{1})^2 + A^2_{1}(k_{-1}-k_{1})(2 m \rho g_{0}\nonumber \\
(3 k_{-1}+ k_{1})+\hbar^2 k_{1}(k_{-1}+k_{1})(k_{-1}+ 3 k_{1}))- 2\rho k_{1}(k_{-1}+k_{1})(-4 m \rho g_{0}+\hbar^2\nonumber \\
( k^2_{-1}+4 k_{-1}k_{1}+k^2_{1})))-2 m \hbar^2 A_{-1}A^5_{1}g^2_{1}(k^2_{-1}+4 k_{-1}k_{1}+k^2_{1})^2)(4 (-1)^n \nonumber \\
(A^2_{1}-\rho)k_{1}(k_{-1}+k_{1})(m A^2_{1}g_{0}(5 k_{1}-3k_{-1})+k_{1}(-6 m g_{0}\rho+ \hbar^2( k^2_{-1}+ \nonumber \\
4 k_{-1}k_{1}+ k^2_{1})))+ A^2_{1}(m A^2_{1} g_{0}(-7 k^3_{-1}-k^2_{1}k_{1}+23 k_{-1}k^2_{1}+ k^3_{1})+ k_{1}\nonumber \\
(- 8 m \rho g_{0}(2 k^2_{-1}+9k_{-1}k_{1}+k^2_{1}) + \hbar^2 (k_{-1}+k_{1})^2(2 k^2_{-1}+9k_{-1}k_{1}+k^2_{1}))))-\nonumber \\
A^3_{-1}A^3_{1}(4 m \epsilon +\hbar^2(k^2_{-1}-k^2_{1})^2)(-64 (1+(-1)^n) m^3 A^6_{1}g^3_{1}(g_{0}+g_{1})(k_{-1}+k_{1})\nonumber \\
+ 2 m \hbar^2 A^2_{1}g^2_{1}(4 m \rho g_{0}((-2+6(-1)^n)k^3_{1}-3 (2+5(-1)^n)k^2_{-1}k_{1}-6 (2+ \nonumber \\
3(-1)^n)k_{-1}k^2_{1}+(-4 +3(-1)^n)k^3_{1})+\hbar^2(k_{-1}+k_{1})(k^4_{-1}+2(5+4(-1)^n)\nonumber \\
k^3_{-1}k_{1}+ (17+32(-1)^n)k^2_{-1}k^2_{1} + 2(9+4(-1)^n)k_{-1}k^3_{1}+2 k^4_{1}))+\hbar^4 k_{-1}k_{1}\nonumber \\
(k_{-1}+k_{1})(g_{0}(4 m \epsilon +\hbar^2(k^2_{-1}-k^2_{1})^2)^2-16(-1)^n m \rho g^2_{1}(-6 m \rho g_{0}+ \hbar^2 \nonumber \\
( k^2_{-1}+4k_{-1}k_{1}+k^2_{1})))+8 m^2 A^4_{1}g^2_{1}(8(1+(-1)^n) m \rho g^2_{1}(k_{-1}+k_{1})+\nonumber \\
g_{0}(8(1+(-1)^n) m \rho g_{1}(k_{-1}+k_{1})- \hbar^2 (k_{-1}-k_{1})((5+6(-1)^n)k^2_{-1}+ \nonumber \\
(5+3(-1)^n)k_{-1}k_{1} - (4+3(-1)^n)k^2_{1}))))-2 m A^5_{-1}A_{1}g^2_{1}(4 m \epsilon + \hbar^2(k^2_{-1} - \nonumber \\
k^2_{1})^2)(-64 (1+(-1)^n) m^2 A^6_{1} g_{1}(g_{0}+g_{1})(k_{-1}+k_{1})+ 4 (-1)^n\rho \hbar^2 k_{-1}k_{1} (k_{-1}\nonumber \\
+k_{1})(6 m g_{0}-\hbar^2 (k^2_{-1}+4k_{-1}k_{1}+k^2_{1}))+ \hbar^2 A^2_{1}(4 m \rho g_{0}(k_{-1}-k_{1})((-4+ \nonumber \\
3(-1)^n)k^2_{-1}- (4+3(-1)^n)k_{1}k_{-1}-2 (-1+3(-1)^n)k^2_{1})+\hbar^2(k_{-1}+k_{1})\nonumber \\
(2 k^4_{-1}+2(7+2(-1)^n)k^3_{-1}k_{1}+(32(1+16(-1)^n)k^2_{-1}k^2_{1}+2(3+2(-1)^n) \nonumber \\
k_{-1}k^3_{1} + k^4_{1}))+ 2 m A^4_{1}(k_{-1}+k_{1})(32(1+(-1)^n)m \rho g^2_{1}+ g_{0}(32(1+(-1)^n)\nonumber \\
m\rho g_{1}-3 \hbar^2((3+2(-1)^n)k^2_{-1}-2 ((2+(-1)^n)k_{1}k_{-1} + (5+4(-1)^n)k^2_{1}))))\nonumber \\
-2 m A^7_{-1}A_{1}g^2_{1}(4 m \epsilon +\hbar^2(k^2_{-1}-k^2_{1})^2)(32(1+(-1)^n) m^2 A^4_{1}g_{1}(g_{0}+g_{1})\nonumber \\
(k_{-1}+k_{1})+\hbar^2 (\hbar^2(k_{-1}-k_{1})(k_{-1}+k_{1})(k^2_{-1}+7k_{-1}k_{1}+k^2_{1})+ 4 m g_{0}\nonumber \\
(-2k^3_{-1}+(2+(-1)^n)k^2_{-1}k_{1}+(2+ (-1)^n)k_{-1}k^2_{1}+ 3 (-1)^n k^3_{1}))+ 4 m A^2_{1}\nonumber \\
(8(1+(-1)^n)m \rho g^2_{1}(k_{-1}+k_{1})+(8(1+(-1)^n) m \rho g_{1}(k_{-1}+k_{1})-\hbar^2 ( k^3_{-1}\nonumber \\
+(2+(-1)^n)k^2_{-1}k_{1}+(7+4(-1)^n)k_{-1}k^2_{1}+(2 + 3 (-1)^n) k^3_{1}))))+\hbar^2 A^4_{-1}\nonumber \\
A^2_{1}g_{1}(24 m^3 A^6_{1}g_{0}g^2_{1}(k_{-1}-k_{1})^3-8m^2 A^4_{1}g_{1}(k_{-1}-k_{1})^2(\hbar^2 g_{1}(7k_{-1}-3k_{1})\nonumber \\
k_{1}(k_{-1}+k_{1})+ g_{0}(-3 \hbar^2(k_{-1}-k_{1})^2(k_{-1}+k_{1})+2 m \rho g_{1}(k_{-1}+5k_{1})))-\nonumber \\
m A^2_{1}(k_{-1}-k_{1})^2(k_{-1}+k_{1})(-24 m \rho \hbar^2 g^2_{1}(k^2_{-1}+6k_{-1}k_{1}+k^2_{1})+5 g_{0} \nonumber \\
(48 m^2 \rho^2 g^2_{1}+(4 m \epsilon +\hbar^2(k^2_{-1}-k^2_{1})^2)^2))+k_{1}k_{-1}(k_{-1}+k_{1})(-12 m \rho g_{0}\nonumber \\
(16 m^2 \rho^2 g^2_{1}+(4 m \epsilon +\hbar^2(k^2_{-1}-k^2_{1})^2)^2)+\hbar^2(k^2_{-1}+4k_{-1}k_{1}+k^2_{1})(48 m^2 \rho^2 g^2_{1}\nonumber \\
+(4 m \epsilon +\hbar^2(k^2_{-1}-k^2_{1})^2)^2)))+\hbar^2 A^2_{-1}A^2_{1}g_{1}(4 m^2 A^4_{1}g_{1}(k_{-1}-k_{1})^2(-\hbar^2 g_{1}\nonumber \\
(k_{-1}+k_{1})(k^2_{-1}+3k_{-1}k_{1}-8k^2_{1})+2 g_{0}(3 m \rho g_{1}(3k_{-1}-k_{1})+\hbar^2(k_{-1}-k_{1})^2 \nonumber \\
(k_{-1}+k_{1})))+k_{1}k_{-1}(k_{-1}+k_{1})(-12 m \rho g_{0}(16 m^2 \rho^2 g^2_{1}+(4 m \epsilon + \hbar^2(k^2_{-1}-\nonumber \\ k^2_{1})^2)^2)+\hbar^2(k^2_{-1}+4k_{-1}k_{1}+k^2_{1})(48 m^2 \rho^2 g^2_{1} + (4 m \epsilon +\hbar^2(k^2_{-1}-k^2_{1})^2)^2))-\nonumber \\
m A^2_{1}(k_{-1}-k_{1})(-8 m \rho \hbar^2 g^2_{1}(k_{-1}+k_{1})( k^3_{-1}+k^2_{-1}k_{1}-19 k_{-1}k^2_{1} - 3 k^3_{1})+ \nonumber \\ g_{0}(k_{-1}-k_{1})(3 \hbar^4 k^9_{-1}+ 2 \hbar^4 k^8_{-1}k_{1}-12\hbar^4 k^7_{-1}k^2_{1}-8 \hbar^4 k^6_{-1}k^3_{1}+6k^5_{-1}(4 m \epsilon \hbar^2 \nonumber \\
+3\hbar^4 k^4_{1})+4 k^4_{-1}(4 m \epsilon \hbar^2 k_{1}+3\hbar^4 k^5_{1})-12 k^3_{-1}(4 m \epsilon \hbar^2 k^2_{1}+\hbar^4k^6_{1})- \nonumber \\
8 k^2_{-1}(4 m\epsilon \hbar^2 k^3_{1}+\hbar^4k^7_{1})+ 2 k_{1}(72 m^2 \rho^2 g^2_{1}+(4 m\epsilon +\hbar^2k^4_{1})^2)+k_{-1}(176 m^2 \nonumber\\
\rho^2 g^2_{1} + 3(4 m \epsilon +\hbar^2k^4_{1})^2))))+m \hbar^2 A^6_{-1} g_{1}(16 m \rho^2 g^2_{1}k_{-1}k_{1} (k_{-1}+k_{1})\nonumber \\
(-4 m \rho g_{0} + \hbar^2(k^2_{-1}+4k_{-1}k_{1}+k^2_{1}))- 8 m A^4_{1}g_{1}(k_{-1}+k_{1})^2(- \hbar^2 g_{1}k_{-1}\nonumber \\
(3k_{-1}-7k_{1}) (k_{-1}+k_{1}) + g_{0}(-3 \hbar^2 (k_{-1}-k_{1})^2(k_{-1}+k_{1})+2 m \rho g_{1}(5k_{-1}+ \nonumber \\
k_{1})))- A^2_{1}(k_{-1}-k_{1})(-8 m \rho \hbar^2 g_{1}(k_{-1}+k_{1})(3 k^3_{-1}+19k^2_{-1}k_{1}-k_{-1}k^2_{1}- k^3_{1})\nonumber \\
+g_{0}(k_{-1}-k_{1})(2 \hbar^4 k^9_{-1}+3 \hbar^4 k^8_{-1}k_{1}-8 \hbar^4 k^7_{-1}k^2_{1}-12 \hbar^4 k^6_{-1}k^3_{1}+4 k^5_{-1}(4 m \epsilon \hbar^2 \nonumber \\
+3\hbar^2k^4_{1}) + 6 k^4_{-1}(4 m \epsilon \hbar^2 k_{1}+ 3\hbar^4 k^5_{1})- 8 k^3_{-1}(4 m \epsilon \hbar^2 k^2_{1}+ \hbar^4 k^6_{1}) -12 k^2_{-1}\nonumber \\
(4 m \epsilon \hbar^2 k^3_{1}+\hbar^4 k^7_{1})+2 k_{1}(72 m^2\rho^2 g^2_{1} + (4 m \epsilon +\hbar^2k^4_{1})^2)+k_{1}(176 m^2 \rho^2 g^2_{1} + \nonumber \\
3(4 m \epsilon + \hbar^2k^4_{1})^2)))))
\label{eq:22}
\end{eqnarray}%
\begin{eqnarray} C^{n=0,1}_{k^7, \omega}= -\frac{\hbar^{9}}{32 m^7 g_{1}(A^2_{-1}+A^2_{1})^3}(-8 m^2 \hbar^2
A^{10}_{-1}g^2_{1}(k_{-1}-k_{1})^2(8 g_{0}(k_{-1}+k_{1})-g_{1}(3k_{-1}\nonumber \\
+k_{1}))- 2 m A^7_{-1} A_{1} g_{1}(4 m \epsilon +\hbar^2(k^2_{-1}-k^2_{1})^2)(g_{1}(-32(1+(-1)^n) m A^2_{1}g_{1}\nonumber \\
(k_{-1}+k_{1})+\hbar^2 (k_{-1}-k_{1})(5 k^2_{-1}+14 k_{-1}k_{1}+ 5k^2_{1}))+4 g_{0}(2(1+(-1)^n)\nonumber \\
m g_{1}(\rho - 7A^2_{1})k_{-1}+3(\rho - 3A^2_{1})k_{1})+\hbar^2(k_{-1}+k_{1})(3k^2_{-1}-(7+(-1)^n)\nonumber \\
k_{-1}k_{1} - 3 (-1)^n k^2_{1}))) + 4 m \hbar^2 A^8_{-1}g_{1} (k_{-1}-k_{1})(g_{0}(k_{-1}+k_{1})(8 m(3 \rho- \nonumber \\
2 A^2_{1})g_{1}(k_{-1}-k_{1})+\hbar^2 k_{1}(3 k^2_{-1}+ 12 k_{-1}k_{1}+ 5k^2_{1}))+ 2 g_{1}(-\hbar^2 k_{-1}(k_{-1}+ \nonumber \\
k_{1})(5 k^2_{-1}+ 12 k_{-1}k_{1} + 3k^2_{1})+ m g_{1}(5(\rho+A^2_{1})k^2_{1}+2(7\rho-3A^2_{1})k_{1}k_{-1}+\nonumber \\
(5\rho+A^2_{1})k^2_{1})))- 2 A_{-1}A^5_{1}(4 m \epsilon +\hbar^2(k^2_{-1}-k^2_{1})^2)(2 g_{0}(4 m^2 g^2_{1}(3\rho ((-1)^n \nonumber \\
\rho-(1+(-1)^n)A^2_{1})k_{-1}+(3(-1)^n\rho^2 +(1+(-1)^n)A^2_{1})(-5\rho + 2 A^2_{1})) \nonumber \\
k_{1})+(-1)^n \hbar^4 k_{-1}k_{1}(k_{-1}+k_{1})(k^2_{-1}+ 4 k_{-1}k_{1}+k^2_{1})+2 m \hbar^2 g_{1}(k_{-1}+k_{1})\nonumber \\
(A^2_{1}k_{1}(k_{-1}+3 k_{1})-(-1)^n k_{-1}(A^2_{1}(3 k_{-1}- 7 k_{1})+ 8\rho k_{1})))+ \hbar^2 g_{1}(m A^2_{1}g_{1} \nonumber \\
(7 k^3_{-1}+27 k^2_{-1}k_{1}+45 k_{-1}k^2_{1}+ 17 k^3_{1})+4 (-1)^n (k_{-1}+k_{1})(-3 m(\rho-A^2_{1})g_{1}\nonumber \\
(k_{-1}+k_{1})^2+\hbar^2 k_{1}k_{-1} (k^2_{-1}+4k_{-1}k_{1}+k^2_{1}))))+ 4 A^6_{1}g_{1}(\hbar^2\hbar^4 k_{1}k_{-1}(k_{-1}+ \nonumber \\
k_{1})^3 (k^2_{-1}+6k_{-1}k_{1}+k^2_{1})-2 m \hbar^2 g_{1}k_{1}(k_{-1}+k_{1})((4\rho-3A^2_{1})k^3_{-1}+(16\rho- \nonumber \\
9A^2_{1})k^2_{-1}k_{1}+(4\rho-7A^2_{1})k_{-1}k^2_{1}+5 A^2_{1}k^3_{1})+ 2 m^2 g^2_{1}((6\rho^2- 5\rho A^2_{1}+A^4_{1})\nonumber \\
k^3_{-1}+(18\rho^2-9\rho A^2_{1}+A^4_{1})k^2_{-1}k_{1}+(18\rho^2+9\rho A^2_{1}-5A^4_{1})k^2_{1}k_{-1}+(6\rho^2\nonumber \\
+5\rho A^2_{1}+3A^4_{1})k^3_{1}-m g_{0}(k_{-1}+k_{1})(16 m^2 \rho^3 g^2_{1}+8 m \hbar^2 g_{1}(-3\rho A^2_{1}(k_{-1}- \nonumber \\
k_{1})^2+2 A^4_{1}(k_{-1}-k_{1})^2-4\rho^2 k_{-1}k_{1})+ \hbar^4 k_{-1} (4\rho k_{1}(k^2_{-1}+4k_{-1}k_{1}+k^2_{1})+\nonumber \\
A^2_{1}(k_{-1}-k_{1})(5 k^2_{-1}+12 k_{-1}k_{1}+ 3k^2_{1}))))- 2 A^5_{-1}A_{1}(4 m \epsilon +\hbar^2(k^2_{-1}-k^2_{1})^2)\nonumber \\
(g_{1}(-64 (1+(-1)^n) m^2 A^4_{1} g^2_{1}(k_{-1}+k_{1})+ m \hbar^2A^2_{1} g_{1}(17 k^3_{-1}+45 k^2_{-1}k_{1}+ \nonumber \\
27 k_{-1}k^2_{1}+7k^3_{1}+ 12(-1)^n (k_{-1}+k_{1})^3)+ 4 (-1)^n \hbar^2(k_{-1}+k_{1})(-3 m \rho g_{1}\nonumber \\
(k_{-1}+k_{1})^2 + \hbar^2 k_{1}k_{-1}(k^2_{-1}+4k_{-1}k_{1}+k^2_{1})))+ 2 g_{0} (4 m^2 g^2_{1}((3(-1)^n \rho^2 - \nonumber \\
(1+(-1)^n)A^2_{1}(\rho+14 A^2_{1}))k_{-1}+(3 (-1)^n \rho^2 +(1+(-1)^n)A^2_{1}(\rho-16 A^2_{1}))\nonumber \\
k_{1})+(-1)^n \hbar^4 k_{1}k_{-1} (k_{-1}+k_{1})(k^2_{-1}+4k_{-1}k_{1}+k^2_{1})-2 m \hbar^2 g_{1}(k_{-1}+k_{1})\nonumber \\
(8 (-1)^n \rho k_{-1}k_{1}+A^2_{1}(3(-2+(-1)^n) k^2_{-1}-(-13+5(-1)^n)k_{1}k_{-1}+ \nonumber \\
3(1+2(-1)^n)k^2_{1}))))-4 A^6_{1}g_{1}(m g_{0}(k_{-1}+k_{1})(16 m^2 \rho^3 g^2_{1}-32 m \hbar^2 g_{1}\nonumber \\
(3\rho A^2_{1}(k_{-1}-k_{1})^2+A^4_{1}(k_{-1}-k_{1})^2 +\rho^2k_{-1}k_{1})+ \hbar^4 (4 \rho k_{1}k_{-1}(k^2_{-1}+ \nonumber \\
4k_{-1}k_{1}+k^2_{1})+A^2_{1}(k_{-1}-k_{1})(5 k^3_{-1}+3 k^2_{-1}k_{1}-33 k_{-1}k^2_{1}-15 k^3_{1})))-\nonumber \\
\hbar^2(\hbar^4 k_{1}k_{-1}(k_{-1}+k_{1})^3(k^2_{-1}+6k_{-1}k_{1}+k^2_{1})+4 m^2 g^2_{1}((3\rho^2+ 5\rho A^2_{1}- \nonumber \\
A^4_{1}) k^3_{1}+(9\rho^2+9\rho A^2_{1}-A^4_{1})k^2_{-1} k_{1}+(9\rho^2-9\rho A^2_{1}+5A^4_{1})k^2_{1} k_{-1}+\nonumber \\
(3\rho^2-5\rho A^2_{1}-A^4_{1})k^3_{1})-2 m \hbar^2 g_{1}(k_{-1}+k_{1}) + 4\rho k_{1}k_{-1}(k^2_{-1}+4k_{-1}k_{1}\nonumber \\
+k^2_{1})+ A^2_{1}(k_{-1}-k_{1})(15 k^3_{-1}+33 k^2_{-1}k_{1}-3 k_{-1}k^2_{1}- 5 k^3_{1})))) + A^3_{-1}A^3_{1}\nonumber \\
(4 m \epsilon +\hbar^2(k^2_{-1}-k^2_{1})^2)(2 g_{1}(32(1+(-1)^n) m^2 A^4_{1} g^2_{1}(k_{-1}+k_{1})- m \hbar^2 \nonumber \\
A^2_{1}g_{1}(19 k^3_{-1}+ 63 k^2_{-1}k_{1}-81 k_{-1}k^2_{1}+ 29 k^3_{1}+24(-1)^n(k_{-1}+k_{1})^3)-\nonumber \\
8 (-1)^n \hbar^2(k_{-1}+k_{1})(-3 m \rho g_{1}(k_{-1}+k_{1})^2 + \hbar^2 k_{1}k_{-1}(k^2_{-1}+4k_{-1}k_{1}\nonumber \\
+k^2_{1})))+ g_{0}(\hbar^4k^9_{-1}+\hbar^4k^8_{-1}k_{1}-4\hbar^4k^7_{-1}k^2_{1}-4\hbar^4k^6_{-1}k^3_{1}+2\hbar^2k^4_{-1}k_{1}\nonumber \\
(4 m\epsilon-4(-1)^n \hbar^2+3\hbar^2 k^4_{1})+ k^5_{-1}(8 m\epsilon \hbar^2+6 \hbar^4 k^4_{1})+k_{1}(-16 m^2 \nonumber \\
(6(-1)^n)\rho^2-(1+(-1)^n) A^2_{1}(7\rho+5A^2_{1}))g^2_{1}+24 (-2+(-1)^n) m \hbar^2 \nonumber \\
A^2_{1}g_{1}k^2_{1}+(4 m\epsilon + \hbar^2 k^4_{1})^2)+k_{-1}(-16 m^2(16 (-1)^n \rho^2-(1+(-1)^n) A^2_{1}\nonumber \\
(5\rho+ 7A^2_{1})g^2_{1}+ 8 m \hbar^2(16 (-1)^n \rho-(1+10(-1)^n)A^2_{1})g_{1}k^2_{1}-8(-1)^n\hbar^4 k^4_{1}\nonumber \\
+(4 m\epsilon + \hbar^2 k^4_{1})^2)- 4\hbar^4 k^3_{-1}(-6 (-1+ 2(-1)^n)m A^2_{1}g_{1}+ k^2_{1}(4 m\epsilon +\nonumber \\
10(-1)^n \hbar^2k^4_{1}))-4\hbar^2 k_{1}k^2_{-1}(-2 m (16 (-1)^n \rho+(2-7(-1)^n)A^2_{1})g_{1}+\nonumber \\
k^2_{1}(4 m\epsilon + 10(-1)^n \hbar^2k^4_{1})))))+ A^4_{-1}A^2_{1}(\hbar^2 g_{1}(3 \hbar^4 k^{11}_{-1}+9\hbar^4 k^{10}_{-1}k_{1}- \nonumber \\
3 \hbar^4 k^{9}_{-1}k^2_{1}-33\hbar^4 k^{8}_{-1}k^3_{1}+ 6 \hbar^2 k^{6}_{-1}k_{1}(2(6 m \epsilon + \hbar^2)+7 \hbar^2 k^4_{1}+ 6 k^7_{-1} \nonumber \\
(4 m\epsilon \hbar^2-3\hbar^4 k^4_{1})+3 k^3_{-1}(16 m^2 \epsilon^2 +16 m^2(3 \rho^2-A^4_{1}) g^2_{1}+16 m \hbar^2 (-10 \rho \nonumber \\
+ 7 A^2_{1})g_{1} k^2_{1}+(40 m \epsilon \hbar^2 + 88\hbar^4) k^4_{1}-11 \hbar^4 k^8_{-1}) + k^2_{-1}k_{1}(16 m^2(9 \epsilon^2 + \nonumber \\
(27 \rho^2+ 5A^2_{1})g^2_{1}) + 48 m \hbar^2 (-10\rho+7A^2_{1})g^2_{1}k^2_{1}+ 12 \hbar^2 (2 m \epsilon +9 \hbar^2)k^4_{1}-\nonumber \\
3 \hbar^4 k^8_{1})+ k_{-1}k^2_{1}( 16 m^2(9 \epsilon^2 +(27\rho-A^2_{1})g^2_{1})- 24 m \hbar^2(4 \rho+9 A^2_{1})g_{1}k^2_{1}+ \nonumber \\
12 \hbar^2 (6 m \epsilon +\hbar^2) k^4_{1}+9 \hbar^4 k^8_{1}+k^3_{1}(16 m^2(9 \rho^2 -  A^4_{1})g^2_{1}-120 m \hbar^2 A^2_{1}g_{1}k^2_{1}+ \nonumber \\
3 (4 m \epsilon +\hbar^2 k^4_{-1})^2) - 6 \hbar^2 k^4_{-1} k_{1}(4 m (4\rho+9 A^2_{1})g_{1}+k^2_{1}(20  m \epsilon - 44 \hbar^2 +\nonumber \\
3 \hbar^2 k^4_{1}))+ 6 \hbar^2 k^5_{-1} (-20 m A^2_{1} g_{1}+ k^2_{1}(4  m \epsilon + 18 \hbar^2 + 7 \hbar^2 k^4_{1})))- 4 g_{0}(k_{-1}+ \nonumber \\
k_{1})(48 m^3 \rho^3 g^3_{1}- 16 m^2 \hbar^2 g^2_{1}( 9\rho A^2_{1}(k_{-1}-k_{1})^2 + 2 A^4_{1}(k_{-1}- k_{1})^2 + \nonumber \\
6\rho^2 k_{-1}k_{1}- 2 k_{1}k_{-1}(4 m \epsilon +\hbar^2(k^2_{-1}-k^2_{1})^2)^2+3 m g_{1}( 16 m^2 \epsilon^2 \rho+\nonumber \\
\hbar^2 (\rho \hbar^2 k^8_{-1} + 4\hbar^2 (\rho+A^2_{1})k^3_{-1}k_{1}-2 k_{-1}k_{1}(4 m \epsilon +\hbar^2(k^2_{-1}-k^2_{1})^2)^2+ \nonumber \\
3 m g_{1}(16 m^2 \epsilon^2 \rho + \hbar^2(\rho \hbar^2 k^8_{-1}+4 \hbar^2 (\rho + A^2_{1})k^3_{-1}k_{1}-k^4_{1}(8 m \epsilon \rho + \hbar^2  \nonumber \\
(5 A^2_{1}+\rho)k^4_{1}))+k^4_{-1}(8 m \epsilon \rho + \hbar^2 (5 A^2_{1}+6\rho k^4_{1}))))))+
A^2_{-1}A^4_{1}(\hbar^2 g_{1} \nonumber \\
(3 \hbar^4 k^{11}_{-1}+9\hbar^4 k^{10}_{-1}k_{1}- 3 \hbar^4 k^{10}_{-1}k^{2}_{1}-
 3 \hbar^4 k^{9}_{-1}k^{2}_{1}-33 \hbar^4 k^{8}_{-1}k^{3}_{1}+ 6 \hbar^2 k^{6}_{-1}k_{1} \nonumber \\
(2(6 m \epsilon + \hbar^2)+7 \hbar^2 k^4_{1})+ 6 k^7_{-1}(4 m \epsilon - 3\hbar^4 k^4_{1})+ k^3_{-1} (48 m^2 \epsilon^2 +
8 m^2  \nonumber \\
(18 \rho^2 - 10 \rho A^2_{1} - A^4_{1}) g^2_{1} + 16 m\hbar^2(-30\rho+ 19A^2_{1})g_{1}k^2_{1}+24 \hbar^2 (-5 m\epsilon +  \nonumber \\
11 \hbar^2)k^4_{1}-33\hbar^4 k^8_{1})+ k^2_{-1}k_{1}( 144 m^2 \epsilon^2 + 8 m^2(54 \rho^2 - 18 \rho A^2_{1} + 7 A^4_{1})g^2_{1}+\nonumber \\
48 m \hbar^2 (3A^2_{1}-10\rho)g_{1}k^2_{1}+ 12(2 m \epsilon \hbar^2 + 9\hbar^4) k^4_{1}- 3 \hbar^4 k^8_{1})+k_{-1}k^2_{1} \nonumber \\
(144 m^2 \epsilon^2 + 8 m^2(54 \rho^2 + 18 \rho A^2_{1} - 11 A^4_{1})g^2_{1}-24 m \hbar^2(4 \rho+ 11 A^2_{1})g_{1} k^2_{1}+ \nonumber \\
12 (6 m \epsilon \hbar^2 +
\hbar^4) k^4_{1}+9 \hbar^4 k^8_{1}) +k^3_{1}(8 m^2(18 \rho^2 - 10 \rho A^2_{1} + 5 A^4_{1})g^2_{1} - \nonumber \\
120 m \hbar^2 A^2_{1} g_{1}k^2_{1} + 3 (4 m \epsilon + \hbar^2 k^4_{1})^2)- 6 \hbar^2 k^4_{-1} k_{1}
(4 m (4 \rho + A^2_{1}) g_{1}+ k^2_{1} \nonumber \\
(20 m \epsilon -44\hbar^2 + 3\hbar^2 k^4_{1}))+ 2 \hbar^2 k^5_{-1}(-20 m A^2_{1} g_{1} + 3 k^2_{1} (4 m \epsilon + 18\hbar^2 +  \nonumber \\
7\hbar^2 k^4_{1})))- 2 k_{-1}k_{1} (4 m \epsilon +\hbar^2(k^2_{-1}-k^2_{1})^2)^2 + m g_{1}( 48 m^2 \epsilon^2 \rho +
\hbar^2 (3 \rho \hbar^2 k^8_{-1} \nonumber \\
+6 \hbar^2(2\rho+3A^2_{1})k^3_{1}k_{1}- 12\rho \hbar^2 k^6_{-1}k^2_{1}+ 2 \hbar^2(6\rho-A^2_{1})k_{-1}k^3_{1}-12 k^2_{-1}k^2_{1} \nonumber \\
(4 m\epsilon \rho - 4\rho \hbar^2+3\hbar^2 A^2_{1}+\rho \hbar^2 k^4_{1})+ k^4_{1}(24 m\epsilon \rho +
\hbar^2(5A^2_{1}+3\rho k^4_{1}))+ \nonumber\\
3 k^4_{-1} (8 m \epsilon \rho + \hbar^2(5A^2_{1}+6\rho k^4_{1})))))))
\label{eq:23}
\end{eqnarray}%
\begin{eqnarray} C^{n=0,1}_{k^9, \omega}= -\frac{\hbar^{11}}{64 m^7 g_{1}(A^2_{-1}+A^2_{1})^2}(16 m A^6_{-1}g_{1}(k_{-1}-k_{1})
(m \rho g^2_{1}-2 \hbar^2 g_{1}(k_{-1}+k_{1})\nonumber \\
(2 k_{-1}+k_{1})+\hbar^2 g_{0}(k_{-1}+k_{1})(k_{-1}+2k_{1})+4 m A^5_{-1}A_{1}g_{1}(g_{1}(-k_{-1}+k_{1})\nonumber \\
+ 2(1+(-1)^n)g_{0} (k_{-1}+3k_{1}))(4 m \epsilon +\hbar^2(k^2_{-1}-k^2_{1})^2)+ 2A^4_{1}g_{1}(8 m^2 \rho g^2_{1}\nonumber \\
((3\rho-A^2_{1})k_{-1}+(3\rho+A^2_{1})k_{1})-8 m \hbar^2 g_{1}(k_{-1}+k_{1})((2\rho-2A^2_{1})k^2_{-1}+ \nonumber \\
2(3\rho-A^2_{1})k_{-1}k_{1}+ (3\rho+4A^2_{1})k^2_{1})+\hbar^4 (k_{-1}+k_{1})(5 k^4_{-1}+20 k^3_{-1}k_{1}+\nonumber \\
46k^2_{-1}k^2_{1}+20k^3_{1}k_{1}+ 5 k^4_{1})+ 4 m g_{0}(k_{-1}+k_{1}) (8 m \rho^2 g_{1}+\hbar^2 (-(3\rho+4A^2_{1})\nonumber \\
k^2_{-1}+2(-3\rho+A^2_{1})k_{-1}k_{1}+(-3\rho+2A^2_{1})k^2_{1})))+2 A^4_{-1}g_{1}(8 m^2\rho g^2_{1}((3\rho+\nonumber \\
A^2_{1})k_{-1}+ (3\rho-A^2_{1})k_{1})-24 m \hbar^2 g_{1}(k_{-1}+k_{1})((\rho+2A^2_{1})k^2_{-1}+2(\rho-A^2_{1})\nonumber \\
k_{-1}k_{1}+\rho k^2_{1}+\hbar^4 (k_{-1}+k_{1}) (5k^4_{-1}20 k^3_{-1}k_{1}+46k^2_{-1}k^2_{1}+20k^3_{1}k_{1}+ 5 k^4_{1})- \nonumber \\
4 m g_{0}(k_{-1}+k_{1})(-8 m \rho^2 g_{1}+ 3\hbar^2 (\rho k^2_{-1}+ 2(\rho-A^2_{1})k_{-1}k_{1}+(\rho+2A^2_{1})\nonumber \\
k^2_{1})))-2 A^3_{-1}A_{1}(k_{-1}+k_{1})(4 m \epsilon +\hbar^2(k^2_{-1}-k^2_{1})^2)(12 m A^2_{1}g^2_{1}+(-1)^n\nonumber \\
(6 g_{1}(2 m(-\rho+A^2_{1})g_{1}+\hbar^2(k_{-1}+k_{1})^2)+g_{0}(- 16 m \rho g_{1}+ 3 \hbar^2 (k_{-1}+\nonumber \\
k_{1})^2)))+2 A_{-1}A^3_{1}(4 m \epsilon +\hbar^2(k^2_{-1}-k^2_{1})^2)(-2 m A^2_{1}g_{1}(2 g_{0} (k_{-1}+3 k_{1})+\nonumber \\ g_{1}(5k_{-1}+7k_{1}))- (-1)^n(6 g_{1}(2 m(A^2_{1}-\rho)g_{1}+\hbar^2(k_{-1}+k_{1})^2)+ g_{0}\nonumber \\
(3 \hbar^2(k_{-1}+k_{1})^3+ 4 m g_{1}((A^2_{1}-4\rho)k_{-1}+(3A^2_{1}-4\rho)k_{1}))))+A^2_{-1}A^2_{1} \nonumber \\
(4 g_{0}(k_{-1}+k_{1})(32 m^2 \rho^2 g^2_{1}-12 m \hbar^2 g_{1}((\rho+A^2_{1})k^2_{-1}(2\rho-A^2_{1})k_{-1}k_{1}+\nonumber \\
\rho k^2_{1}+(4 m \epsilon +\hbar^2(k^2_{-1}-k^2_{1})^2)^2)^2)+ g_{1}(3 \hbar^4 k^9_{-1}+ 16 m^2(3 \epsilon^2 + \rho (6\rho+ \nonumber \\
A^2_{1})g^2_{1})k_{1} + 3\hbar^4 k^8_{-1}k_{1}- 12\hbar^4 k^7_{-1}k^2_{1}-96 m \hbar^2(\rho-A^2_{1}) g_{1}k^3_{1}- 12 \hbar^4 k^6_{-1}k^3_{1}\nonumber \\
+4 \hbar^2(6 m \epsilon +5\hbar^2)k^5_{1}+3 \hbar^4 k^9_{1}+ 2 \hbar^2k^4_{-1}k_{1} (12 m \epsilon +50 \hbar^2 +9\hbar^2(k^4_{1})+2 k^5_{-1}\nonumber \\
(12 m \epsilon \hbar^2 +10 \hbar^4 +9 \hbar^4 k^4_{1})+ k_{-1}(16 m^2 (3 \epsilon^2 +\rho(6\rho-A^2_{1})g^2_{1})- 288 m \rho \hbar^2 \nonumber \\ g_{1}k^2_{1}+ 4 \hbar^2 (6m \epsilon +25 \hbar^2)k^4_{1}+3 \hbar^4 k^8_{1}-12 \hbar^2 k^3_{-1}(8 m \rho g_{1}+k^2_{1}(4 m \epsilon - \nonumber \\ 22 \hbar^2 +\hbar^2 k^4_{-1}k_{1}))- 12\hbar^2 k^2_{-1}k_{1}(8 m(3\rho-A^2_{1})g_{1}+k^2_{1}\nonumber \\
(4 m \epsilon -22 \hbar^2 +\hbar^2 k^4_{-1}k_{1})))))
\label{eq:24}
\end{eqnarray}%
\begin{eqnarray} C^{n=0,1}_{k^{11}, \omega}= -\frac{\hbar^{13}}{64 m^7 g_{1}(A^2_{-1}+A^2_{1})}(- 4 m A^4_{-1}(g_{0}-2g_{1})
g_{1}(k_{-1}-k_{1})+ 3 (-1)^n A_{-1}A_{1}\nonumber \\
(g_{0}+2g_{1})(k_{-1}+k_{1})(4 m \epsilon +\hbar^2 k^4_{-1}-2 \hbar^2 k^2_{-1}k^2_{1} + \hbar^2 k^4_{1}) + 2 A^2_{1}g_{1}\nonumber \\ (k_{-1}+k_{1})(6 m \rho g_{0} + 12 m \rho g_{1}- \hbar^2(7 k^2_{-1}+10k_{-1}k_{1}+ 7k^2_{1}))+2 A^2_{1}g_{1}\nonumber \\
(2 m g_{0}((3\rho+A^2_{1})k_{-1}+(3\rho-A^2_{1})k_{1})+4 m g_{1}((3\rho-A^2_{1})k_{-1}  +(3\rho+ \nonumber \\ 
A^2_{1})k_{1})- \hbar^2 (7 k^3_{-1} + 17 k^2_{-1}k_{1}+17 k_{-1}k^2_{1}+ 7 k^3_{1})))
\label{eq:25}
\end{eqnarray}%
\begin{eqnarray} C^{n=0,1}_{k^{13}, \omega}=& -\frac{\hbar^{15}(k_{-1}+k_{1})}{16 m^7}
\label{eq:26}
\end{eqnarray}%
\begin{eqnarray} C^{n=0,1}_{k^{6}, \omega^0}= \frac{\hbar^{6}}{16 m^8 g_{1}(A^2_{-1}+A^2_{1})^3}(-4 m^3 \hbar^4 A^{12}g^3_{1}
(k_{-1}-k_{1})^3 (3k_{-1}+k_{1})-4 m^2 \hbar^4 A^9_{-1}\nonumber \\
A_{1} g_{0}g^2_{1}(k_{-1}-k_{1})^2 (k^2_{-1}-3k_{-1}k_{1}-k^2_{1})(4 m \epsilon +\hbar^2(k^2_{-1}-k^2_{1})^2)-\nonumber \\
4 m^2 \hbar^4 A^{10}_{-1}g^3_{1}(k_{-1}-k_{1})^2 (\hbar^2k^2_{-1}(k_{-1}+k_{1})^2 + 2 m g_{0}((-2 \rho+A^2_{1})k^2_{-1}\nonumber \\
+2 (3\rho-A^2_{1})k_{-1}k_{1} + (2\rho+A^2_{1})k^2_{1}))+4 m \hbar^4 A^6_{1}g^3_{1}((2\rho-A^2_{1})k_{-1}+ \nonumber \\
A^2_{1} k_{1})^2 ( m A^2_{1}g_{0}(k_{-1}-k_{1})(k_{-1}+ 3k_{1})- k^2_{1}(-4 m g_{0} \rho + \hbar^2 (k_{-1}+k_{1})^2))+\nonumber \\
4 m \hbar^4 A_{-1}A^5_{1}g^2_{1}(4 m \epsilon +\hbar^2(k^2_{-1}- k^2_{1})^2)(A^2_{1}(k_{-1}+k_{1})(-m A^2_{1}g_{0}\nonumber \\
( k^3_{-1}+2k^2_{-1}k_{1}-6 k_{-1}k^2_{1}+ k^3_{1}))+k_{-1}k^2_{1}(-8 m \rho g_{0} +\hbar^2(k^2_{-1}-k^2_{1})^2))+ \nonumber \\
2 (-1)^n (\rho-A^2_{1}) k^2_{-1}(- m A^4_{1}g_{0}(k^2_{-1}+2k_{-1}k_{1}-5k^2_{1})-k^2_{1}(-6 m \rho g_{0} +\nonumber \\
\hbar^2(k^2_{-1}-k^2_{1})^2)))- 4 m^2 \hbar^4 A^8_{-1}g^2_{1}(k_{-1}-k_{1})(\hbar^2 k^2_{-1}(k_{-1}+k_{1})^2- \nonumber \\ 
m \rho g_{0}(k_{-1}-k_{1})(4k_{-1}+k_{1}))+A^2_{1}(k_{-1}-k_{1})( \hbar^2 g_{1}k_{-1}(3k_{-1}-2 k_{1}) \nonumber \\ (k_{-1}+k_{1})^2+g_{0}(\hbar^2(k_{-1}-k_{1})^2 + 4 m \rho g_{1}(-5 k^2_{-1}+7 k_{-1}k_{1}+k^2_{1})-\nonumber \\
A^3_{-1}A^3_{1}(4 m \epsilon +\hbar^2(k^2_{-1}-k^2_{1})^2) (192(1+(-1)^n)m^4 A^8_{1} g_{0}g^4_{1}-\nonumber \\ 
4 m \hbar^4 A^2_{1}g^2_{1}(\hbar^2 k_{1}k_{-1}(k_{-1}+k_{1})^2( k^2_{-1}+(1+4(-1)^n)k_{-1}k_{1}+2k^2_{1})+ \nonumber \\ 
2 m \rho g_{0}(2(-1)^n k^4_{-1}+ 4(1+(-1)^n)k^3_{-1}k_{1}-21 (-1)^n k^2_{-1}k^2_{1} +\nonumber \\ 
2(-4+(-1)^n)k^3_{1}k_{-1} + (-1)^n k^4_{1}))- 32(1+(-1)^n) m^3 A^6_{1}g^3_{1}(-\hbar^2 g_{1}\nonumber \\ 
(k_{-1}+k_{1})^2+g_{0}(12 m \rho g_{1}-\hbar^2 (k_{-1}+k_{1})^2))+ \hbar^4 k^2_{-1}k^2_{1}(-16 (-1)^n m \rho g^2_{1}\nonumber \\ 
(6 m \rho g_{0}-\hbar^2 (k_{-1}+k_{1})^2)+g_{0}(4 m \epsilon + \hbar^2(k^2_{-1}-k^2_{1})^2)^2)+8 m^2 A^4_{1}g^2_{1}\nonumber \\ 
(-4(1+(-1)^n) m \rho \hbar^2 g^2_{1}(k_{-1}+k_{1})^2 + g_{0}(24(1+(-1)^n) m^2 \rho^2 g^2_{1}-\nonumber \\ 
4 (1+(-1)^n) m \rho \hbar^2 g_{1}(k_{-1}+k_{1})^2 + \hbar^4(k_{-1}-k_{1}) (2(1+(-1)^n)k^3_{-1}+\nonumber \\ 
2(2+3(-1)^n) k^2_{-1}k_{1}-(-1+3(-1)^n)k^2_{1}k_{-1}-(3+(-1)^n)k^3_{1}))))- \nonumber \\ 
4 m A^5_{-1}A_{1}g^2_{1}(4 m \epsilon +\hbar^2(k^2_{-1}-k^2_{1})^2)(96(1+(-1)^n) m^3 A^8_{1}g_{0}g^2_{1}-2 (-1)^n \nonumber \\ 
\rho \hbar^4 k^2_{1}k^2_{-1}(6 m \rho g_{0} - \hbar^2 (k^2_{-1}-k^2_{1})^2)- \hbar^4 A^2_{1}
(\hbar^2 k_{1}k_{-1}(k_{-1}+k_{1})^2 (2 k^2_{-1}+\nonumber \\ 
 (-1+2(-1)^n)k_{-1}k_{1}+k^2_{1})+ 2 m \rho g_{0}(k_{-1}-k_{1})((-1)^n k^3_{-1}+(-8+ \nonumber \\ 
3(-1)^n) k^2_{-1}k_{1}-2(-2 + 3(-1)^n)k^2_{1}k_{-1}-2(-1)^n k^3_{1}))- 16(1+(-1)^n)\nonumber \\ 
 m^2 A^6_{1} g_{1} ((-\hbar^2 g_{1}(k_{-1}+k_{1})^2+ g_{0}(12 m \rho g_{1}-\hbar^2 (k_{-1}+k_{1})^2))+2 m A^4_{1}\nonumber \\ 
 (-8(1+(-1)^n) m \rho \hbar^2 g^2_{1}(k_{-1}+k_{1})^2 + g_{0}(48(1+(-1)^n) m^2 \rho^2 g^2_{1}-\nonumber \\ 
(-8(1+(-1)^n) m \rho \hbar^2 g^2_{1}(k_{-1}+k_{1})^2 + \hbar^4((3+(-1)^n)k^4_{-1}+ (-3+ \nonumber \\ 
 2(-1)^n)k^3_{-1}k_{1}-3(-1+(-1)^n)k^2_{-1}k^2_{1}+(-1+4(-1)^n)k_{-1}k^3_{1}+ 2(2+\nonumber \\ 
(-1)^n)k^4_{1}))))-4 m A^7_{-1}A_{1} g^2_{1}(4 m \epsilon +\hbar^2(k^2_{-1}-k^2_{1})^2)(48(1+(-1)^n) m^3 \nonumber \\ 
A^6_{1}g_{0} g^2_{1}-8(1+(-1)^n) m^2 A^4_{1} g_{1}(-\hbar^2 g_{1}(k_{-1}+k_{1})^2 + g_{0}(12 m \rho g_{1}-\hbar^2\nonumber \\ 
(k_{-1}+k_{1})^2))+\hbar^4 k_{1}(-2 (-1)^n) m\rho g_{0}k_{1}(k_{-1}+k_{1})^2 + k^2_{-1}(8 m \rho g_{0}\nonumber \\ 
 (k_{-1}-2k_{1})-\hbar^2(k_{-1}-k_{1})(k_{-1}+k_{1})^2))+2 m A^2_{1}(-4(1+(-1)^n) m \rho \hbar^2 \nonumber \\ 
g^2_{1}(k_{-1}+k_{1})^2+g_{0}(24 (1+(-1)^n) m^2 \rho^2 g^2_{1}-4(1+(-1)^n) m \rho \hbar^2 g_{1}\nonumber \\ 
(k_{-1}+k_{1})^2+ \hbar^4 (2 k^4_{-1}-6 k^3_{-1}k_{1}+ 7 k^2_{-1}k^2_{1}+k^4_{1}+(-1)^nk^2_{1}(k_{-1}+k_{1})^2)))) \nonumber \\ 
+ \hbar^4 A^4_{1} A^2_{1} g_{1}(-4 m^3 A^6_{1}g_{0}g^2_{1}(k_{-1}-k_{1})^3 (5k_{-1}+7k_{1})+4 m^2 A^4_{1}g_{1}(k_{-1}-k_{1})^2\nonumber \\ 
(- \hbar^2 g_{1}(k_{-1}+k_{1})^2(k^2_{-1}+6k_{-1}k_{1}+3k^2_{1})+ g_{0}(-3\hbar^2(k_{-1}+k_{1})^2+8 m \rho g_{1}\nonumber \\ 
(2 k^2_{-1}+9k_{-1}k_{1}+k^2_{1})))+k^2_{-1}k^2_{1}(12 m \rho g_{0}(16 m^2 \rho^2 g^2_{1}+(4 m \epsilon +\hbar^2(k^2_{-1}-\nonumber \\ k^2_{1})^2)^2)-\hbar^2 (k_{-1}+k_{1})^2( 48 m^2 \rho^2 g^2_{1}(4 m \epsilon +\hbar^2(k^2_{-1}-k^2_{1})^2)^2))+ m A^2_{1}\nonumber \\ (k_{-1}-k_{1})^2(-48 m \rho \hbar^2 g^2_{1}k_{1}k_{-1}(k_{-1}-k_{1})^2+g_{0}(k^2_{-1}+8k_{-1}k_{1}+k^2_{1})\nonumber \\ 
( 48 m^2 \rho^2 g^2_{1}(4 m \epsilon +\hbar^2(k^2_{-1}-k^2_{1})^2)^2))+m  A^6_{-1} g_{1}(16 m^2 A^6_{1}g_{0}g^2_{1}(k_{-1}-k_{1})^4\nonumber \\ 
+16 m \rho^2 g^2_{1}k^2_{1}k^2_{-1}(4 m \rho g_{0}- \hbar^2 (k_{-1}-k_{1})^2)+4 m A^4_{1}g_{1}(k_{-1}-k_{1})^2(- \hbar^2 g_{1}\nonumber \\ (k_{-1}+k_{1})^2(3 k^2_{-1}- 6k_{-1}k_{1}+k^2_{1}) + g_{0}(-3 \hbar^2 (k_{-1}-k_{1})^2 + 8 m g_{1}(4 k^2_{-1}\nonumber \\ 
-3k_{-1}k_{1}+2k^2_{1})))+ A^2_{1}(k_{-1}-k_{1})(-16 m \rho \hbar^2 g^2_{1}k_{-1}(3k_{-1}-k_{1}) k_{1}(k_{-1}+\nonumber \\ 
k_{1})^2 + g_{0}(k_{-1}-k_{1})(16 m^2 \rho^2 g^2_{1} (k^2_{-1}+16k_{-1}k_{1}+3k^2_{1})+ k_{1}(4k_{-1}+k_{1})\nonumber \\ 
(4 m \epsilon +\hbar^2(k^2_{-1}-k^2_{1})^2)^2)))-\hbar^4 A^2_{-1}A^4_{1} g_{1}(8 m^3 A^6_{1}g_{0}g^2_{1}(k_{-1}-k_{1})^4 + \nonumber \\ 
4 m^2 A^4_{1} g_{1}(k_{-1}-k_{1})^2(-\hbar^2g_{1}(2k_{-1}-3k_{1})k_{1}(k_{-1}+k_{1})^2+ g_{0}(4 m \rho g_{1} (k^2_{-1}+\nonumber \\ 
7 k_{-1}k_{1}- 5k^2_{1})+ \hbar^2 (k_{-1}-k_{1})^2))+ k^2_{-1}k^2_{1}(-12 m \rho g_{0}(16 m^2 \rho^2 g^2_{1}+(4 m \epsilon + \nonumber \\ \hbar^2(k^2_{-1}-k^2_{1})^2)^2)+ \hbar^2 (k_{-1}+k_{1})^2 (48 m^2 \rho^2 g^2_{1}+(4 m \epsilon +\hbar^2(k^2_{-1}-k^2_{1})^2)^2) \nonumber \\ 
-m A^2_{1}(k_{-1}-k_{1})(-16 m \rho \hbar^2 g^2_{1} k_{-1}(k_{-1}-3 k_{1}) k_{1}(k_{-1}+k_{1})^2 + g_{0}(k_{-1}- \nonumber \\ 
k_{1})(\hbar^4 k^{10}_{-1}+4\hbar^4 k^{9}_{-1}k_{1} + 16 m^2 \rho^2 g^2_{1} k^2_{1}- 4 \hbar^4 k^{8}_{-1}k^{2}_{-1}-16 \hbar^4 k^{7}_{-1}k^{3}_{-1} + k^6_{-1} \nonumber \\ 
(8 m \epsilon \hbar^2 + 6\hbar^4 k^4_{1})+8 k^5_{-1}(4 m \epsilon \hbar^2 k_{1}+ 3\hbar^4 k^5_{1})-4 k^4_{-1}
(4 m \epsilon \hbar^2 k^2_{1} + \hbar^4 k^6_{1})- \nonumber \\ 
16 k^3_{-1}(4 m \epsilon \hbar^2 k^3_{1}+\hbar^4 k^7_{1}) + k^2_{-1}(48 m^2 \rho^2 g^2_{1}+(4 m \epsilon + 
\hbar^2 k^4_{1})^2)+\nonumber \\ 
4 k_{1}k_{-1}(64 m^2 \rho^2 g^2_{1} + (4 m \epsilon + \hbar^2 k^4_{1})^2)))))
\label{eq:27}
\end{eqnarray}%
\begin{eqnarray} C^{n=0,1}_{k^{8}, \omega^0}= \frac{\hbar^{6}}{64 m^8 g_{1}(A^2_{-1}+A^2_{1})^3}( 4 m^2 \hbar^2 A^9_{-1} A_{1}
g_{0}g^2_{1}(k_{-1}-k_{1})^2(4 m \epsilon +\hbar^2(k^2_{-1}-k^2_{1})^2)\nonumber \\ 
- 4 m^2 \hbar^2 A^{10}_{-1}g^2_{1}(k_{-1}-k_{1})^2(-\hbar^2 g_{1}(5 k^2_{-1}+2 k_{-1}k_{1} + k^2_{1})+
g_{0}(4 m \rho g_{1}+\nonumber \\ 
2 \hbar^2 (3 k_{-1}+k_{1})(k_{-1}+3k_{1})))+4 m^2 \hbar A^{8}_{-1}g_{1}(k_{-1}-k_{1})(g_{0}(4 m^2 \rho \nonumber \\ 
(3\rho-A^2_{1}) g^2_{1}(k_{-1}-k_{1})+\hbar^4 k^2_{1}(k_{-1}+k_{1})^2(3k_{-1}+k_{1}) + 2m \hbar^2 g_{1}\nonumber \\ 
(k_{-1}-k_{1})((4 \rho-A^2_{1})k^2_{-1} + 2 (8\rho-7A^2_{1})k_{1}k_{-1} + (4 \rho-A^2_{1})k^2_{1}))+\nonumber \\ 
\hbar^2 g_{1}(m A^2_{1} g_{1}(k_{-1}-k_{1}) (13 k^2_{-1}-6 k_{-1}k_{1}+k^2_{1})+2(-\hbar^2 k^5_{-1}\nonumber \\ 
(-2 m\rho g_{1} + \hbar^2 k^2_{-1})k_{1}+k_{-1}(4 m\rho g_{1} - 7 \hbar^2 k^2_{-1})k^2_{1} + (2 m\rho g_{1} - 3 \hbar^2 k^2_{-1})\nonumber \\ 
k^3_{1}))) + 4 \hbar^2 A^6_{1}g_{1}(\hbar^2(\hbar^4k^2_{-1}k^2_{1}(k_{-1}+k_{1})^4-2 m \hbar^2 g_{1}k^2_{1}(k_{-1} + k_{1})^2\nonumber \\ 
(( 4\rho-3A^2_{1})k^2_{-1}+ 2A^2_{1}k_{-1}k_{1}+ A^2_{1}k^2_{1}+ m^2 g^2_{1}((-2\rho+A^2_{1})^2 k^4_{-1}+ 4\rho \nonumber \\ 
(2\rho-A^2_{1})k^3_{-1}k_{1}+2(12 \rho^2 -6 \rho A^2_{1} + A^4_{1}) k^2_{-1}k^2_{1}+4(2\rho^2+\rho A^2_{1} -2 A^4_{1})\nonumber \\ 
k_{-1}k^3_{1}+(4\rho^2+5 A^4_{1})k^4_{1}))- m g_{0}(4 m^2 \rho g^2_{1}((4 \rho^2 - 3 \rho A^2_{1} + A^4_{1})k^2_{-1}-\nonumber \\ 
2 A^2_{1}(-3\rho + A^2_{1})k_{1}k_{-1}+(4 \rho^2 -3 \rho A^2_{1} + A^4_{1})k^2_{1})+ \hbar^4 k^2_{-1}(k_{-1}+k_{1})^2\nonumber \\ 
(4 \rho^2 + A^2_{1}(k_{-1}-k_{1})(k_{-1}+3 k_{1}))- 2 m \hbar^2 g_{1}((4\rho-3A^2_{1}) k^2_{-1}+2 A^2_{1}k_{-1}k_{1}\nonumber \\ 
+ A^2_{1} k^2_{1})(4 \rho k^2_{1}+A^2_{1}(k_{-1}-k_{1})(k_{-1}+3k_{1}))))-4 A^5_{-1}A_{1}(4 m \epsilon +\hbar^2(k^2_{-1}-\nonumber \\ 
k^2_{1})^2)(-64 (1+(-1)^n)m^3 A^6_{1}g^3_{1}(3 g_{0}+g_{1})+ (-1)^n \hbar^2 (g_{0}(-16 m \rho \hbar^2 g_{1}\nonumber \\ 
k^2_{1}k^2_{-1} + \hbar^4 k^2_{1}k^2_{-1}(k_{-1}+k_{1})^2 + 12 m^2 \rho^2 g^2_{1}(k^2_{-1}+k^2_{1}))-
2\hbar^2 g_{1}(-\hbar^2 k^2_{-1} k^2_{1}\nonumber \\
 (k_{-1}+k_{1})^2+ m\rho g_{1}(k^4_{1}+2k^3_{-1}k_{1}+6 k^2_{-1}k^2_{1}+2k^3_{1}k_{-1}+k^4_{1})))+ m \hbar^2 A^2_{1}g_{1}\nonumber \\ 
(\hbar^2 g_{1}((1+2(-1)^n)k^4_{-1}+(13+4(-1)^n) k63_{-1}k_{1} + 3(1+4(-1)^n)k^2_{-1}k^2_{1}+ \nonumber \\ 
(7+4(-1)^n) k^3_{1}k_{-1}+2(-1)^n k^4_{1})- 2 g_{0}(m \rho g_{1}(k_{-1}-k_{1})((-4+5(-1)^n)\nonumber \\ 
k_{-1}-(-8+(-1)^n)k_{1})+ \hbar^2 ((-2+(-1)^n) k^4_{-1}+2(-2+(-1)^n)k^3_{-1}k_{1}- \nonumber \\ 
(-13+5(-1)^n)k^2_{-1}k^2_{1} + 2 (1+2(-1)^n) k_{-1}k^3_{1} + (1+2 (-1)^n)k^4_{1})))+\nonumber \\ 
2 m^2 A^4_{1}g^2_{1}(8(1+(-1)^n)g_{1}(4 m \rho g_{1} - \hbar^2 (k_{-1}+k_{1})^2)+g_{0}(96(1+(-1)^n) m \rho g_{1}\nonumber \\ 
- \hbar^2(4(k_{-1}+2k_{1})(3k_{-1}+2k_{1})+ (-1)^n (k_{-1}+k_{1})(13 k_{-1}+17 k_{1})))))- \nonumber \\ 
4 \hbar^2 A_{-1}A^5_{1} (4 m \epsilon +\hbar^2(k^2_{-1}-k^2_{1})^2)(\hbar^2 g_{1}(2 (-1)^n \hbar^2 k^2_{-1}k^2_{1}(k_{-1}+k_{1})^2+
m g_{1}\nonumber \\ 
(A^2_{1}(k_{-1}+k_{1}) (k^3_{-1}+2 k^2_{-1}k_{1}+7 k^2_{1}k_{-1}+2 k^3_{1})+ 2(-1)^n (-\rho+A^2_{1})(k^4_{1}+\nonumber \\ 
2k^3_{-1}k_{1}+6 k^2_{-1}k^2_{1}+2k^3_{1}k_{-1}+k^4_{1})))+g_{0}( m A^2_{1}g_{1}(2 \hbar^2 g_{1}(2 \hbar^2 k^2_{1}(k_{-1}+k_{1})^2 \nonumber \\ 
+ m g_{1}((-8\rho+A^2_{1})k^2_{-1}-2(4\rho+A^2_{1})k_{1}k_{-1}+(9A^2_{1}-16\rho)k^2_{1}))+(-1)^n \nonumber \\ 
(\hbar^4 k^2_{-1}k^2_{1}(k_{-1}+k_{1})^2+ 2 m^2 (\rho-A^2_{1})g^2_{1}((6\rho-A^2_{1})k^2_{-1}+2 A^2_{1}k_{1}k_{-1}+\nonumber \\ 
(6\rho-5 A^2_{1})k^2_{1})- 2 m \hbar^2 g_{1} k^2_{1}(8 \rho k^2_{1}+ A^2_{1}(k^2_{-1}+2k_{-1}k_{1}-7 k^2_{1})))))- \nonumber \\ 
4 m A^7_{-1} A_{1} g_{1}(4 m \epsilon +\hbar^2(k^2_{-1}-k^2_{1})^2)(-32 (1+(-1)^n) m^2 A^4_{1} g^2_{1}(3g_{0}+g_{1})+\nonumber \\
2 (1+(-1)^n) m A^2_{1} g_{1}(4 g_{1}(4 m \rho g_{1}- \hbar^2 (k_{-1}+k_{1})^2)+g_{0}(48 m \rho g_{1}-\hbar^2 \nonumber \\ 
(7 k^2_{-1}+14 k_{-1}k_{1}+ 11 k^2_{1})+\hbar^2 (-\hbar^2 g_{1} k_{1}(-5k^3_{-1}+3 k^2_{-1}k_{1}+k^2_{1}k_{-1}+k^3_{1})\nonumber \\
+ 2 g_{0}(m \rho g_{1}(4+(-1)^n)k^2_{-1}+2(-2+(-1)^n) k_{-1}k_{1}+ (8+5(-1)^n)k^2_{1})+\hbar^2 \nonumber \\
(-(-1)^n k^2_{1}(k_{-1}+k_{1})^2+ k^2_{-1}(k^2_{-1}+2k_{-1}k_{1}-7k^2_{1})))))+A^3_{-1}A^3_{1}(4 m \epsilon + \nonumber \\
\hbar^2(k^2_{-1}-k^2_{1})^2)(2 g_{1}(32(1+(-1)^n) m^2 A^4_{1} g^2_{1}(k_{-1}+k_{1})- m \hbar^2 A^2_{1}g_{1}\nonumber \\
(19 k^3_{-1}+ 63 k^2_{-1}k_{1}-81 k_{-1}k^2_{1}+ 29 k^3_{1}+24(-1)^n(k_{-1}+k_{1})^3)- 8 (-1)^n \nonumber \\ 
\hbar^2(k_{-1}+k_{1})(-3 m \rho g_{1}(k_{-1}+k_{1})^2 + \hbar^2 k_{1}k_{-1}(k^2_{-1}+4k_{-1}k_{1}+k^2_{1})))+ \nonumber \\ g_{0}(\hbar^4k^9_{-1}+\hbar^4k^8_{-1}k_{1}-4\hbar^4k^7_{-1}k^2_{1}-4\hbar^4k^6_{-1}k^3_{1}+2\hbar^2k^4_{-1}k_{1}(4 m\epsilon- \nonumber \\
4(-1)^n \hbar^2+3\hbar^2 k^4_{1})+ k^5_{-1}(8 m\epsilon \hbar^2+6 \hbar^4 k^4_{1})+k_{1}(-16 m^2(6(-1)^n)\rho^2-\nonumber \\ 
(1+(-1)^n) A^2_{1}(7\rho+5A^2_{1})) g^2_{1}+24 (-2+(-1)^n) m \hbar^2 A^2_{1}g_{1}k^2_{1}+(4 m\epsilon + \nonumber \\ 
\hbar^2 k^4_{1})^2)+k_{-1}(-16 m^2(16 (-1)^n \rho^2-(1+(-1)^n) A^2_{1}(5\rho + 7A^2_{1})g^2_{1}+ \nonumber \\ 
8 m \hbar^2(16 (-1)^n \rho-(1+10(-1)^n)A^2_{1})g_{1}k^2_{1}-8(-1)^n\hbar^4 k^4_{1}+(4 m\epsilon +  \nonumber \\ 
\hbar^2 k^4_{1})^2)- 4\hbar^4 k^3_{-1}(-6 (-1+ 2(-1)^n)m A^2_{1}g_{1}+ k^2_{1}(4 m\epsilon + 10(-1)^n \hbar^2k^4_{1})) \nonumber \\ 
-4\hbar^2 k_{1}k^2_{-1}(-2 m (16 (-1)^n \rho+(2-7(-1)^n)A^2_{1})g_{1}+k^2_{1}(4 m\epsilon + 10(-1)^n \nonumber \\ 
\hbar^2k^4_{1})))))+ A^4_{-1}A^2_{1}(\hbar^2 g_{1}(3 \hbar^4 k^{11}_{-1}+9\hbar^4 k^{10}_{-1}k_{1}-3 \hbar^4 k^{9}_{-1}k^2_{1}
-33\hbar^4 k^{8}_{-1}k^3_{1}  \nonumber \\ 
+ 6 \hbar^2 k^{6}_{-1}k_{1}(2(6 m \epsilon + \hbar^2)+7 \hbar^2 k^4_{1}+6 k^7_{-1}(4 m\epsilon \hbar^2-3\hbar^4 k^4_{1})+ \nonumber \\ 
3 k^3_{-1}(16 m^2 \epsilon^2 + 16 m^2(3 \rho^2-A^4_{1}) g^2_{1}+16 m \hbar^2 (-10 \rho + 7 A^2_{1})g_{1} k^2_{1}+ \nonumber \\ 
(40 m \epsilon \hbar^2 + 88\hbar^4) k^4_{1}-11 \hbar^4 k^8_{-1}) + k^2_{-1}k_{1}(16 m^2(9 \epsilon^2 + (27 \rho^2+ 5A^2_{1})g^2_{1}) +  \nonumber \\
48 m \hbar^2 (-10\rho+7A^2_{1})g^2_{1}k^2_{1}+ 12 \hbar^2 (2 m \epsilon +9 \hbar^2)k^4_{1}-3 \hbar^4 k^8_{1})+ k_{-1}k^2_{1} \nonumber \\ 
( 16 m^2(9 \epsilon^2 +(27\rho-A^2_{1})g^2_{1})-24 m \hbar^2(4 \rho+9 A^2_{1})g_{1}k^2_{1}+ 12 \hbar^2 (6 m \epsilon +\hbar^2) \nonumber \\ 
k^4_{1}+ 9 \hbar^4 k^8_{1}+k^3_{1}(16 m^2(9 \rho^2 -  A^4_{1})g^2_{1}-120 m \hbar^2 A^2_{1}g_{1}k^2_{1}+ 3 (4 m \epsilon + \nonumber \\ 
\hbar^2 k^4_{-1})^2) - 6 \hbar^2 k^4_{-1} k_{1}(4 m (4\rho+9 A^2_{1})g_{1}+
k^2_{1}(20  m \epsilon - 44 \hbar^2 + 3 \hbar^2 k^4_{1}))+ \nonumber \\  
6 \hbar^2 k^5_{-1}(-20 m A^2_{1} g_{1}+ k^2_{1}(4  m \epsilon + 18 \hbar^2 + 7 \hbar^2 k^4_{1})))- 4 g_{0}(k_{-1}+k_{1})(48 m^3 \rho^3 \nonumber \\  
g^3_{1}-16 m^2 \hbar^2 g^2_{1}( 9\rho A^2_{1}(k_{-1}-k_{1})^2 + 2 A^4_{1}(k_{-1}- k_{1})^2 +6\rho^2 k_{-1}k_{1}- 2 k_{1}k_{-1} \nonumber \\ 
(4 m \epsilon +\hbar^2(k^2_{-1}-k^2_{1})^2)^2 + 3 m g_{1}( 16 m^2 \epsilon^2 \rho+ \hbar^2(\rho \hbar^2 k^8_{-1} + 4\hbar^2 (\rho+A^2_{1}) \nonumber \\ 
k^3_{-1}k_{1}-2 k_{-1}k_{1}(4 m \epsilon +\hbar^2(k^2_{-1}-k^2_{1})^2)^2+3 m g_{1}(16 m^2 \epsilon^2 \rho + \hbar^2
(\rho \hbar^2 k^8_{-1} \nonumber \\ 
+4 \hbar^2 (\rho + A^2_{1})k^3_{-1}k_{1}-k^4_{1}(8 m \epsilon \rho + \hbar^2 (5 A^2_{1}+\rho)k^4_{1}))+k^4_{-1}
(8 m \epsilon \rho + \hbar^2 \nonumber \\ 
(5 A^2_{1}+6\rho k^4_{1}))))))+A^2_{-1}A^4_{1}(\hbar^2 g_{1}(3 \hbar^4 k^{11}_{-1}+9\hbar^4 k^{10}_{-1}k_{1}- 
3 \hbar^4 k^{10}_{-1}k^{2}_{1}-  \nonumber \\ 
3 \hbar^4 k^{9}_{-1}k^{2}_{1}-33 \hbar^4 k^{8}_{-1}k^{3}_{1}+ 6 \hbar^2 k^{6}_{-1}k_{1}(2(6 m \epsilon +
\hbar^2)+7 \hbar^2 k^4_{1})+ 6 k^7_{-1}(4 m \epsilon \nonumber \\ 
 - 3\hbar^4 k^4_{1})+ k^3_{-1} (48 m^2 \epsilon^2 + 8 m^2 (18 \rho^2 - 10 \rho A^2_{1} - A^4_{1}) g^2_{1} + 16 m \hbar^2 (-30\rho+ \nonumber \\
19A^2_{1})g_{1}k^2_{1} +24 \hbar^2 (-5 m\epsilon + 11 \hbar^2)k^4_{1}-33 \hbar^4 k^8_{1}) + 
 k^2_{-1}k_{1}( 144 m^2 \epsilon^2 + 8 m^2 \nonumber \\ 
(54 \rho^2 - 18 \rho A^2_{1} + 7 A^4_{1})g^2_{1}+48 m \hbar^2 (3A^2_{1}-10\rho)g_{1}k^2_{1}+ 12(2 m \epsilon \hbar^2 + 9\hbar^4) \nonumber \\ 
k^4_{1}-3 \hbar^4 k^8_{1})+k_{-1}k^2_{1}(144 m^2 \epsilon^2 + 8 m^2(54 \rho^2 + 18 \rho A^2_{1} - 11 A^4_{1})g^2_{1}-24 m  \nonumber \\ 
\hbar^2(4 \rho+ 11 A^2_{1})g_{1} k^2_{1}+ 12 (6 m \epsilon \hbar^2 + \hbar^4) k^4_{1}+9 \hbar^4 k^8_{1}) +k^3_{1}(8 m^2(18 \rho^2 - \nonumber \\ 
10 \rho A^2_{1} + 5 A^4_{1})g^2_{1} - 120 m \hbar^2 A^2_{1} g_{1}k^2_{1} + 3 (4 m \epsilon + \hbar^2 k^4_{1})^2)- 6 \hbar^2 k^4_{-1} k_{1} 
 (4 m \nonumber \\ 
(4 \rho + A^2_{1}) g_{1}+ k^2_{1}(12 \rho^2 - 12 \rho A^2_{1} + A^4_{1})k^2_{1}-8 \hbar^2 k^2_{1}k^2_{-1}
(4 m \epsilon + \hbar^2 (k^2_{-1}- \nonumber \\
k^2_{1})^2)-8 m^2 \hbar^2 g^2_{1}(48 \rho^2 k^2_{-1}k^2_{1}+ 16 \rho A^2_{1}(k_{-1}-k_{1})^2 (k^2_{-1}+ 4 k_{-1}k_{1}+k^2_{1})- \nonumber \\ 
A^4_{1}(k_{-1}-k_{1})^2 (k^2_{-1}+ 14 k_{-1}k_{1}+k^2_{1})) + m g_{1}(3 \hbar^2(4\rho-A^2_{1})k^{10}_{-1}+ 6 \hbar^4 A^2_{1} \nonumber \\ k^9_{-1}k_{1}+ 9 \hbar^4(-4\rho+A^2_{1})k^8_{-1}k^2_{1}-24 \hbar^4 A^2_{1} k^7_{-1}k^3_{1}+ 12 \hbar^2 A^2_{1}k^5_{-1}k_{1}(4 (m \epsilon+ \nonumber \\
\hbar^2) +3 \hbar^2 k^4_{1})+ 3 \hbar^2 k^4_{1})+ 6\hbar^2 k^6_{-1}(16 m\epsilon \rho+2(-2 m\epsilon+\hbar^2) A^2_{1}+\hbar^2(4\rho- \nonumber \\
A^2_{1})k^4_{1})+ 6 \hbar^2 k^4_{-1}k^2_{1}(8\rho(-2 m\epsilon+\hbar^2) A^2_{1}+\hbar^2 (4\rho-A^2_{1})k^4_{1})+ 2 A^2_{1}k_{1}k_{-1} \nonumber \\ (48 m^2 \epsilon^2 +8(3 m \epsilon \hbar^2)k^4_{1}+3 \hbar^4 k^8_{1})+k^2_{-1}(48 m^2 \epsilon^2 (4\rho-A^2_{1})+ 4 \hbar^2 (12\rho \nonumber \\ 
(-2 m\epsilon +\hbar^2)+ (6 m\epsilon -7 \hbar^2)A^2_{1})k^4_{1}+9\hbar^4(-4\rho+A^2_{1})k^8_{1})-8 \hbar^2 k^3_{-1}k^3_{1} \nonumber \\ 
(-12 \rho \hbar^2 + A^2_{1}(12 m \epsilon + 8 \hbar^2+3\hbar^2 k^4_{1}))+ k^2_{1}(12\rho( 4 m \epsilon+ \hbar^2 k^4_{1})^2 - A^2_{1} \nonumber \\ (48 m^2 \epsilon^2 - 4 \hbar^2(\hbar^2-6 m \epsilon) k^4_{1}+ 3\hbar^4 k^8_{1}))))))
\label{eq:28}
\end{eqnarray}%
\begin{eqnarray} C^{n=0,1}_{k^{10}, \omega^0}= \frac{\hbar^{10}}{256 m^8 g_{1}(A^2_{-1}+A^2_{1})^3}(-16 m^2 \hbar^2 A^{10}_{-1}
g^3_{1}(k_{-1}-k_{1})- 4 m \hbar^2 A^8_{-1}g_{1}(k_{-1}-k_{1}) \nonumber \\ 
(2 g_{1}(2 m A^2_{1}g_{1}(k_{-1}-k_{1})+8 m \rho g_{1}k_{1}- \hbar^2(k_{-1}+3k_{1})(5 k^2_{-1}+2 k_{-1}k_{1} \nonumber \\ 
+k^2_{1}))+ g_{0}(16 m \rho g_{1}(-k_{-1}+k_{1})+\hbar^2 (3 k_{-1}+k_{1})(k^2_{-1}+2 k_{-1}k_{1}+ 5k^2_{1}))) \nonumber \\ 
+8 m A^7_{-1}A_{1}g_{1}(4 m \epsilon +\hbar^2(k^2_{-1}-k^2_{1})^2)(-2 g_{1}(8(1+(-1)^n) m A^2_{1} g_{1}+  \nonumber \\ 
\hbar^2k_{1}(-k_{-1}+k_{1})+ g_{0}(4(1+(-1)^n) m (\rho-7 A^2_{1}) g_{1}-\hbar^2((3+(-1)^n) \nonumber \\ 
k^2_{-1}+2(-1+(-1)^n)k_{-1}k_{1}+(7+5(-1)^n)k^2_{1})))+4 A^6_{-1} g_{1}(\hbar^2(4 m^2 g^2_{1} \nonumber \\ 
((-5\rho^2 + 4 \rho A^2_{1} + 2 A^4_{1})k^2_{-1}-2(\rho^2 + 8 \rho A^2_{1} +2 A^4_{1})k_{1}k_{-1}+(-5\rho^2 +  \nonumber \\ 
12 \rho A^2_{1} +2 A^4_{1})-\hbar^4 (k_{-1}+k_{1})^2 (k^4_{1}+2k^3_{-1}k_{1}+10 k^2_{-1}k^2_{1}+2k^3_{1}k_{-1}+ 
k^4_{1})+  \nonumber \\ 
8 m \hbar^2 g_{1}((\rho+3A^2_{1})k^4_{-1}+2(\rho+4A^2_{1})k^3_{-1}k_{1}+2(3\rho-5A^2_{1})k^2_{-1}k^2_{1}+ 2\rho k_{-1}k^3_{1} \nonumber \\ +(\rho-A^2_{1})k^4_{1}))+4 m g_{0}(4 m^2 \rho^3 g^2_{1}-8 m \rho \hbar^2 g_{1}((\rho-2 A^2_{1})k^2_{1}+4 A^2_{1}k_{1}k_{-1}+ \nonumber \\ 
(\rho-2 A^2_{1})k^2_{1})+ \hbar^4((\rho-A^2_{1})k^4_{-1}+2 \rho k^3_{-1}k_{1}+ 2(3\rho-5A^2_{1})k^2_{-1}k^2_{1}+2(\rho+ \nonumber \\ 
4 A^2_{1})k_{-1}k^3_{1}+(\rho+3A^2_{1})k^4_{1})))-4 A^6_{1}g_{1}( m g_{0}(-16 m^2 \rho^3 g^2_{1} + 16 m \rho \hbar^2 g_{1} \nonumber \\ ((2\rho-A^2_{1})k^2_{-1}+2 A^2_{1}k_{-1}k_{-1}+(2\rho-A^2_{1}) k^2_{1})+ \hbar62(-(4\rho+5 A^2_{1})k^4_{-1}- \nonumber \\ 
 4(2\rho+3A^2_{1}) k^3_{-1}k_{1}+2 (-12 \rho+5 A^2_{1})k^2_{1}k^2_{-1}+ 4 (-2\rho+A^2_{1})k_{-1}k^3_{1}+(-4\rho+ \nonumber \\ 
3 A^2_{1})k^4_{1}))+ \hbar^2( 4 m^2 g^2_{1} ((5 \rho^2 - 4 \rho A^2_{1} + A^4_{1}) k^2_{1}+2 (\rho^2 + 
2 \rho A^2_{1} - A^4_{1})k_{1}k_{-1}+  \nonumber \\
(5\rho+A^2_{1})k^2_{1}+ \hbar^4 (k_{-1}+k_{1})^2  (k^4_{-1}+2k^3_{-1}k_{1}+10 k^2_{-1} k^2_{1}+2k^3_{1}k_{-1}+k^4_{1})- \nonumber \\ 
2 m \hbar^2 g_{1}((4\rho-3A^2_{1}) k^4_{-1}+4 (2\rho-A^2_{1})k^3_{-1}k_{1}+2 (12 \rho-5 A^2_{1})k^2_{-1}k^2_{1}+  \nonumber \\ 
4 (2\rho+3A^2_{1}) k_{-1}k^3_{1}+ (4\rho+5 A^2_{1})k^4_{1})))+ 4 A_{-1}A^5_{1}(4 m \epsilon +\hbar^2(k^2_{-1}-k^2_{1})^2) \nonumber \\ 
(g_{0}(4 m^2(3(-1)^n \rho^2 + (1+(-1)^n)A^2_{1} (A^2_{1}- 4\rho)) g^2_{1}- 2 m \hbar^2 g_{1}(8(-1)^n \rho - \nonumber \\ 
(1+3(-1)^n)A^2_{1})k^2_{-1}+2 (1+(-1)^n)A^2_{1}k_{-1}k_{1} + (8 (-1)^n \rho -(5+ 7(-1)^n) \nonumber \\ 
A^2_{1}) k^2_{1})+ (-1)^n \hbar^4 (k^4_{-1}+2k^3_{-1}k_{1}+6 k^2_{-1}k^2_{1}+2k^3_{1}k_{-1}+k^4_{1}))+ 2 \hbar^2 g_{1}((-1)^n \nonumber \\ 
\hbar^2(k^4_{-1}+2k^3_{-1}k_{1}+6 k^2_{-1}k^2_{1} + 2k^3_{1}k_{-1}+k^4_{1})+ m g_{1}(-(-1)^n (\rho-A^2_{1}) \nonumber \\ 
(5 k^2_{-1}+2k_{-1}k_{1}+ 5k^2_{1})+A^2_{1}(3 k^2_{-1}+4k_{-1}k_{1}+ 5k^2_{1}))))- A^3_{1}A^3_{-1}(4 m \epsilon + \nonumber \\ 
\hbar^2(k^2_{-1}-k^2_{1})^2)(g_{0}(-96 m^2 ((-1)^n \rho^2-(1+(-1)^n)A^2_{1} (\rho+2 A^2_{1}))g^2_{1}+ \nonumber \\ 
8 m \hbar^2 g_{1}((A^2_{1}+(-1)^n (16\rho-5A^2_{1}))k^2_{-1}+ 6(1+(-1)^n)A^2_{1}k_{1}k_{-1}+(16  \nonumber \\ 
(-1)^n \rho -3(1+3(-1)^n)A^2_{1}) k^2_{1})-8 (-1)^n \hbar^4(k^4_{-1}+ 2k^3_{-1}k_{1}+6 k^2_{-1}k^2_{1}+ \nonumber \\ 
2 k^3_{1}k_{-1}+k^4_{1})+(4 m \epsilon +\hbar^2(k^2_{-1}-k^2_{1})^2)^2)+16 g_{1}(8(1+(-1)^n) m^2 A^4_{1} g^2_{1}  \nonumber \\ 
- m \hbar^2 A^2_{1}g_{1}((3+5(-1)^n) k^2_{-1}+(5+2(-1)^n)k_{1}k_{-1}+(4+5(-1)^n)k^2_{1})- \nonumber \\ 
(-1)^n \hbar^2(-m \rho g_{1}(5 k^2_{-1}+2 k_{-1}k_{1}+5 k^2_{1})+\hbar^2(k^4_{-1}+2k^3_{-1}k_{1}+6 k^2_{-1}k^2_{1}+ \nonumber \\ 2k^3_{1}k_{-1}+k^4_{1}))))+4 A^5_{-1}A_{1}(4 m \epsilon +\hbar^2(k^2_{-1}-k^2_{1})^2)(g_{0}(12 m^2((-1)^n \rho^2 -  \nonumber \\ 
9(1+(-1)^n) A^4_{1}) g^2_{1}- 2 m \hbar^2 g_{1}((8(-1)^n \rho - (-5+(-1)^n) A^2_{1})k^2_{-1}+6 \nonumber \\ 
(1+(-1)^n) A^2_{1}k_{1}k_{-1}+(8(-1)^n \rho +3 (3+(-1)^n) A^2_{1})k^2_{1})+(-1)^n \hbar^4 \nonumber \\ 
(k^4_{-1}+2k^3_{-1}k_{1}+6 k^2_{-1}k^2_{1}+2k^3_{1}k_{-1}+k^4_{1})) + 2g_{1}(-32(1+(-1)^n) m^2  \nonumber \\ 
A^4_{1} g^2_{1}+ m \hbar^2 A^2_{1}g_{1}((3+5(-1)^n)k^2_{-1}+2(4+(-1)^n) k_{1}k_{-1}+(1+ \nonumber \\ 
5(-1)^n)k^2_{1})+(-1)^n \hbar^2(-m \rho g_{1}(5 k^2_{-1}+2k_{-1}k_{1}+5k^2_{1}) + \hbar^2 (k^4_{-1}+ \nonumber \\ 
2k^3_{-1}k_{1}+6 k^2_{-1}k^2_{1}+2k^3_{1}k_{-1}+k^4_{1}))))+A^4_{-1}A^2_{1}(-\hbar^2 g_{1}(5 \hbar^4 k^{10}_{-1}+ \nonumber \\ 
2 \hbar^4 k^9_{-1}k_{1} +16 m^2( 5\epsilon^2 +(15 \rho^2- 2(6 \rho A^2_{1}+A^4_{1}))g^2_{1})k^2_{1}-15 \hbar^4 k^8_{-1}k^2_{1}- \nonumber \\ 
8 \hbar^4 k^7_{-1}k^3_{1}-48 m \hbar^2(2\rho+A^2_{1})g_{1}k^4_{1}+4(10 m \epsilon \hbar^2 +3 \hbar^4)k^6_{1} + 5\hbar^4 k^{10}_{1} + \nonumber \\ 4 \hbar^2 k^5_{-1}k_{1}(4 m\epsilon + 12 \hbar^2+ 3\hbar^2 k^4_{1}) + 2 k^6_{-1}
(20 m \epsilon \hbar^2 + 6\hbar^4 + 5 \hbar^4 k^4_{1})- \nonumber \\
8 \hbar^2 k^3_{-1}k_{1}(24 m(\rho+A^2_{1})g_{1}+ 4(m\epsilon -9\hbar^2)k^2_{1}+\hbar^2 k^6_{1})+ k^2_{-1}(16 m^2(5 \epsilon^2+ \nonumber \\ 
(15 \rho^2- 2(16 \rho A^2_{1}+A^4_{1})) g^2_{1})+96 m \hbar^2 (-6\rho+5A^2_{1})g_{1}k^2_{1}+ 20 \hbar^2(-2 m\epsilon \nonumber \\  
+9\hbar^2)k^4_{1}-15\hbar^4 k^8_{1})+ 2k_{1}k_{-1}(16 m^2(\epsilon^2 +(3\rho^2 + 12 \rho A^2_{1} +2 A^4_{1})g^2_{1})- \nonumber \\  
 96 m \hbar^2(\rho+A^2_{1})g_{1}k^2_{1}+ 8\hbar^2(m \epsilon+3 \hbar^2)k^4_{1}+\hbar^4 k^8_{1})+2\hbar^2 k^4_{-1}(-24 m \nonumber \\ 
(2\rho+A^2_{1}) g_{1}+ 5k^2_{1}(-4 m\epsilon + 18 \hbar^2+\hbar^2 k^4_{1})))+ 4g_{0}(48 m^3\rho^3 g^3_{1}-  \nonumber \\ 
96 m^2 \rho \hbar^2 g^2_{1}((\rho-A^2_{1})k^2_{-1}+ 2 A^2_{1}k_{-1}k_{1}+(\rho-A^2_{1})k^2_{1}- 2 (k^2_{-1}+k^2_{1}) \nonumber \\ 
(4 m \epsilon +\hbar^2(k^2_{-1}-k^2_{1})^2)+ 3 m g_{1}(16 m^2 \epsilon^2 \rho+ \hbar^2(\rho \hbar^2 k^8_{-1}
+8\hbar^2(\rho+A^2_{1}) \nonumber \\  
k^3_{-1}k_{1}- 4\rho \hbar^2 k^6_{-1}k^2_{1} + 8\hbar^2(\rho+A^2_{1})k_{-1}k^3_{1}+k^4_{1}(4\rho(2 m \epsilon+\hbar^2)+
\hbar^2(\rho k^4_{1} \nonumber \\
+2 A^2_{1}))-4 k^2_{-1}k^2_{1}(4 m \epsilon \rho - 6\rho \hbar^2 + \hbar^2(\rho k^4_{1}+5 A^2_{1}))+ 2 k^4_{-1}(2 \rho(2 m \epsilon+ \nonumber \\ \hbar^2)+\hbar^2 (A^2_{1}+3\rho k^4_{1}))))))-A^2_{-1}A^4_{1}(\hbar^4 g_{1}(5 \hbar^4 k^{10}_{-1}+2\hbar^4 k^9_{-1}k_{1}+16 m^2  \nonumber \\ 
(5\epsilon^2 + ( 15\rho^2 - 4 \rho A^2_{1} + A^4_{1})g^2_{1})k^2_{1}-15 \hbar^4 k^8_{-1}k^2_{1}-8 \hbar^4 k^7_{-1}k^3_{1}-96 m\hbar^2 \nonumber\\
(\rho+A^2_{1}) g_{1}k^4_{1}+4 (10 m \epsilon \hbar^2 - 3\hbar^2) k^6_{1}+5 \hbar^4 K^{10}_{1}+4 \hbar^2 k^5_{-1}k_{1}(4 m \epsilon+ \nonumber \\ 
12 \hbar^2 + 3 \hbar^2 k^4_{1}) +2 k^6_{-1}(20 m \epsilon \hbar^2 + 6\hbar^4 + 5\hbar^4 k^4_{1})-
8 \hbar^2 k^3_{-1}k_{1}(24 m \rho g_{1} + \nonumber \\
4(m \epsilon - 9\hbar^2) k^2_{1}+\hbar^2 k^6_{1})+ k^2_{-1}(16 m^2(5 \epsilon^2( 15\rho^2 - 12 \rho A^2_{1} + A^4_{1})g^2_{1})+ \nonumber \\ 
64 m \hbar^2(5A^2_{1}-9\rho)g_{1}k^2_{1}+20 \hbar^2(9\hbar^2-2 m \epsilon)k^4_{1}-15 \hbar^4 k^8_{1})+ 2k_{1}k_{-1}(16 m^2  \nonumber \\ 
(\epsilon^2+ (3\rho^2 + 8 \rho A^2_{1} - A^4_{1}) g^2_{1})-32 m \hbar^2(3\rho+4A^2_{1})g_{1}k^2_{1}+8 \hbar^2(m \epsilon+3\hbar^2) \nonumber \\ 
k^4_{1}+ \hbar^4 k^8_{1})+ 2\hbar^2 k^4_{-1}(16 m (-3 \rho+A^2_{1})g_{1}+ 5k^2_{1}(-4 m \epsilon + 18 \hbar^2 + \hbar^2 k^4_{1}))) \nonumber \\ 
- 4 g_{0}(48 m^3 \rho^3 g^3_{1} - 32 m^2 \rho \hbar^2 g^2_{1}((-2A^2_{1}+3\rho)k^2_{-1}+ 4 A^2_{1}k_{1}k_{-1}+ (3\rho- \nonumber \\ 
2A^2_{1})k^2_{1})-2 (k^2_{-1}+k^2_{1})(4 m \epsilon +\hbar^2(k^2_{-1}-k^2_{1})^2)^2 + m g_{1}(48 m^2 \epsilon^2 \rho+ \nonumber \\ 
\hbar^2(3 \rho \hbar^2k^8_{-1}+8 \hbar^2(3\rho+4A^2_{1}) k^3_{-1}k_{1}- 12 \rho \hbar^2 k^6_{-1}k^2_{1}+ 
24\rho \hbar^2 k_{-1}k^3_{1}+6 k^4_{-1} \nonumber \\
(2 \rho(2 m \epsilon + \hbar^2) + \hbar^2(3\rho k^4_{1}+2A^2_{1}))-4 k^2_{-1}k^2_{1}(10 \hbar^2 A^2_{1}+3\rho 
(4 m \epsilon -6 \hbar^2 \nonumber \\
+ \hbar^2 k^4_{1}))+ k^4_{1}(-4 \hbar^2 A^2_{1}+3 \rho(8 m \epsilon + 4 \hbar^2 +\hbar^2 k^4_{1}))))))
\label{eq:29}
\end{eqnarray}%
\begin{eqnarray} C^{n=0,1}_{k^{12}, \omega^0}= \frac{\hbar^{12}}{256 m^8 g_{1}(A^2_{-1}+A^2_{1})^2}(4 m \hbar^2 A^6_{-1}g_{-1}
(k_{-1}-k_{1})(g_{0}(3k_{-1}+k_{1})- 2 g_{1}(k_{-1}+\nonumber \\
3k_{1}))+8(1+(-1)^n)m A^5_{-1} A_{1} g_{0}g_{1}(4 m \epsilon +\hbar^2(k^2_{-1}-k^2_{1})^2)+A^3_{-1}A_{1}\nonumber \\
(4 m \epsilon +\hbar^2(k^2_{-1}-k^2_{1})^2)(8 m((-1)^n \rho - (1+(-1)^n) A^2_{-1}) g^2_{1} - 2 (-1)^n\nonumber \\
\hbar^2 g_{1}(5 k^2_{-1}+2k_{-1}k_{1}+ 5k^2_{1})+(-1)^n g_{0}(16 m \rho g_{1}-\hbar^2(5 k^2_{-1}+2k_{-1}k_{1}+ \nonumber \\
5k^2_{1}))) + A^4_{1} g_{1}(16 m^2 \rho^2 g^2_{1}-8 m \hbar^2 g_{1}((5\rho-3A^2_{1}) k^2_{-1} + 2(\rho+A^2_{1})k_{-1}k_{1} +\nonumber \\
(5 \rho+A^2_{1})k^2_{1})+ \hbar^4 (9k^4_{-1}+20k^3_{-1}k_{1}+38 k^2_{-1}k^2_{1}+20k^3_{1}k_{-1} + 9k^4_{1})+ 4 m g_{0}\nonumber \\
(8 m \rho^2 g_{1}+\hbar^2(5(A^2_{1}-\rho)k^2_{-1}-2(3 A^2_{1}+\rho)k_{-1}k_{1}(A^2_{1}-5 \rho)k^2_{1}))) + A_{-1}A^3_{1}\nonumber \\
(4 m \epsilon +\hbar^2(k^2_{-1}- k^2_{1})^2)(- 8 m A^2_{1}g_{1}(g_{0}+g_{1})-(-1)^n (g_{0}(A^2_{1}-2 \rho)g_{1}+ \hbar^2\nonumber \\
(5 k^2_{-1}+2k_{-1}k_{1}+5 k^2_{1}))+2 g_{1}(4 m(A^2_{1}-\rho) g_{1}+ \hbar^2(5 k^2_{-1}+2k_{-1}k_{1}+\nonumber \\
5k^2_{1}))))+A^2_{-1}A^2_{1}(2 g_{0}(32 m^2 \rho^2 g^2_{1}-2 m \hbar^2 g_{1}((\rho-A^2_{1}) k^2_{-1}+2(2\rho+3A^2_{1})\nonumber \\
k_{1}k_{-1}+5(2\rho-A^2_{1})k^2_{1})+(4 m \epsilon +\hbar^2(k^2_{-1}-k^2_{1})^2)^2)+ g_{1}(16 m^2 \epsilon^2+\nonumber \\
32 m^2 \rho^2 g^2_{1}-8 m \hbar^2 g_{1}(5(2\rho-A^2_{1})k^2_{-1}+2(2\rho+ 3A^2_{1})k_{1}k_{-1}+(10\rho- \nonumber \\
A^2_{1}) k^2_{1})+ \hbar^2 (\hbar^2 k^8_{-1} + 40\hbar^2k^3_{-1}k_{1}-4 \hbar^2 k^6_{-1}k^2_{1}- 40\hbar^2 k^3_{1}k_{-1}-4 k^2_{-1}k^2_{1}\nonumber \\
(4 m \epsilon -19 \hbar^2 +\hbar^2 k^4_{1})+ k^4_{1}(8 m \epsilon +18 \hbar^2 +\hbar^2 k^4_{1})+2 k^4_{-1}\nonumber \\
(4 m \epsilon +9 \hbar^2 +3\hbar^2 k^4_{1})))))
\label{eq:30}
\end{eqnarray}%
\begin{eqnarray} C^{n=0,1}_{k^{14}, \omega^0}= \frac{\hbar^{14}}{256 m^8 g_{1}(A^2_{-1}+A^2_{1})}((-1)^n A_{-1}A_{1}
(g_{0}+2 g_{1})(4 m \epsilon +\hbar^2(k^2_{-1}-k^2_{1})^2)+ \nonumber \\
2 (A^2_{-1}+A^2_{1}) g_{1}  (2 m \rho (g_{0}+2 g_{1})- \hbar^2(3 k^2_{-1}+2k_{-1}k_{1}+3 k^2_{1}))))
\label{eq:31}
\end{eqnarray}%
\begin{eqnarray} C^{n=0,1}_{k^{16}, \omega^0}=& \frac{\hbar^{16}}{256 m^8}
\label{eq:32}
\end{eqnarray}%

\clearpage
\newpage
\begin{center}
\textbf{\large Supplemental Material for: 'Continuous-wave solutions and modulational instability in spinor
condensates of positronium'}
\end{center}
\setcounter{equation}{0}
\setcounter{figure}{0}
\setcounter{table}{0}
\setcounter{page}{1}
\makeatletter
\renewcommand{\theequation}{\arabic{equation}}
\renewcommand{\thefigure}{\arabic{figure}}
\renewcommand{\bibnumfmt}[1]{[#1]}
\renewcommand{\citenumfont}[1]{#1}
\section{Supplement 2}
The sixth-order characteristic polynomial in the case of  spinor ortho Positronium is
\begin{eqnarray}
0 = \ (\hbar \omega )^{6}+(\hbar \omega )^{5}C_{k,\omega ^{5}}^{n=0}k+
(\hbar \omega )^{4}\sum_{j=1}^{3}C_{k^{2j-2},\omega ^{4}}^{n=0}k^{2j-2}
+(\hbar \omega )^{3}\sum_{j=1}^{3}C_{k^{2j-1},\omega
^{3}}^{n=0}k^{2j-1}+\nonumber \\
(\hbar \omega )^{2}\sum_{j=1}^{4}C_{k^{2j},\omega
^{2}}^{n=0}k^{2j} + (\hbar \omega )\sum_{j=1}^{4}C_{k^{2j+1},\omega
}^{n=0}k^{2j+1}+\sum_{j=1}^{5}C_{k^{2j+2},\omega ^{0}}^{n=0}k^{2j+2}.
\label{eq:1}
\end{eqnarray}%
The coefficients $C^{n=0}_{k^\alpha, \omega^\beta}$ shown explicitly below
contain the variables like the nonlinearities $g_0$ and $g_1$,
the amplitudes $A_{1}$ and $A_{-1}$ and the wavenumbers of the of the $M=\pm1$ components.
$\hbar$ is the reduced Planck constant and $m$ is the mass of the Positronium atom.
\begin{eqnarray} C^{n=0}_{k, \omega^5} = -\frac{3 \hbar^7}{m}(k_{1}+k_{-1})
\label{eq:2}
\end{eqnarray}%
\begin{eqnarray} C^{n=0}_{k^0, \omega^4} = -\frac{ \hbar^4}{16 m^2 (A_{1}+A_{-1})^2}(8 m A^4_{-1} g_{1}+A^2_{1}(8 m A^2 g_{1}-
\hbar^2(k^2_{-1}-k^2_{1})^2)+ A^2_{-1}\nonumber \\
(48 m A^2_{1} g_{1}-\hbar^2(k^2_{-1}-k^2_{1})^2)+2 A_{-1}A_{1}(16 m A^2_{1}g_{1}+\hbar^2(k_{-1}-k_{1})^2))\nonumber \\
(8 m (A_{-1}+A_{1})^2 g_{1}-\hbar^2(k^2_{-1}-k^2_{1})^2 )
\label{eq:3}
\end{eqnarray}%
\begin{eqnarray} C^{n=0}_{k^2, \omega^4} = -\frac{ \hbar^6}{4 m^2 g_{1} (A_{1}+A_{-1})^2}(-4 m A^4_{-1} g_{1}(g_{0}+2g_{1})-
16 m A^3_{-1}A_{1}g_{1}(g_{0}+2g_{1}) +\nonumber \\
2 A^2_{-1} g_{1}( -12 m A^2_{1}(g_{0}+2g_{1}) + \hbar^2(7 k^2_{-1}+16 k_{-1}k_{1}+ 7 k^2_{1})) + 2 A^2_{1} g_{1}\nonumber \\
(-2 m A^2_{1} (g_{0}+2g_{1})+ \hbar^2(7 k^2_{-1}+16 k_{-1}k_{1}+ 7 k^2_{1})) + A_{1}A_{-1}(-16 m \nonumber \\
A^2_{1}g_{1}(g_{0}+2g_{1}) + \hbar^2(g_{0}(k_{-1}-k_{1})^2 + 3 g_{1}(9 k^2_{-1}+22 k_{-1}k_{1}+ 9 k^2_{1}))))
\label{eq:4}
\end{eqnarray}%
\begin{eqnarray} C^{n=0}_{k^4, \omega^4} = -\frac{3 \hbar^8}{4 m^2}
\label{eq:5}
\end{eqnarray}%
\begin{eqnarray} C^{n=0}_{k, \omega^3} = \frac{\hbar^5}{16 m^3 (A_{1}+A_{-1})^3} (8 m (A_{1}+A_{-1})^2 g_{1} -
\hbar^2(k^2_{-1}-k^2_{1})^2) (8 m (A_{1}+A_{-1})^4 \nonumber \\
(A_{1}+3A_{-1}) g_{1} k_{-1}-\hbar^2 (A_{-1}-A_{1}) (3A^2_{-1}-A^2_{1}) k^3_{-1}+(8 m (A_{-1}+
A_{1})^4\nonumber \\
(A_{-1}+3 A_{1})g_{1}+(5 A^2_{-1} + A^2_{1})k^2_{-1}) k_{1} - \hbar^2(A_{-1}-A_{1})(A^2_{-1} + 5 A^2_{1})\nonumber \\
k_{-1}k^2_{1} - \hbar^2 (A_{-1}-A_{1})(A^2_{-1} - 3 A^2_{1})k^3_{1})
\label{eq:6}
\end{eqnarray}%
\begin{eqnarray} C^{n=0}_{k^3, \omega^3} = \frac{\hbar^7}{4 m^3 g_{1}(A_{1}+A_{-1})^2}(4 m (A_{-1}+A_{1})^3 g_{1}(A_{1}
(3g_{0}+2g_{1})+A_{1}(g_{0}+6 g_{1})) k_{-1}-\nonumber \\
\hbar^2(9 A^2_{1}g_{1} + 7 A^2_{1} g_{1} + 2 A_{1}A_{-1}(g_{0}+7g_{1}) k^3_{-1} + (4 m (A_{-1}+A_{1})^3 g_{1}(A_{1}\nonumber \\
(3 g_{0}+2 g_{1})+ A_{1}(g_{0}+6 g_{1}))-\hbar^2 (-2 A_{1}A_{-1}(g_{0}- 33 g_{1})+ 29 A^2_{-1} g_{1}+ \nonumber \\
 35 A^2_{1} g_{1}) k^2_{-1}) k_{1}- \hbar^2(-2 A_{1}A_{-1}(g_{0}- 33 g_{1})+ 29 A^2_{1} g_{1}+35 A^2_{-1} g_{1})
k^2_{1}) \nonumber \\
k_{-1} - \hbar^2(7 A^2_{-1} g_{1} + 9 A^2_{1} g_{1}+ 2 A_{1}A_{-1} (g_{0}+7 g_{1})) k^3_{1})
\label{eq:7}
\end{eqnarray}%
\begin{eqnarray} C^{n=0}_{k^5, \omega^3}  = \frac{3 \hbar^9}{2 m^3}(k_{-1}+k_{1})
\label{eq:8}
\end{eqnarray}%
\begin{eqnarray} C^{n=0}_{k^2, \omega^2}  = \frac{\hbar^4}{64 m^4 g_{1}(A_{1}+A_{-1})^4}(8 m (A_{-1}+A_{1})^2 g_{1}- \hbar^2
(k_{-1}-k_{1})^2)(32 m^2 A^8_{-1}g_{0} g^2_{1} +  \nonumber \\
256 m^2 A^7_{-1} A_{1} g_{0}g^2_{1}+ A^4_{1}g_{1} (8 m A^2_{1}g_{1}- \hbar^2 (k_{-1}-k_{1})^2) (4 m A^2_{1}g_{0}-\hbar^2 \nonumber \\
( k^2_{-1}+10 k_{-1}k_{1}+13 k^2_{1})) + 4 m A^6_{-1}g_{1}(224 m A^2_{1}g_{0}g_{1} - \hbar^2(g_{0} (k_{-1}- k_{1})^2 \nonumber \\
+ 2 g_{1}(13k^2_{-1}+10 k_{-1}k_{1}+ k^2_{1})))+8 m A^5_{-1}A_{1}g_{1}(224 m A^2_{1} g_{0} g_{1} - \hbar^2(2g_{0} \nonumber \\
(k_{-1}- k_{1})^2 + g_{1} (59 k^2_{-1}+74 k_{-1}k_{1} +11 k^2_{1}))) + A^4_{-1}g_{1}(2240 m^2 A^4_{1}g_{0}g_{1} \nonumber \\
- \hbar^4(k_{-1}-k_{1})^2 (13k^2_{-1}+10 k_{-1}k_{1}+ k^2_{1})- 4 m \hbar^2 A^2_{1}(7 g_{0}(k_{-1}-k_{1})^2 + \nonumber \\
2 g_{1}(107k^2_{-1}+206 k_{-1}k_{1}+47 k^2_{1}) + A^3_{-1}A_{1}(1792 m^2 A^4_{1}g_{0}g^2_{1} + \hbar^4(g_{0}+ \nonumber \\
g_{1})(k_{-1}-k_{1})^4 -16 m \hbar^2 A^2_{1}g_{1}(2 g_{0}(k_{-1}-k_{1})^2 + g_{1}(49 k^2_{-1}+142k_{-1}k_{1}+ \nonumber \\
49k^2_{1})))+A_{-1}A^3_{1}(256 m^2 A^4_{1} g_{0}g^2_{1}+ \hbar^4 (g_{0}+g_{1})(k_{-1}-k_{1})^4 -8 m \hbar^2 A^2_{1}g_{1} \nonumber \\
(2 g_{0}(k_{-1}-k_{1})^2 + g_{1}(11 k^2_{-1}+74 k_{-1}k_{1}+ 59 k^2_{1})))+2 A^2_{-1}A^2_{1}(448 m^2 A^4_{1} \nonumber \\
g_{0}g^2_{1} - \hbar^4 (k_{-1}-k_{1})^2 (g_{0}(k_{-1}-k_{1})^2+ 6 g_{1}(k_{-1}+k_{1})^2)- 2 m \hbar^2 A^2_{1}g_{1} \nonumber \\
(7 g_{0}(k_{-1}-k_{1})^2 + 2 g_{1}(47 k^2_{-1}+206k_{-1}k_{1}+107 k^2_{1}))))
\label{eq:9}
\end{eqnarray}%
\begin{eqnarray} C^{n=0}_{k^4, \omega^2} = \frac{\hbar^6}{32 m^4 g_{1}(A_{1}+A_{-1})^4}(64 m^2 A^8_{-1}g^2_{1}(g_{0}+g_{1})+
512 m^2 A^7_{-1}A_{1}(g_{0}+g_{1})+  \nonumber \\
16 m A^6_{-1}g_{1} (112 m A^6_{1}g_{1}(g_{0}+g_{1})-\hbar^2(2 g_{1}(7 k^2_{-1}+4k_{-1}k_{1}+k^2_{1})))+ \nonumber \\
A^4_{1}g_{1}(64 m^2 A^4_{1}g_{1}(g_{0}+g_{1})+\hbar^4(11k^4_{-1}+124k^3_{-1}k_{1}+234 k^2_{-1}k^2_{1}+ \nonumber \\
76k^3_{1}k_{-1}+35 k^4_{1})- 16 m \hbar^2 A^2_{1}(g_{0}(7 k^2_{-1}+4k_{-1}k_{1}+k^2_{1})+ 2 g_{1}( k^2_{-1}+ \nonumber \\
4 k_{-1}k_{1}+ 7 k^2_{1}))) + A^4_{-1}g_{1}(4480 m^2 A^4_{1} g_{1}(g_{0}+g_{1})+ \hbar^4 (35 k^4_{-1}+ \nonumber \\
76 k^3_{-1}k_{1}+234 k^2_{-1}k^2_{1}+124 k^3_{1}k_{-1}+11 k^4_{1})-16 m \hbar^2 A^2_{1}(5 g_{0}(7 k^2_{-1}+ \nonumber \\
16 k_{-1}k_{1}+ 13 k^2_{1})+ 2 g_{1}(61 k^2_{-1}+ 88 k_{-1}k_{1}+ 31 k^2_{1}))) + 8 m A^5_{-1}A_{1}g_{1} \nonumber \\
( 448 m A^2_{1}g_{1}(g_{0}+g_{1}) - \hbar^2(2 g_{1}(65 k^2_{-1}+62k_{-1}k_{1}+ 17 k^2_{1})+ g_{0} (19 k^2_{-1} \nonumber \\
+58k_{-1}k_{1}+ 67k^2_{1})))+ 4A^2_{-1} A^2_{1}( 448 m^2 A^4_{1}g^2_{1}(g_{0}+g_{1}) - 4 m \hbar^2 A^2_{1}g_{1} \nonumber \\
(5 g_{0}(13 k^2_{-1}+16 k_{-1}k_{1}+ 7 k^2_{1})+2 g_{1}(31 k^2_{-1}+88k_{-1}k_{1}+ 61 k^2_{1}))+ \hbar^4 \nonumber \\
(4g_{0}(k_{-1}-k_{1})^2 ( k^2_{-1}+4k_{-1}k_{1}+k^2_{1}) + g_{1} (23 k^4_{-1}+172 k^3_{-1}k_{1}+330 k^2_{-1}k^2_{1} \nonumber \\
+172 k^3_{1}k_{-1}+ 23 k^4_{1}))) + 2 A^3_{-1}A_{1}(1792 m^2 A^4_{1}g^2_{1}(g_{0}+g_{1})-8 m \hbar^2 A^2_{1}g_{1} \nonumber \\
(5 g_{0}(13 k^2_{-1}+22 k_{-1}k_{1}+13 k^2_{1})+ 2 g_{1} (59 k^2_{-1}+122 k_{-1}k_{1}+ 59 k^2_{1})) + \nonumber \\
\hbar^4( g_{0}(k_{-1}-k_{1})^2 (5 k^2_{-1}+14 k_{-1}k_{1}+ 5 k^2_{1})+g_{1} (47 k^4_{-1}+196 k^3_{-1}k_{1}+  \nonumber \\
450 k^2_{-1}k^2_{1}+244 k^3_{1}k_{-1}+23 k^4_{1})))+ 2 A^3_{1}A_{-1}(256 m^2 A^4_{1} g^2_{1}(g_{0}+g_{1}) - \nonumber \\
4 m \hbar^2 A^2_{1}g_{1}(g_{0}(67 k^2_{-1}+58k_{-1}k_{1}+19 k^2_{1})+ 2 g_{1}(17 k^2_{-1}+62k_{-1}k_{1}+ \nonumber \\
65 k^2_{1})) + \hbar^4 (g_{0} (k_{-1}-k_{1})^2 (5 k^2_{-1} + 14 k_{-1}k_{1}+ 5 k^2_{1})+ g_{1}(23 k^4_{-1}+ \nonumber \\
244 k^3_{-1}k_{1}+450 k^2_{-1}k^2_{1}+196 k^3_{1} k_{-1}+ 47 k^4_{1}))))
\label{eq:10}
\end{eqnarray}%
\begin{eqnarray} C^{n=0}_{k^6, \omega^2} = -\frac{\hbar^8}{8 m^4 g_{1}(A_{1}+A_{-1})^2}(-4 m A^4_{-1}g_{1}(g_{0}+2 g_{1})-
16 m A^3_{-1}A_{1}g_{1}(g_{0}+2 g_{1})+ \nonumber \\
2 A^2_{-1}g_{1}(-12 m A^2_{1}(g_{0}+2g_{1})+ \hbar^2(5 k^2_{-1}+8 k_{-1}k_{1}+5 k^2_{1})) + 2 A^2_{1}g_{1} \nonumber \\
(-2 m A^2_{1}(g_{0}+2g_{1})+ \hbar^2 (5 k^2_{-1}+8 k_{-1}k_{1}+5 k^2_{1})) + A_{1}A_{-1}(-16 m A^2_{1} \nonumber \\
g_{1}(g_{0}+2 g_{1}) + \hbar^2 (g_{0}(k_{-1}-k_{1})^2 + g_{1}(19 k^2_{-1}+34 k_{-1}k_{1}+19 k^2_{1}))))
\label{eq:11}
\end{eqnarray}%
\begin{eqnarray} C^{n=0}_{k^8, \omega^2} = \frac{3 \hbar^{10}}{16 m^4}
\label{eq:12}
\end{eqnarray}%
\begin{eqnarray} C^{n=0}_{k^3, \omega} = -\frac{\hbar^5}{32 m^5 g_{1}(A_{1}+A_{-1})^5}(8 m (A_{1}+A_{-1})^2 g_{1}-
\hbar^2 (k_{-1}-k_{1})^2)(16 m^2 A^9_{-1} g_{0}g^2_{1} \nonumber \\
(k_{-1}+k_{1})+ 144 m^2 A^8_{-1}A_{1}g_{0}g^2_{1}(k_{-1}+k_{1}) + A^5_{1}g_{1}(-8 m A^2_{1}g_{1}+\hbar^2 \nonumber \\
(k_{-1}-k_{1})^2)(k_{-1}+k_{1})(-2 m A^2_{1}g_{0}+\hbar^2 k_{1}(k_{-1}+3 k_{1})) - 2 m A^7_{-1}g_{1} \nonumber \\
(k_{-1}+k_{1})(-288 m A^2_{1} g_{0}g_{1}+ \hbar^2(g_{0}(k_{-1}-k_{1})^2+ 4g_{1}k_{-1}(3k_{-1}+k_{1}))) + \nonumber \\
2 m A^6_{-1}A_{1}g_{1}(672 m A^2_{1}g_{0}g_{1}(k_{-1}+k_{1})- \hbar^2 (g_{0}(k_{-1}-k_{1})^2(9k_{-1}+k_{1})+ \nonumber \\
8 g_{1}k_{-1}(7 k^2_{-1}+15 k_{-1}k_{1}+6 k^2_{1})))+A_{-1}A^4_{1}(144 m^2 A^4_{1}g_{0}g^2_{1}(k_{-1}+k_{1}) + \nonumber \\
\hbar^2 (g_{0}(k_{-1}-k_{1})^2 k_{1}(g_{0} (k_{-1}-k_{1})^2 +2 g_{1} (k^2_{-1}+ k_{-1}k_{1}+2 k^2_{1}))-2 m \hbar^2 \nonumber \\
A^2_{1}g_{1}(g_{0}(k_{-1}-k_{1})^2(k_{-1}+ 9 k_{1}) + 8 g_{1} k_{1}(6 k^2_{-1}+15 k_{-1}k_{1}+ 7 k^2_{1})))+ \nonumber \\
A^5_{-1}g_{1}(2016 m^2 A^4_{1}g_{0}g_{1}(k_{-1}+k_{1}) +\hbar^4 k_{-1}(k_{-1}-k_{1})^2(k_{-1}+k_{1})(3k_{-1}+ \nonumber \\
k_{1})-2 m \hbar^2 A^2_{1}(g_{0}(k_{-1}- k_{1})^2(23 k_{-1}-k_{1}) + 8 g_{1}k_{1}(13 k^3_{-1}+44 k^2_{-1}k_{1}+ \nonumber \\
26 k_{-1}k^2_{1}+ k^3_{1})))-A^4_{-1}A_{1}(2016 m^2 A^4_{1}g_{0}g^2_{1}(k_{-1}+k_{1}) + \hbar^4 k_{-1}(k_{-1}- \nonumber \\
k_{1})^2(g_{0}(k_{-1}-k_{1})^2+ 2g_{1}(2 k^2_{-1}+ k_{-1}k_{1}+ k^2_{1}))-2 m \hbar^2 A^2_{1}g_{1}(g_{0} (k_{-1}- \nonumber \\
k_{1})^2 (23k_{-1}+7k_{1})+4 g_{1}(24 k^3_{-1}+133 k^2_{-1}k_{1}+112 k_{-1}k^2_{1}+ 11 k^3_{1}))) + \nonumber \\
A^2_{-1}A^3_{1}(576 m^2 A^4_{1}g_{0}g^2_{1}(k_{-1}+k_{1})-\hbar^4 (k_{-1}+k_{1})^2 (g_{0}(k_{-1}-k_{1})^2 k_{1} + g_{1} \nonumber \\
(k^3_{-1}+8 k^2_{-1}k_{1}+5 k_{-1}k^2_{1}+ 2 k^3_{1}))-2 m \hbar^2 A^2_{1}g_{1}(-g_{0}(k_{-1}-23 k_{1}) (k_{-1}- \nonumber \\
k_{1})^2 + 8 g_{1}( k^3_{-1}+26 k^2_{-1}k_{1}+44 k_{-1}k^2_{1}+ 13 k^3_{1}))) + A^3_{-1}A^2_{1}(1344 m^2 A^4_{1}g_{0} \nonumber \\
g^2_{1}(k_{-1}+k_{1})-\hbar^4(k_{-1}-k_{1})^2(g_{0}k_{-1}(k_{-1}-k_{1})^2 + g_{1}(2 k^3_{-1}+ 5 k^2_{-1}k_{1}+ \nonumber \\
8 k_{-1}k^2_{1}+ k^3_{1}))-2 m \hbar^2 A^2_{1}g_{1}(g_{0}(k_{-1}-k_{1})^2(7k_{-1}-23k_{1})+4 g_{1}(11 k^3_{-1}+ \nonumber \\
112 k^2_{-1}k_{1}+133 k_{-1}k^2_{1}+ 24 k^3_{1}))))
\label{eq:13}
\end{eqnarray}%
\begin{eqnarray} C^{n=0}_{k^5, \omega} = -\frac{\hbar^7}{64 m^5 g_{1}(A_{1}+A_{-1})^4}(64 m^2 A^8_{-1}g^2_{1}(2g_{0}(k_{-1}+
k_{1})+g_{1}(3k_{-1}+k_{1}))+ \nonumber \\
128 m^2 A^7_{-1}A_{1}g^2_{1}(8 g_{0}(k_{-1}+k_{1})+ g_{1}(11 k_{-1}+5 k_{1}))+A^4_{-1}g_{1}(8960 m^2  \nonumber \\ A^4_{1}g_{1}(g_{0}+g_{1})(k_{-1}+k_{1}) - 16 m \hbar^2 A^2_{1}((31 g_{0}+101 g_{1})k^3_{-1}+(47 g_{0}+ \nonumber \\
185 g_{1}) k^2_{-1}k_{1}+(113g_{0}+167g_{1})k_{-1}k^2_{1}+(49 g_{0}+27 g_{1})k^3_{1})+ \hbar^4(27 k^5_{-1} \nonumber \\
+5 k^4_{-1}k_{1} + 142 k^3_{-1}k^2_{1}+ 170 k^2_{-1}k^3_{1}+ 39 k_{-1}k^4_{1}+ k^5_{1}))+ 16 m A^3_{-1}A_{1} g_{1} \nonumber \\
(-\hbar^2((10g_{0}+63 g_{1}) k^3_{-1}+(8 g_{0}+71 g_{1})k^2_{-1}k_{1}+(50 g_{0}+53 g_{1})k^2_{1}k_{-1}+ \nonumber \\
(28 g_{0}+5 g_{1})k^3_{1}) + 56 m A^2_{1}g_{1}(8 g_{0}(k_{-1}+k_{1})+ g_{1}(9k_{-1}+7k_{1})))+8 A^3_{-1}A_{1} \nonumber \\
(-4 m \hbar^2 A^2_{1} g_{1}(k_{-1}+k_{1})((25 g_{0} + 38 g_{1})k^2_{-1}+6(5 g_{0}+14 g_{1})k_{-1}k_{1}+(25 g_{0} \nonumber \\
+38 g_{1})k^2_{1}) +\hbar^4 k_{-1}(( g_{0}+7 g_{1})k^4_{-1}+12 g_{1}k^3_{-1}k_{1}+2( g_{0}+37 g_{1})k^2_{-1}k^2_{1}+ \nonumber \\
4(-2 g_{0}+19 g_{1}) k_{-1}k^3_{1}+(5 g_{0}+23 g_{1}) k^4_{1})+112 m^2 A^4_{1}g^2_{1}(8 g_{0}(k_{-1}+k_{1})+ \nonumber \\
g_{1}(7k_{-1}+9k_{1})))+ 8 A_{-1}A^3_{1}(-2 m \hbar^2 A^2_{1}g_{1}((28 g_{0}+ 5 g_{1})k^3_{1}+(50 g_{0}+53 g_{1}) \nonumber \\
k_{1}k^2_{-1}+(8 g_{0}+71 g_{1})k^2_{1}k_{-1} + (10 g_{0}+63 g_{1})k^3_{1})+ \hbar^4 k_{1}((5 g_{0}+23 g_{1})k^4_{-1}+ \nonumber \\
4(-2 g_{0}+19 g_{1})k^3_{-1}k_{1}+2(g_{0}+37 g_{1})k^2_{1}k^2_{-1}+12 g_{1}k_{-1}k^3_{1}+(g_{0}+7 g_{1})k^4_{1}) + \nonumber \\
16 m^2 A^4_{1}g^2_{1}(8g_{0}(k_{-1}+k_{1})+g_{1}(5k_{-1}+11k_{1})))+A^4_{1}g_{1}(\hbar^4(k^5_{-1}+39 k^4_{-1}k_{1}+ \nonumber \\
170 k^3_{-1}k^2_{1}+ 142 k^2_{-1}k^3_{1}+5 k_{-1}k^4_{1}+ 27 k^5_{1})+64 m^2 A^4_{1}g_{1} (2g_{0}(k_{-1}+k_{1})+g_{1} \nonumber \\
(k_{-1}+3k_{1}))- 16 m \hbar^2 A^2_{1}(g_{0}(k_{-1}+k_{1})(7k^2_{-1}+k^2_{1})+g_{1}(k_{-1}+3 k_{1})(k^2_{-1}+ \nonumber \\
2k_{-1}k_{1}+ 5 k^2_{1})))+16 m A^6_{-1}g_{1}(56 m A^2_{1}g_{1}(4 g_{0}(k_{-1}+k_{1})+g_{1}(5k_{-1}+3k_{1})- \nonumber \\
\hbar^2 (g_{1}(3k_{-1}+k_{1})(5 k^2_{-1}+2 k_{-1}k_{1}+k^2_{1})+g_{0}(k_{-1}+k_{1})(k^2_{-1}+7k^2_{1})))+ \nonumber \\
8 A^2_{-1}A^2_{1}(-2 m \hbar^2 A^2_{1}g_{1}(49 g_{0}+27 g_{1})k^3_{-1}+(113 g_{0}+167 g_{1})k_{1}k^2_{-1}+(47 g_{0}+ \nonumber \\
185 g_{1})k^2_{1}k_{-1} + (31 g_{0}+101 g_{1})k^3_{1})+112 m^2 A^4_{1}g^2_{1}(4g_{0}(k_{-1}+k_{1})+g_{1}(3k_{-1} \nonumber \\
+5k_{1}))+ \hbar^4 (k_{-1}+k_{1})(8 g_{0}k_{-1}(k_{-1}-k_{1})^2 k_{1}+ g_{1}(3k^4_{-1}+28k^3_{-1}k_{1}+ \nonumber \\
82 k^2_{-1}k^2_{1}+28k^3_{1}k_{-1}+3 k^4_{1})
\label{eq:14}
\end{eqnarray}%
\begin{eqnarray} C^{n=0}_{k^7, \omega} = \frac{\hbar^9}{16 m^5 g_{1}(A_{1}+A_{-1})^2}(-4 m (A_{-1}+A_{1})^3 g_{1}
(A_{1}(3 g_{0}+2 g_{1})+A_{-1}(g_{0}+6 g_{1})) \nonumber \\
k_{-1}+ \hbar^2(9A^2_{-1}g_{1}+7A^2_{1}g_{1}+ 2 A_{-1}A_{1}(g_{0}+7 g_{1}))k^3_{-1}-(4 m (A_{1}+A_{-1})^3 \nonumber \\
g_{1}(A_{-1}(3g_{0}+2 g_{1})+ A_{1}(g_{0}+6 g_{1}))-\hbar^2(-2 A_{-1}A_{1}(g_{0}-17 g_{1})+  \nonumber \\
 13 A^2_{-1}g_{1}+19 A^2_{1}g_{1})k^2_{-1})k_{1} + \hbar^2 (-2 A_{-1}A_{1}(g_{0}-17 g_{1})+19 A^2_{-1}g_{1}+ \nonumber \\
13 A^2_{1}g_{1}) k_{-1}k^2_{1} + \hbar^2(7 A^2_{-1}g_{1}+9 A^2_{1}g_{1}+
 2 A_{-1}A_{1}(g_{0}+7 g_{1}))k^3_{1})
\label{eq:15}
\end{eqnarray}%
\begin{eqnarray} C^{n=0}_{k^9, \omega} = -\frac{3\hbar^{11}}{16 m^5}(k_{1}+k_{-1})
\label{eq:16}
\end{eqnarray}%
\begin{eqnarray} C^{n=0}_{k^4, \omega^0} = \frac{\hbar^{6}}{64 m^6 g_{1}(A_{1}+A_{-1})^6}(8 m (A_{1}+A_{-1})^2 g_{1}-
\hbar^2 (k_{-1}-k_{1})^2)(8 m^2 A^{10}_{-1}g_{0}g^2_{1} \nonumber \\
(k_{-1}+k_{1})^2 + 8 m^2 A^9_{-1}A_{1}g_{0}g^2_{1} (13 k^2_{-1}+14 k_{-1}k_{1}+13 k^2_{1}) + A^6_{1}g_{1} \nonumber \\
(-8 m A^2_{1}g_{1}+ \hbar^2(k_{-1}-k_{1})^2)(k_{-1}+k_{1})^2 (-m A^2_{1}g_{0}+\hbar^2 k^2_{1})+ m A^8_{-1}g_{1} \nonumber \\
(-\hbar^2(8 g_{1}k^2_{-1}+g_{0}(k_{-1}-k_{1})^2)(k_{-1}+k_{1})^2+24 m A^2_{1}g_{0}g_{1}(23 k^2_{-1}+ \nonumber \\
14 k_{-1}k_{1}+23 k^2_{1}))+m A^7_{-1}A_{1}g_{1}(96 m A^2_{1}g_{0}g_{1}(17 k^2_{-1}+6 k_{-1}k_{1}+ 17 k^2_{1}) \nonumber \\
- \hbar^2 (g_{0}(k_{-1}-k_{1})^2 (19 k^2_{-1}+2 k_{-1}k_{1}+3 k^2_{1})+8  g_{1} k^2_{-1}(5 k^2_{-1}+14 k_{-1}k_{1} \nonumber \\
+13k^2_{1}))) +A_{-1}A^5_{1}(8 m^2 A^4_{1}g_{0}g^2_{1}(13 k^2_{-1}+14 k_{-1}k_{1}+ 13 k^2_{1})+\hbar^4(k_{-1}- \nonumber \\
k_{1})^2 k^2_{1}(g_{0}(k_{-1}-k_{1})^2 + g_{1}(3 k^2_{-1}+2k_{-1}k_{1}+3 k^2_{1}))-m \hbar^2 A^2_{1}g_{1}(8 g_{1}k^2_{1} \nonumber \\
(13 k^2_{-1}+14 k_{-1}k_{1}+5 k^2_{1})+ g_{0}(k_{-1}-k_{1})^2 (3 k^2_{-1}+2 k_{-1}k_{1}+19 k^2_{1})))+ \nonumber \\
A^6_{-1}g_{1}(336 m^2 A^4_{1}g_{0}g_{1}(9 k^2_{-1}+2 k_{-1}k_{1}+9 k^2_{1})+ \hbar^2(k^3_{-1}-k_{-1}k^2_{1})^2 - \nonumber \\ 
4 m \hbar^2 A^2_{1}(2 g_{0}(k_{-1}-k_{1})^2(9 k^2_{-1}-2k_{-1}k_{1}+k^2_{1})+ g_{1}(21 k^4_{-1}+76k^3_{-1}k_{1}+ \nonumber \\
122 k^2_{-1}k^2_{1}+4k^3_{1}k_{-1}+k^4_{1}))) + A^5_{-1}A_{1}(336 m^2 A^4_{1}g_{0}g^2_{1}(11 k^2_{-1}+2k_{-1}k_{1}+ \nonumber \\
11k^2_{1})+ \hbar^4 k^2_{-1}(k_{-1}-k_{1})^2 (g_{0}(k_{-1}-k_{1})^2+g_{1}(3 k^2_{-1}+2k_{-1}k_{1}+ 3k^2_{1}))- \nonumber \\ 
m\hbar^2A^2_{1}A^2_{1}g_{1}(g_{0}(k_{-1}- k_{1})^2(117 k^2_{-1}- 50 k_{-1}k_{1}+ 37 k^2_{1})+8 g_{1}(12k^4_{-1}+ \nonumber \\
52k^3_{-1}k_{1}+143 k^2_{-1}k^2_{1}+14k^3_{1}k_{-1}+3k^4_{1})))+A^4_{-1}A^3_{1}g_{1}(\hbar^4 k_{-1}(k_{-1}-2k_{1}) \nonumber \\
(k^2_{-1}-k^2_{1})^2 + 336 m^2 A^4_{1}g_{0}g_{1}(9 k^2_{-1}+2k_{-1}k_{1}+9k^2_{1})-2 m \hbar^2 A^2_{1}( g_{0}(k_{-1}- \nonumber \\
k_{1})^2(47 k^2_{-1}- 34 k_{-1}k_{1}+ 47 k^2_{1})+8 g_{1}(4 k^4_{-1}+19 k^3_{-1}k_{1}+94 k^2_{-1}k^2_{1}+ \nonumber \\
19 k^3_{1}k_{-1}+4 k^4_{1})))+ A^3_{-1}A^3_{1}(96 m^2 A^4_{1} g_{0}g^2_{1}(17 k^2_{-1}+6k_{-1}k_{1}+17 k^2_{1})- \nonumber \\
2 \hbar^4 k_{-1}(k_{-1}-k_{1})^2 k_{1}(g_{0} (k_{-1}-k_{1})^2 +g_{1}(3 k^2_{-1}+2k_{-1}k_{1}+3k^2_{1})) - m \hbar^2 \nonumber \\
A^2_{1}g_{1}(g_{0}(k_{-1}- k_{1})^2(37 k^2_{-1}-50 k_{-1}k_{1}+ 117 k^2_{1})+ 8 g_{1}(3k^4_{-1}+14 k^3_{-1}k_{1}+ \nonumber \\
143 k^2_{-1}k^2_{1} + 52k^3_{1}k_{-1}+12 k^4_{1})))+ A^2_{-1}A^4_{1}g_{1}(-\hbar^4(2k_{-1}-k_{1})k_{1}(k^2_{-1}-k^2_{1})^2 \nonumber \\
+24 m^2 A^4_{1}g_{0}g_{1}(23 k^2_{-1}+14 k_{-1}k_{1}+23 k^2_{1})-4 m \hbar^2 A^2_{1}(2 g_{0}(k_{-1}-k_{1})^2 (k^2_{-1} \nonumber \\
-2 k_{-1}k_{1}+9 k^2_{1})+ g_{1} (k^4_{-1}+4k^3_{-1}k_{1}+122 k^2_{-1}k^2_{1}+76k^3_{1}k_{-1}+21k^4_{1}))))
\label{eq:17}
\end{eqnarray}%
\begin{eqnarray} C^{n=0}_{k^6, \omega^0} = \frac{\hbar^{6}}{256 m^6 g_{1}(A_{1}+A_{-1})^6}(-256 m^3 A^{12}_{-1}g_{0}g^3_{1}
-3072 m^3 A^{13}_{-1}A_{1}g_{0}g^3_{1}-64 m^2 A^{10}_{-1}g^2_{1} \nonumber \\
(264 m A^2_{1} g_{0}g_{1}-\hbar^2((3g_{0}+5 g_{1})k^2_{-1}+2(g_{0}+ g_{1})k_{-1}k_{1}+(3g_{0}+ g_{1})k^2_{1})) \nonumber \\
- 64 m^2 A^{9}_{-1}A_{1}g^2_{1}(880 m A^2_{1}g_{0}g_{1}- \hbar^2((39g_{0}+47 g_{1})k^2_{-1} + 2(g_{0}+ 9 g_{1})k_{1}k_{-1} \nonumber \\
+3(13g_{0}+5 g_{1})k^2_{1}))- 4 m A^8_{-1}g_{1}(31680 m^2 A^4_{1}g_{0}g^2_{1}-16 m \hbar^2 A^2_{1}g_{1}((207g_{0}+ \nonumber \\
197 g_{1})k^2_{-1}+2(-27g_{0}+37 g_{1})k_{1}k_{-1}+(207g_{0}+89 g_{1})k^2_{1})+ \hbar^4((5g_{0}+ \nonumber \\
52 g_{1})k^4_{-1} - 4(g_{0}-8 g_{1})k^3_{-1}k_{1}+2(7g_{0}+20 g_{1})k^2_{-1}k^2_{1}+28 g_{0}k_{-1}k^3_{1}+(21 g_{0} \nonumber \\
+4 g_{1})k^4_{1}))-16 m A^7_{-1}A_{1}g_{1}(12672 m^2 A^4_{1}g_{0}g^2_{1}-16 m \hbar^2 A^2_{1}g_{1}((153g_{0}+121 g_{1}) \nonumber \\
k^2_{-1}+2(-33g_{0}+23 g_{1})k_{-1}k_{1}+(153g_{0}+73 g_{1})k^2_{1})+ \hbar^4((23g_{0}+85 g_{1})k^4_{-1} \nonumber \\
+4(-11 g_{0}+8 g_{1})k^3_{-1}k_{1}+2(43g_{0}+69 g_{1})k^2_{-1}k^2_{1}+ 4(7g_{0}-2 g_{1})k_{-1}k^3_{1}+ \nonumber \\
(35 g_{0}+9 g_{1})k^4_{1}))+ A^6_{-1} g_{1}(-236544 m^3 A^6_{1}g_{0}g^2_{1} + 896 m^2 \hbar^2 A^4_{1}g_{1}((81g_{0}+ \nonumber \\
55 g_{1})k^2_{1}+2(-21 g_{0}+11 g_{1})k_{1}k_{-1}+(81g_{0}+43 g_{1})k^2_{1})- 16 m \hbar^4 A^2_{1}((101g_{0} \nonumber \\
+230 g_{1})k^4_{-1}+4(-49 g_{0}+12 g_{1})k^3_{-1}k_{1}+2(211g_{0}+302 g_{1})k^2_{-1}k^2_{1}+4(3g_{0} \nonumber \\
-8 g_{1})k_{-1}k^3_{1}+(109g_{0}+46 g_{1})k^4_{1})+\hbar^6(21k^6_{-1}-18k^5_{-1}k_{1}+55k^4_{-1}k^2_{1}+ \nonumber \\
116 k^3_{-1}k^3_{1}+83 k^2_{-1}k^4_{1}-2k_{-1}k^5_{1}+k^6_{1}))+ A^6_{1}g_{1}(-256 m^3 A^6_{1}g_{0}g^2_{1}+64 m^2 \nonumber \\ 
\hbar^2 A^4_{1}g_{1}((3g_{0}+ g_{1})k^2_{-1}+2(g_{0}+ g_{1})k_{1}k_{-1}+(3g_{0}+5 g_{1})k^2_{-1})- 4 m \hbar^4 A^2_{1} \nonumber \\
((21g_{0}+4 g_{1})k^4_{-1}+28g_{0}k^3_{-1}k_{1}+2(7g_{0}+20 g_{1})k^2_{-1}k^2_{1}-4 (g_{0}-8 g_{1})k_{-1}k^3_{1} \nonumber \\
+(5g_{0}+52 g_{1})k^4_{1})+ \hbar^6(k^6_{-1}-2k^5_{-1}k_{1}+83k^4_{-1}k^2_{1}+116 k^3_{-1}k^3_{1}+55 k^2_{-1}k^4_{1} \nonumber \\
-18 k_{-1}k^5_{1}+21 k^6_{1}))+ A^5_{-1}A_{1}(-202752 m^3 A^6_{1}g_{0}g^3_{1}+896 m^2 \hbar^2 A^4_{1}g^2_{1}((99g_{0}+ \nonumber \\
59 g_{1})k^2_{-1}+2(-27 g_{0}+13 g_{1})k_{-1}k_{1}+(99 g_{0}+59 g_{1})k^2_{1})-16 m \hbar^4 A^2_{1}g_{1} \nonumber \\
((213g_{0}+335 g_{1})k^4_{-1}+ 4(-91g_{0}+6 g_{1})k^3_{-1}k_{1} + 2 (485 g_{0}+667 g_{1})k^2_{-1}k^2_{1}- \nonumber \\
12(11g_{0}+4 g_{1})k_{-1}k^3_{1} + (209g_{0}+147 g_{1})k^4_{1})+ \hbar^6(g_{0}(k_{-1}-k_{1})^2(17 k^4_{-1}- \nonumber \\
36 k^3_{-1}k_{1}+86 k^2_{-1}k^2_{1}-4k^3_{1}k_{-1}+k^4_{1})+g_{1}(91k^6_{-1}-106k^5_{-1}k_{1}+477k^4_{-1}k^2_{1} \nonumber \\
+532 k^3_{-1}k^3_{1}+549 k^2_{-1}k^4_{1}-10k_{-1}k^5_{1}+3 k^6_{1})))+A^4_{-1}A^2_{1}(-126720 m^3 A^6_{1}g_{0}g^3_{1} \nonumber \\ 
+ 896 m^2 \hbar^2 A^4_{1}g^2_{1}((81 g_{0}+43 g_{1})k^2_{-1}+2(-21g_{0}+11 g_{1})k_{1}k_{-1}+(81g_{0}+ \nonumber \\
55 g_{1})k^2_{1})-8 m \hbar^4 A^2_{1}g_{1}((527g_{0}+572 g_{1})k^4_{-1}-4(167 g_{0}+12 g_{1})k^3_{-1}k_{1}+ \nonumber \\
26(97g_{0}+132 g_{1})k^2_{-1}k^2_{1}-4(167 g_{0}+12 g_{1})k_{-1}k^3_{1}+(257 g_{0}+572 g_{1})k^4_{1})+  \nonumber \\
\hbar^4 (32 g_{0}k^2_{-1}(k_{-1}-k_{1})^2(k^2_{-1}- 2 k_{-1}k_{1}+9 k^2_{1}) +g_{1}(135 k^6_{-1}-158 k^5_{-1}k_{1}+ \nonumber \\
1285k^4_{-1}k^2_{1}+ 1228 k^3_{-1}k^3_{1}+1345 k^2_{-1}k^4_{1}-14 k_{-1}k^5_{1}+19 k^6_{1}))) - 2A^3_{-1}A^3_{1} \nonumber \\
(28160 m^3 A^6_{1} g_{0}g^3_{1}-128 m^2 \hbar^2 A^4_{1}g^2_{1}((153g_{0}+73 g_{1}) k^2_{-1}+2(-33g_{0}+23 g_{1}) \nonumber \\
k_{-1}k_{1} + (153 g_{0}+121 g_{1}) k^2_{1})+8 m \hbar^4 A^2_{1}g_{1}((209g_{0}+147 g_{1})k^4_{-1}- 12 (11g_{0} \nonumber \\
+4 g_{1}) k^3_{-1}k_{1}+2(485 g_{0}+667 g_{1})k^2_{-1}k^2_{1}+4(-91 g_{0}+6 g_{1})k_{-1}k^3_{1}+(213g_{0}+ \nonumber \\
335 g_{1})k^4_{1})- \hbar^6(g_{0}(k_{-1}-k_{1})^2(9k^4_{-1}-20 k^3_{-1}k_{1}+214 k^2_{-1}k^2_{1}-20k^3_{1}k_{-1}+ \nonumber \\
9k^4_{1})+ g_{1}(43 k^6_{-1}-50 k^5_{-1}k_{1}+901 k^4_{-1}k^2_{1}+ 772 k^3_{-1}k^3_{1}+901 k^2_{-1}k^4_{1}- \nonumber \\
50 k_{-1}k^5_{1}+43 k^6_{1}))) + A_{-1}A^5_{1}(-3072 m^3 A^6_{1}g_{0}g^3_{1}+ 64 m^2 \hbar^2 A^4_{1}g^2_{1}(3(13 g_{0}+ \nonumber \\
5 g_{1})k^2_{-1} + 2(g_{0}+9 g_{1})k_{1}k_{-1}+(39 g_{0}+47 g_{1})k^2_{1})-16 m \hbar^4 A^2_{1}g_{1}((35 g_{0}+ \nonumber \\
 9 g_{1})k^4_{-1}+4(7g_{0}-2 g_{1})k^3_{-1}k_{1}+2(43g_{0}+69 g_{1}) k^2_{-1}k^2_{1} + 4(-11 g_{0}+8 g_{1}) \nonumber \\
k_{-1}k^3_{1} + (23g_{0}+ 85 g_{1})k^4_{1})+ \hbar^6 (g_{0}(k_{-1}-k_{1})^2(k^4_{-1}- 4k^3_{-1}k_{1}+86 k^2_{-1}k^2_{1} \nonumber \\
-36k^3_{1}k_{-1}+17 k^4_{1}) + g_{1}(3k^6_{-1}-10k^5_{-1}k_{1}+549k^4_{-1}k^2_{1}+532 k^3_{-1}k^3_{1}+ \nonumber \\
477 k^2_{-1}k^4_{1}-106 k_{-1}k^5_{1}+91 k^6_{1}))) + A^2_{-1}A^2_{1}(-16896 m^3 A^6_{1} g_{0}g^3_{1}+ 64 m^2 \nonumber \\
\hbar^2 A^4_{1} g^2_{1}((207 g_{0}+89 g_{1}) k^2_{-1}+2(-27 g_{0}+37 g_{1})k_{-1}k_{1}+(207 g_{0}+197 g_{1}) \nonumber \\
k^2_{1})-16 m \hbar^4 A^2_{1}g_{1}((109 g_{0}+46 g_{1})k^4_{-1}+ 4 (3g_{0}-8 g_{1})k^3_{-1}k_{1}+2 (211 g_{0}+ \nonumber \\
302 g_{1})k^2_{-1}k^2_{1}+4 (-49g_{0}+12 g_{1})k_{-1}k^3_{1}+(101 g_{0}+230 g_{1}) k^4_{1})+\hbar^6 (32g_{0} \nonumber \\
(k_{-1}-k_{1})^2 k^2_{1}(9 k^2_{-1}-2 k_{-1}k_{1}+k^2_{1})+ g_{1}(19k^6_{-1}-14k^5_{-1}k_{1}+1345 k^4_{-1}k^2_{1} \nonumber \\
+ 1228 k^3_{-1}k^3_{1}+1285 k^2_{-1}k^4_{1}-158 k_{-1}k^5_{1}+ 135 k^6_{1}))))
\label{eq:18}
\end{eqnarray}%
\begin{eqnarray} C^{n=0}_{k^8, \omega^0} = -\frac{\hbar^{8}}{256 m^6 g_{1}(A_{1}+A_{-1})^4}(64 m^2 A^8_{-1}g^2_{1}
(2g_{0}+ g_{1})+512 m^2 A^7_{-1}A_{1} g^2_{1}(2g_{0}+ g_{1}) \nonumber \\
+16 m A^6_{-1}g_{1}(112 m A^2_{1}g_{1}(2 g_{0}+ g_{1})- \hbar^2(2 g_{0}(k^2_{-1}+3 k^2_{1})+g_{1}(11 k^2_{-1}+ \nonumber \\
2 k_{-1}k_{1}+ 3 k^2_{1})))+A^4_{1}g_{1}(64 m^2 A^2_{1}g_{1}(2g_{0}+g_{1}) + \hbar^4(21 k^4_{-1}+ 28 k^3_{-1}k_{1} \nonumber \\
+110 k^2_{-1}k^2_{1}- 4 k^3_{1}k_{-1}+ 37 k^4_{1}) - 16 m \hbar^2 A^2_{1}(2g_{0}(3 k^2_{-1}+k^2_{1})+ g_{1}(3 k^2_{-1} \nonumber \\
+2k_{-1}k_{1}+15 k^2_{1})+ g_{0}(19 k^2_{-1}-6 k_{-1}k_{1}+ 35 k^2_{1}))) + A^4_{-1}g_{1}(4480 m^2  \nonumber \\
A^4_{1}g_{1}(2g_{0}+g_{1})+ \hbar^4(37 k^4_{-1}-4 k^3_{-1}k_{1}+110 k^2_{-1}k^2_{1}+ 28 k^3_{1}k_{-1}+ 21 k^4_{1}) \nonumber \\
- 16 m \hbar^2 A^2_{1}(g_{0}(62k^2_{-1}-24 k_{-1}k_{1}+82 k^2_{1})+g_{1}(141 k^2_{-1} -2 k_{-1}k_{1} + \nonumber \\
101 k^2_{1})))+ 4 A^3_{-1}A_{1}(896 m^2 A^4_{1}g^2_{1}(2g_{0}+g_{1})- 8 m \hbar^2 A^2_{1}g_{1}(2g_{1}(41 k^2_{-1}- \nonumber \\
2 k_{-1}k_{1} + 41 k^2_{1})+g_{0}(49 k^2_{-1}- 18 k_{-1}k_{1}+ 49 k^2_{1}) + \hbar^4 (g_{0}(k_{-1}-k_{1})^2 \nonumber \\
(5 k^2_{-1}- 2 k_{-1}k_{1}+ 5 k^2_{1}) + g_{1}(31k^4_{-1}+4 k^3_{-1}k_{1}+114 k^2_{-1}k^2_{1}+20 k^3_{1}k_{-1}+ \nonumber \\
23 k^4_{1})))+4 A_{-1}A^3_{1}(128 m^2 A^4_{1}g^2_{1}(2g_{0}+g_{1})-4 m \hbar^2 A^2_{1}g_{1}(g_{0}(35 k^2_{-1}+ \nonumber \\
6 k_{-1}k_{1}+ 19 k^2_{1}) + 2 g_{1}(15 k^2_{-1}+2 k_{-1}k_{1}+31 k^2_{1})) +\hbar^4(g_{0}(k_{-1}-k_{1})^2 \nonumber \\ 
(5 k^2_{-1} -2 k_{-1}k_{1}+5 k^2_{1}) + g_{1}(23 k^4_{-1}+20 k^3_{-1}k_{1}+114 k^2_{-1}k^2_{1}+4 k^3_{1}k_{-1} \nonumber \\
+31 k^4_{1})))+ 2 A^2_{-1}A^2_{1}(896 m^2 A^4_{1}g^2_{1}(2 g_{0}+g_{1}) - 8 m \hbar^2 A^2_{1}g_{1}(g_{0}(82 k^2_{-1}- \nonumber \\
24 k_{-1}k_{1}+ 62 k^2_{1})+g_{1}(101 k^2_{-1} -2 k_{-1}k_{1}+ 141 k^2_{1}))+ \hbar^4(16 g_{0}(k_{-1}-k_{1})^2 \nonumber \\
(k^2_{-1}+k^2_{1})+g_{1}(77 k^4_{-1}+44 k^3_{-1}k_{1}+334 k^2_{-1}k^2_{1}+44 k^3_{1}k_{-1}+77 k^4_{1}))))
\label{eq:19}
\end{eqnarray}%
\begin{eqnarray} C^{n=0}_{k^{10}, \omega^0} = \frac{\hbar^{10}}{64 m^6 g_{1}(A_{1}+A_{-1})^2}(-4 m A^4_{-1}g_{1}
(g_{0}+2g_{1})- 16 m A^3_{-1}A_{1}g_{1}(g_{0}+2g_{1})+ \nonumber \\
6 A^2_{-1}g_{1}(-4 m A^2_{1}(g_{0}+2g_{1})+\hbar^2(k^2_{-1}+ k^2_{1}))+2 A^2_{1}g_{1}(-2 m A^2_{1}(g_{0}+ \nonumber \\
2g_{1})+ 3\hbar^2 (k^2_{-1}+k^2_{1}))+A_{-1}A_{1}(-16 m A^2_{1}g_{1}(g_{0}+2g_{1}) +\hbar^2( g_{0} \nonumber \\
(k_{-1}-k_{1})^2 + g_{1}(11 k^2_{-1}+2 k_{-1}k_{1}+ 11 k^2_{1}))))
\label{eq:20}
\end{eqnarray}%
\begin{eqnarray} C^{n=0}_{k^{12}, \omega^0} = -\frac{\hbar^{12}}{64 m^6}
\label{eq:20}
\end{eqnarray}%

\clearpage
\newpage
\begin{center}
\textbf{\large Supplemental Material for: 'Continuous-wave solutions and modulational instability in spinor
condensates of positronium'}
\end{center}
\setcounter{equation}{0}
\setcounter{figure}{0}
\setcounter{table}{0}
\setcounter{page}{1}
\makeatletter
\renewcommand{\theequation}{\arabic{equation}}
\renewcommand{\thefigure}{\arabic{figure}}
\renewcommand{\bibnumfmt}[1]{[#1]}
\renewcommand{\citenumfont}[1]{#1}
\section{Supplement 3}
The fourth-order characteristic polynomial with nil $M=\pm 1$ fields is
\begin{eqnarray}
0 = \ (\hbar \omega )^{4}+(\hbar \omega )^{3}C_{k,\omega ^{3}}k+(\hbar
\omega )^{2}\sum_{j=1}^{3}C_{k^{2j-2},\omega ^{2}}k^{2j-2}\nonumber \\
+(\hbar \omega )\sum_{j=1}^{3}C_{k^{2j-1},\omega
}k^{2j-1}+\sum_{j=1}^{4}C_{k^{2j},\omega ^{0}}k^{2j}  \label{eq:1}
\end{eqnarray}%
The coefficients $C^{n=0}_{k^\alpha, \omega^\beta}$ shown explicitly below
contain the variables like the nonlinearities $g_0$ and $g_1$,
the amplitudes $A_{0}$ and $A_{p}$ and the wavenumbers of the of the constituting components.
$\hbar$ is the reduced Planck constant and $m$ is the mass of the Positronium atom.
\begin{eqnarray}
C_{k, \omega^3} = -\frac{2 \hbar^5}{m}(k_{0}+k_{p})
\label{eq:2}
\end{eqnarray}%
\begin{eqnarray}
C_{k^0, \omega^2} = -16 \hbar^2 A^2_{0}A^2_{p}g^2_{1}
\label{eq:3}
\end{eqnarray}%
\begin{eqnarray}
C_{k^2, \omega^2} = -\frac{\hbar^4}{m^2}(-m (A^2_{0}+A^2_{p})(g_{0}+2 g_{1})+\hbar^2(k^2_{0}+4 k_{0}k_{p}+k^2_{p}))
\label{eq:4}
\end{eqnarray}%
\begin{eqnarray}
C_{k^4, \omega^2} = -\frac{ \hbar^6}{2 m^2}
\label{eq:5}
\end{eqnarray}%
\begin{eqnarray}
C_{k, \omega} = \frac{16 \hbar^3 A^2_{0}A^2_{p}g^2_{1}}{m}(k_{0}+k_{p})
\label{eq:6}
\end{eqnarray}%
\begin{eqnarray}
C_{k^3, \omega} = \frac{2 \hbar^5}{m^3}(-\hbar^2 k_{0}k_{p}(k_{0} + k_{p}) + m A^2_{0}(g_{1}k_{0}+\nonumber \\
(g_{0}+g_{1})k_{p})+ m A^2_{p}(g_{0}k_{0}+ g_{1}(k_{0} + k_{p})))
\label{eq:7}
\end{eqnarray}%
\begin{eqnarray}
C_{k^2, \omega^0} = -\frac{4 \hbar^2 A^2_{0}A^2_{p}g_{1}}{ m^2}(-4 m g_{0}g_{1} (A^2_{0}+ A^2_{p}) +
\hbar^2 (g_{0}(k_{0}- k_{p})^2 + g_{1}(k_{0} + k_{p})^2
\label{eq:8}
\end{eqnarray}%
\begin{eqnarray}
C_{k^4, \omega^0} = -\frac{\hbar^4}{ m^4}(m^2 A^4_{0}g_{1}(g_{0}+g_{1})+ (m A^2_{p}g_{1}-\hbar^2 k^2_{0})
(m A^2_{p}(g_{0}+g_{1})-\hbar^2 k^2_{p}) +\nonumber \\
m A^2_{0}(2 m g_{1} A^2_{p}(3g_{0}+g_{1})-\hbar^2 (g_{1}k^2_{0}+(g_{0}+g_{1})k^2_{p}))
\label{eq:9}
\end{eqnarray}%
\begin{eqnarray}
C_{k^6, \omega^0} = \frac{\hbar^6}{4 m^4}(m(A^2_{0}+A^2_{p})(g_{0}+2 g_{1}- \hbar^2 (k^2_{0}+k^2_{p}))
\label{eq:10}
\end{eqnarray}%
\begin{eqnarray}
C_{k^6, \omega^0} = \frac{\hbar^8}{16 m^4}
\label{eq:11}
\end{eqnarray}%


\end{document}